\documentclass[aps,pra,reprint,superscriptaddress,nofootinbib,notitlepage,onecolumn]{revtex4-2}
\pdfoutput=1 

\usepackage[all]{xy}
\usepackage{graphicx}
\usepackage{amsmath,amsthm,amssymb,amsfonts,setspace}
\usepackage{bm}
\usepackage{mathtools}
\usepackage{braket}
\usepackage{float}
\usepackage{color}
\usepackage[usenames,dvipsnames]{xcolor}
\usepackage{hyperref}  
\usepackage[capitalize,nameinlink,noabbrev]{cleveref}
\usepackage{wrapfig}
\usepackage{tikz}
\usepackage[left=1in,right=1in,top=1in,bottom=1in]{geometry}

\usepackage{mathrsfs}

\hypersetup{colorlinks=true,urlcolor=[rgb]{0,0,0.5},citecolor=[rgb]{0.5,0,0},linkcolor=[rgb]{0,0,0.4}}

\usetikzlibrary{decorations.pathmorphing}

\newtheorem{definition}{Definition}
\newtheorem{theorem}{Theorem}
\newtheorem{lemma}[theorem]{Lemma}
\newtheorem{corollary}[theorem]{Corollary}
\newtheorem{proposition}[theorem]{Proposition}
\newtheorem{fact}[theorem]{Fact}
\newtheorem*{remark}{Remark}
\newtheorem{result}{Result}
\theoremstyle{plain}

\def\ep{\varepsilon}

\def\tr{{\rm tr}}
\def\C{\mathbb{C}}
\def\R{\mathbb{R}}

\def\op{{\cal O}}
\def\p{\partial}

\def\ketbra#1{ |{#1}\rangle\!\langle{#1}| }
\def\iden{\mathbb{I}}

\def\with{\quad {\rm with} \quad}

\def\and{\quad {\rm and} \quad}

\def\ni{\noindent}
\def\nn{\nonumber\\}
\def\ra{\rightarrow}
\def\ie{{\rm i.e.\ }}

\def\CA{{\cal A}}
\def\CB{{\cal B}}
\def\CC{{\cal C}}
\def\CE{{\cal E}}
\def\CH{{\cal H}}

\def\BE{\mathbb{E}}

\def\Ug{\mathbb{U}}
\def\aalpha{a}

\newcommand{\compl}{\mathcal{C}_\ep}
\newcommand{\complhalf}{\mathcal{C}_{\ep/2}}

\newcommand{\Deltat}{\Delta \tau}
\newcommand{\gap}{{\rm gap}}
\newcommand{\Mgates}{M_{{\rm RQC,\Delta t},k}}
\newcommand{\gates}{RQC,\Delta t}

\newcommand{\gdirac}{g_k^\sigma}
\newcommand{\Jdirac}{J_k}
\newcommand{\h}[1]{\boldsymbol{#1}}
\newcommand{\dt}{{\rm d}}

\newcommand{\gset}{\mathcal{G}}
\newcommand{\vol}{{\rm Vol}}
\newcommand{\cupper}{c^o}
\newcommand{\clower}{c_o}

\newcommand{\npack}{{\cal N}_{\rm pack}}
\newcommand{\ncov}{{\cal N}_{\rm cov}}

\newcommand{\ucr}{\mathcal{C}^{>r}_\ep}
\newcommand{\ucre}{\mathcal{C}^{>r}_{\ep}(\nu)}
\newcommand{\scr}{\mathcal{S}^{> r}_\ep}
\newcommand{\scre}{\mathcal{S}^{> r}_{\ep}(\nu_\S)}

\renewcommand{\H}{\mathcal{H}} 
\newcommand{\dg}{\dagger}
\newcommand{\U}{\h{U}} %
\newcommand{\V}{\h{V}} %
\newcommand{\W}{\h{W}} %
\newcommand{\I}{\h{I}} 
\newcommand{\arc}{\mathrm{Arc}} 
\newcommand{\Eig}{\mathrm{Eig}} 

\newcommand{\UU}{\mathcal{U}} 
\renewcommand{\S}{\mathcal{S}} 
\newcommand{\G}{\mathcal{G}} 
\newcommand{\X}{\mathcal{X}} 
\newcommand{\Y}{\mathcal{Y}} 

\newcommand{\avgstate}[1]{M_{#1}^\S}

\newcommand{\ii}{\mathrm{i}} 

\newcommand{\F}{\mathcal{F}} 
\renewcommand{\L}{L}

\newcommand{\x}{\mathbf{x}} 
\newcommand{\y}{\mathbf{y}} 

\newcommand{\Prob}{\mathrm{Pr}} 

\newcommand{\supp}{\mathrm{supp}}
\newcommand{\ot}{\otimes}

\newcommand{\<}{\langle}

\renewcommand{\>}{\rangle}
\newcommand\be{\begin{equation}}
\newcommand\ee{\end{equation}}

\newcommand{\poly}{\mathrm{poly}}
\renewcommand{\d}{\mathrm{D}} 
\renewcommand{\dim}{d} 
\newcommand{\kapgr}{\kappa_{\scalebox{0.7}{$>$}}}
\newcommand{\kapls}{\kappa_{\scalebox{0.7}{$<$}}}

\newcommand{\Grqc}{G}
\newcommand{\nugrqc}{\nu^{(\Grqc)}}
\newcommand{\nugrqct}{\nu^{(\Grqc)}_t}
\newcommand{\nugrqcts}{\nu^{(\Grqc)}_{t,\S}}
\newcommand{\nurqc}{\nu^{(n)}}
\newcommand{\nurqct}{\nu^{(n)}_t}
\newcommand{\nurqcts}{\nu^{(n)}_{t,\S}}
\def\crqc{c_2}
\newcommand{\nuslh}{\nu^{SLH}_t}

\newcommand{\Pp}{\mathbb{P}}
\newcommand{\dilation}{\mathscr S}
\newcommand{\blackdot}{\bullet}
\newcommand{\qv}[1]{\langle #1 \rangle}

\usepackage{bbm}
\newcommand{\indc}{\mathbbm{1}}
\newcommand{\elltwo}{L^2(\UU(\dim))}
\newcommand{\scalar}[2]{\left\langle #1, #2\right\rangle}
\newcommand{\norm}[1]{\left\Vert #1 \right\Vert}
\newcommand{\abs}[1]{\vert #1 \vert}

\begin{document}

\title{Saturation and recurrence of quantum complexity \\ in random local quantum dynamics}

\author{Micha{\l} Oszmaniec}
\email{\!oszmaniec@cft.edu.pl}
\affiliation{Center for Theoretical Physics, Polish Academy of Sciences, Al.\ Lotnik\'ow 32/46, 02-668
Warsaw, Poland}
\affiliation{NASK National Research Institute, Kolska 12, 01-045 Warsaw, Poland}

\author{Marcin Kotowski}
\email{\!mkotowski@cft.edu.pl}
\affiliation{Center for Theoretical Physics, Polish Academy of Sciences, Al.\ Lotnik\'ow 32/46, 02-668
Warsaw, Poland}
\affiliation{NASK National Research Institute, Kolska 12, 01-045 Warsaw, Poland}

\author{Micha{\l} Horodecki}
\email{\!michal.horodecki@ug.edu.pl}
\affiliation{International Centre for Theory of Quantum Technologies, University of Gdansk, Poland}

\author{Nicholas Hunter-Jones}
\email{\!nickrhj@stanford.edu}
\affiliation{Stanford Institute for Theoretical Physics, Stanford, CA 94305, USA\\
Perimeter Institute for Theoretical Physics, Waterloo, ON N2L 2Y5, Canada}

\begin{abstract}
Quantum complexity is a measure of the minimal number of elementary operations required to approximately prepare a given state or unitary channel. Recently, this concept has found applications beyond quantum computing---in studying the dynamics of quantum many-body systems and the long-time properties of AdS black holes. In this context Brown and Susskind \cite{BrownSusskind17} conjectured that the complexity of a chaotic quantum system grows linearly in time up to times exponential in the system size, saturating at a maximal value, and remaining maximally complex until undergoing recurrences at doubly-exponential times. 
In this work we prove the saturation and recurrence of complexity in two models of chaotic time evolutions based on (i) random  local quantum circuits and (ii) stochastic local Hamiltonian evolution. 
Our results advance an understanding of the long-time behaviour of chaotic quantum systems and could shed light on the physics of black hole interiors. From a technical perspective our results are based on establishing new quantitative connections between the Haar measure and high-degree approximate designs, as well as the fact that random quantum circuits of sufficiently high depth converge to approximate designs. 

\end{abstract}

\maketitle

\section{Introduction}\label{sec:intro}
Over the past few years, the notion of complexity has emerged from the confines of theoretical computer science and into the lexicon of modern theoretical physics. Intuitively speaking, the circuit complexity of a quantum state or unitary allows one to distinguish between feasible and intractable computational tasks for quantum computers \cite{Kitaev2002,Watrous2009,aaronson2016complexity}. 
From classifying topological phases of matter \cite{topCOMPLEXITY} to grappling with the dynamics of black hole interiors \cite{SusskindCCBH14,CABH15,susskind2018black,BrownSusskind17,bouland2019computational} and other interacting quantum systems \cite{sykcomplexity,QuenchesCaputa}, 
the quantum complexity of states and unitaries has become a central concept in quantum information and quantum many-body physics. 

The circuit complexity of a state or unitary transformation is defined as the minimal number of elementary operations required to approximately prepare the state from an unentangled initial state, or to implement the unitary to within a set tolerance. Specifically, the elementary operations we consider are a set of universal gates, where each gate only acts on a few qubits.
Despite being a seemingly
innocuous concept, the quantum complexity of a specific state or unitary is extremely difficult to compute \cite{aaronson2016complexity,bouland2019computational}. Naively, it requires one to list
all possible circuits in order to correctly identify the shortest one. 

The complexity of unitary time-evolutions appears connected to some deep ideas in theoretical physics.  The broad interest in understanding of the behaviour of complexity of time-evolving states and unitaries is motivated by the AdS/CFT correspondence, a duality between a theory of quantum gravity in asymptotically Anti-de Sitter space (AdS) in $D$ dimensions and a Conformal Field Theory (CFT) defined on its $(D-1)$-dimensional boundary. Intriguing consequences seem to emerge if this conjectured duality is applied to a specific state on two copies on the CFT \cite{MaldacenaEternal,HMtimeevol}. The holographic picture of this state is a two-sided (or maximally extended) AdS-Schwarzschild black hole. In this geometry, the region behind the horizon of the AdS black hole, a wormhole connecting the two asymptotic regions, appears to grow linearly in time for an exponentially long time, up to a time exponential in the black hole entropy $t\sim e^{S_{\rm BH}}$.
The claim is that the quantum complexity of the dual CFT state is the quantity whose exponentially long linear growth in time describes the growth of the black hole interior \cite{SusskindCCBH14,SusskindEnt14}. More specifically, there have been a number of conjectures for the quantity in the bulk which precisely computes the complexity of the time-evolved thermofield double state of the boundary CFTs. One proposal is that the complexity of the time-evolved state equals spatial volume of region behind the horizon on a maximal time slice of the geometry \cite{SScomp14}. A second proposal conjectures that the action computed on the Wheeler-DeWitt patch, a region of the bulk geometry anchored on the two boundaries, which extends behind the black hole horizon, is equal to the complexity \cite{CABH15}.

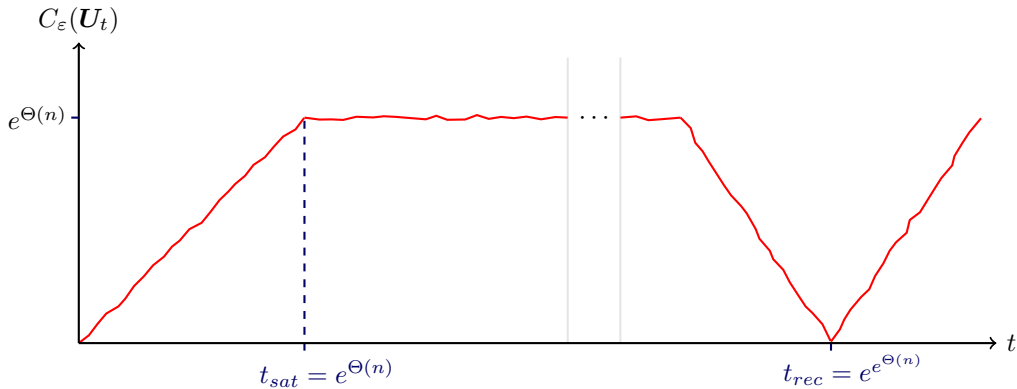
\begin{figure}[t]
    \centering
    \definecolor{db}{rgb}{0,0,0.4}
    \begin{tikzpicture}[scale=1]
    \draw[thick,dashed,color=db] (3,-0.1) -- (3,3);
    \node at (3.3,-0.4) {\textcolor{db}{$t_{sat}=e^{\Theta(n)}$}};
    \node at (10.3,-0.35) {\textcolor{db}{$t_{rec}=e^{e^{\Theta(n)}}$}};
    \draw[thick,dashed,color=db] (10,-0.1) -- (10,0);
    \draw[red,thick, decorate, decoration={random steps,segment length=5pt,amplitude=1pt}] (0,0) -- (3,3);
    \draw[red,thick, decorate, decoration={random steps,segment length=5pt,amplitude=1pt}] (3,3) -- (6.5,3);
    \draw[red,thick, decorate, decoration={random steps,segment length=5pt,amplitude=1pt}] (7.2,3) -- (8,3);
    \draw[red,thick, decorate, decoration={random steps,segment length=5pt,amplitude=1pt}] (8,3) -- (10,0.02);
    \draw[red,thick, decorate, decoration={random steps,segment length=5pt,amplitude=1pt}] (10,0.02) -- (12,3);
    \draw[thick,color=gray,opacity=0.2] (6.5,0) -- (6.5,3.8);
    \draw[thick,color=gray,opacity=0.2] (7.2,0) -- (7.2,3.8);
    \draw[thick,->] (0,0) -- (12.2,0);
    \draw[thick,->] (0,0) -- (0,4);
    \node at (12.4,0) {$t$};
    \node at (0,4.3) {$C_\ep(\h{U}_t)$};
    \node at (6.88,3) {$\cdots$};
    \node at (-0.55,3) {$e^{\Theta(n)}$};
    \draw[thick,dashed,color=db] (-0.1,3) -- (0,3);
    \end{tikzpicture}
    \caption{The conjectured \cite{BrownSusskind17} time evolution of the circuit complexity in an $n$ qubit system. The complexity exhibits a (linear) growth for an exponentially-long time until it saturates at its maximal value in time $t_{sat}=\exp(\Theta(n))$. Afterwards the complexity remains maximal until it undergoes a recurrence at doubly exponential times $t_{rec}= \exp(\exp(\Theta(n)))$. \
    }
    \label{fig:compgrowth}
\end{figure}

More generally in quantum many-body physics, it is  believed that long-time complexity growth is a universal aspect of chaotic systems. Specifically, for unitary evolutions $e^{-iHt}$ generated by a chaotic Hamiltonian $H$, one expects the circuit complexity of the unitary time-evolution operator, as well as that of a state reached by evolving from an initial unentangled state, to grow linearly in time for an exponentially long time.
This behaviour of quantum complexity for a time-independent chaotic evolution has been conjectured by Brown and Susskind \cite{BrownSusskind17}. The quantum complexity of an $n$ qubit state is expected to (i) grow linearly in time for exponential time $t\sim e^{\Theta(n)}$, (ii) saturate at its maximal possible value, and (iii) undergo a recurrence back to trivial complexity at times doubly-exponential in $n$,  $t\sim e^{e^{\Theta(n)}}$; see \cref{fig:compgrowth}. The intuition for the initial growth is that we should not be able to approximate a chaotic many-body evolution by a substantially shorter circuit. But this reasoning needs to break down for deep circuits as the complexity saturates and collisions start to dominate. The evidence for this is the observation that at exponential depths the number of circuits one can generate from a finite set of universal quantum gates is comparable to the volume of the unitary group \cite{BrownSusskind17} (or rather, under the suitable normalization of the Haar measure, the number of pairwise distinct $\ep$-balls in the unitary group). Lastly, recurrence of complexity after double exponential time can be seen as expected consequence of the Poincar\'e recurrence of states and unitaries in isolated quantum systems  \cite{bocchieri57,schulman78,PeresRecurrence1982}.

Proving the Brown-Susskind conjecture for a concrete time-evolution generated by a specific Hamiltonian is unfortunately out of reach using currently available techniques. In the light of this it is interesting to consider its generalization to other models of chaotic quantum evaluations such as \emph{local random quantum circuits} \cite{BHH2016}. Random quantum circuits are known to rapidly scramble quantum information \cite{HaydenPreskill,BrownFawzi2015} and efficiently approximate moments of the Haar measure \cite{HL08,BHH2016,HM18,NHJ19,HHJ20,OSH2020}. Their ability to generate pseudo-randomness is the theoretical basis for recent experimental demonstrations of quantum advantage \cite{Suprem2019}. In the context of strongly-interacting quantum dynamics, random quantum circuits have been utilized to investigate quantum chaos and the spread of entanglement \cite{NahumRuhmanVijayHaah2017EntanglementGrowth,NahumVijayHaah2018OperatorSpreading,vonKeyserlingk2018OperatorHydrodynamics}.

The goal of this work is to provide rigorous results on the long-time behaviour of complexity for random local quantum circuits (RQCs) and stochastic local Hamiltonians (SLH) \cite{Onorati17}.
The latter have been also studied under the name of Brownian quantum circuits (see \cite{FastScrambling}, \cite{brownian-circuits}), also recently in the context of non-rigorous bounds on complexity growth \cite{brownian-complexity}. For the case of RQCs we  consider circuits constructed  from the random application of local 2-site unitaries to a system of $n$ qudits.  In this model the depth (number of gates) of the quantum circuit plays the role of time in the chaotic evolution discussed above.  For the case of SLH we consider quantum evolutions generated by local Hamiltonians that randomly fluctuate in time.

The main results of our work are rigorous proofs of saturation and recurrence of quantum complexity in random unitaries and states generated by the above models. Importantly, our findings hold for both RQC and SLH models irrespectively on their fine details.  We show that the timescales of saturation $t_{sat}$ and recurrence $t_{rec}$ are, respectively, exponential and doubly-exponential in number of qudits $n$ (size of the system), with an essentially optimal dependence on the accuracy parameter $\ep$. In the saturation regime, we can control the fluctuations of complexity by proving that it is exponentially unlikely that complexity is subexponential. Furthermore, we prove that in RQCs the complexity recurrences last an exponential amount of time.  This supports the conjectured connection \cite{BHexpTIME2020} between the evolution of complexity in chaotic systems and the Penrose diagram (tracking the evolution of an AdS wormhole) and points towards even deeper connections between complexity recurrences and regimes in which quasi-classical descriptions of spacetime can be used. 

Previous work has made some progress in understanding the evolution of complexity in random quantum circuits. Specifically in Ref.~\cite{complexitygrowth2019}, by proving bounds on the complexity of structured ensembles called unitary designs, and using the known convergence of random quantum circuits to approximate designs, they were able to lower bound the complexity of random circuits. Notably, they used a strong definition of circuit complexity, which, for quantum states, captures the difficulty of implementing a measurement to distinguish a given quantum state from the maximally mixed state.\footnote{We note that this definition of complexity necessarily assumes dichotomic measurements. If this condition is relaxed it can be shown \cite{avDIST2021} that computational basis measurements preceded by local random circuits of depth $O(n)$ suffice to efficiently distinguish any pure state from maximally mixed state.} More recently, Refs.~\cite{exactcomplexity2021,LiExact2022} used algebraic-geometric techniques to prove a long-time linear growth of the complexity of random circuits for qubit systems, but only for the \emph{exact complexity}, i.e.\ demanding that the target unitary be implemented exactly by quantum circuits capable of implementing arbitrary two qubit gates. Note that this notion of complexity is very stringent and of limited physical relevance as in an arbitrarily small neighborhood of the identity there exist unitary transformations of maximal complexity.\footnote{Indeed, general dimension counting arguments show that \emph{almost all} (with respect to the Haar measure) unitary circuits have maximal exact complexity (scaling as $e^{\Theta(n)}$). As a result, in an arbitrarily small neighborhood of the computationally-trivial identity channel there exist unitary transformations of that maximal exact complexity.} The notion of approximate complexity adopted by us does not exhibit this problem and assigns trivial (zero) complexity to the $\ep$ ball around the identity. 

There are other approaches to proving lower bounds on the complexity of unitary transformations. First, the seminal work of Nielsen and others \cite{NielsenCompGeo1,NielsenCompGeo2} connected complexity to the geodesic distance in a suitably-defined sub-Riemannian geometry. This approach however cannot be used to directly compute (or estimate) complexity due to the immense difficulty of the optimization problem of finding the shortest geodesic. Note, however, that in Refs.~\cite{FermCompl2018,BosComplexity2018} Nielsen's geometric approach was used to compute complexity in free fermionic and bosonic theories. Furthermore, Ref.~\cite{Eisert2021} used this perspective to lower bound (exact) complexity in terms of the entangling power of quantum gates. Additionally, complexity-theoretic assumptions (such as PSPACE $\neq$ BQP) can be used to show that the complexity of specific Hamiltonian evolutions must be super-polynomial at exponentially long times, see e.g. Refs.~\cite{aaronson2016complexity,FastForwarding2017,susskind2018black,bohdanowicz2017universal}.

In contrast to these works, our proof techniques crucially depend on a novel property called {\it approximate equidistribution}, which we introduce for probability measures on the space of unitaries and the space of pure quantum states. The property captures an ensemble's ability to approximate the Haar measure on small discretized scales.
We prove that ensembles of unitaries exhibiting this property are characterized by maximal (exponential) complexity. Furthermore, we show that realizations of random circuits resulting from iterations of equidistributed ensembles typically give recurrences of complexity after doubly-exponential time. Importantly, we use approximate complexity, a notion with clear physical significance. In order to make practical use of these results, we prove that measures exhibiting a spectral gap satisfy the approximate equidistribution property. Together with convergence bounds implying an exponential spectral gap \cite{BHH2016}, this allows us to establish (exponential) upper bounds on equidistribution times for local random quantum circuits and stochastic local Hamiltonians. A careful application of the approximate equidistribution property then allows us to prove that typical recurrences of complexity have exponential duration.


We emphasize that our results hold for very general classes of local random quantum evolutions (discrete or continuous time), which indicates the universality of quantum complexity saturation and recurrences. Many of the results in the literature regarding, for instance, approximate designs, decoupling, and scrambling in random quantum circuits are tailored to specific geometries (often 1D or a complete graph) with Haar-random 2-site unitary gates. Our results apply to local random quantum circuits with any arrangement of qudits defined on a graph $G$ with a Hamiltonian path and with gates chosen randomly from any universal gate set $\G$ (without requiring inverses, as in \cite{BHH2016}).
We can prove the same results for the continuous time stochastic local Hamiltonian (SLH) model \cite{Onorati17}, which further underscores the versatility of our tools and the universality of saturation and recurrence phenomenon. Finally, our methods also deliver insight into the scale of typical fluctuations of complexity in the saturation regime, thus providing a more finely detailed picture than previously available.

{\bf Organization of work}
After fixing notation and presenting some definitions, in \cref{sec:overview} we give an overview of our results regarding late-time complexity, including the novel notion of approximate equidistribution, which we define. We also present a list of possible directions for future work. In \cref{sec:technical} we introduce the technical tools that we require in our proofs, including volumes in the space of unitary channels and states as well as spectral gaps and approximate unitary designs. We also formally define the RQC and SLH models and state their main properties important for the proofs. In \cref{sec:approxEQUIfromDESINGS} we relate approximate equidistribution and spectral gaps, from which it follows that deep random circuits constitute approximately equidistributed distributions. In \cref{sec:MaxcompApproxequi} we prove that typical unitaries and states picked at random from  equidistributed distributions have (essentially) maximal complexity. In \cref{sec:CompRec}, we show that if random quantum evolutions equidistribute at some time scale $\tau$, then their complexity undergoes a recurrence at doubly exponential times. Building on previous sections, we use this to prove that complexity of random quantum circuits or SLH experiences a recurrence to trivial complexity at (and not before) doubly-exponential times. In \cref{sec:recurrTIME} we prove that with high probability, the duration of recurrences dips is exponential in the number of qubits. Finally, in \cref{sec:lineargrowth} we prove a linear complexity growth for exponentially deep circuits. In \cref{app:equidist} we generalize results of \cite{OSH2020} and connect approximate unitary designs (expanders) with approximate equidistribution property for states and unitaries 
 - this allows us to establish our results for RQC without requiring inverses in $\mathcal{G} $. In \cref{app:technicalapp} and \cref{app:rqcproofs} and  we  present proofs of some additional technical results used in the main body of the work. Finally, \cref{app:slh} contains technical statements and proofs necessary to extend results for random quantum circuits to the SLH model.

\subsection*{Setting}

We consider systems of $n$ qudits, with local dimension $q$. The dimension of the corresponding Hilbert space $\H=(\C^q)^{\ot n}$ is then $\dim=q^n$. Every unitary operator $U$ on $\H$ defines a unitary channel $\U$ via conjugation $\U(\rho)=U\rho U^\dg$. Note that all unitary operators $\exp(\ii \varphi) U$, which differ from $U$ by a global phase, define the same unitary channel $\U$. We denote the set of quantum channels on $\H$ by $\UU(\dim)$. Similarly, let $\S(\dim)$ denote the set of pure quantum states in $\H$, i.e.\ rank-one projectors of the form $\psi=\ketbra{\psi}$. We choose to work with the following definitions of distance between unitary channels and quantum states
\begin{equation}\label{eq:distancesDEF}
    \d(\U,\V)= \min_{\varphi\in[0,2\pi)} \|U-\exp(\ii \varphi) V \| _\infty \qquad{\rm and}\qquad \d(\rho,\sigma)=\frac{1}{2}\|\rho-\sigma\|_1\,,
\end{equation}
where $\|\cdot\|_\infty$ and $\|\cdot \|_1$ are the operator norm and 1-norm, respectively. These distances describe optimal statistical distinguishability of unitary channels and states. For quantum states this follows from the operational definition of trace distance while for unitaries it follows from the relation between the operator norm and the diamond norm
\begin{equation}\label{eq:equivDISTANCE}
\d(\U,\V) \leq \|\U-\V\|_\diamond \leq 2\, \d(\U,\V)\,,
\end{equation}
for a proof of this see, for instance, Ref.~\cite{OSH2020}.

We consider circuits formed from a universal discrete gate set $\gset$ of size $|\gset|=\mathrm{poly}(n)$. Assume that each gate in $\gset$ is $\ell$-local, i.e.\ acts non-trivially on at most $\ell$ qudits, where $\ell=O(1)$. Moreover, we will make the assumption that the gates in $\gset$ are universal for any subset of qudits of size $\ell$. Let $\gset^r$ ($r=0,1,2,\ldots$) be the set of all size $r$ circuits built from $\gset$. We can now define the complexity of a unitary and state.

\begin{definition}[Unitary complexity]\label{def:complexityUnitary}
For $\ep \in [0,1]$, we say that a unitary $\U$ has $\ep$-unitary complexity equal to $r$ if and only if $r=\min \left\{l: \exists \V\in\gset^l ~\text{s.t.}~ \d(\U,\V)\leq \ep\right\}$, which we denote as $C_\ep(\U)=r$. 
\end{definition}

\begin{definition}[State complexity]\label{def:complexityState}
For $\ep \in [0,1]$, we say that a pure state $\psi$ has $\ep$-state complexity equal to $r$ if and only if $r=\min \left\{l: \exists \V\in\gset^l ~\text{s.t.}~ \d(\psi,\V(\psi_0))\leq\ep \right\}$, which we denote as $C_\ep(\psi)=r$.
\end{definition}

In the definition of state complexity we assume that $\psi_0$ is chosen to be an unentangled product state, for instance $\psi_0=\ketbra{0}^{\otimes n}$.
Let $\CC^r_\ep=\{\U:  \exists \V\in\gset^l,\ l\leq r\ ~\text{s.t.}~ \d(\U,\V) \leq \ep\}$ denote the set of unitary channels with complexity at \emph{at most} $r$, \ie unitaries that can be approximated to an accuracy $\ep$ by circuits of size at most $r$. Likewise let $\S^r_\ep=\left\{\psi:\ \exists  \V\in\gset^l,\ l\leq r\ ~\text{s.t.}~ \d(\psi,\V(\psi_0)) \leq \ep\right\}$ be the set of states that can be $\ep$-approximated by the states generated by circuits of size at most $r$ acting on an initial state $\psi_0$. 

In this work we are interested in understanding the complexity of states and unitaries generated by random quantum circuits. Specifically, let $\nu$ be a probability measure on $\UU(\dim)$ that corresponds to one layer of a random quantum circuit, e.g.\ a single time step. Then by $\nu_t$ and $(\nu_{t})_\S$ we denote measures describing the distributions of unitary transformations and quantum states realized by depth $t$ random circuit:
\begin{equation}\label{eq:ranomUNITARIESandCIRCUITS}
    \h{U}_t=\h{V}^{(t)} \h{V}^{(t-1)} \ldots \h{V}^{(1)} \ \ ,\ \ \psi_{t}=\h{V}^{(t)} \h{V}^{(t-1)} \ldots \h{V}^{(1)}(\psi_0)\,,
\end{equation}
where unitaries $\h{V}^{(i)}$ are i.i.d.\ and distributed according to measure $\nu$.  Later we shall consider two versions of $\nu$: one corresponding to local circuits in one dimension, which will be denoted by $\nurqc$ and a more general one, corresponding to circuits on connected graphs $G$, denoted by $\nugrqc$. We shall also denote $(\nu_t)_\S$ by $\nu_{t,\S}$, hence also $(\nurqct)_\S$ by $\nurqcts$, and $(\nugrqcts)_\S$ by $\nugrqcts$. 

In addition to measures coming from random quantum circuits, we will also consider analogous measures coming from a continuous time random walk on the unitary group, namely the SLH (Stochastic Local Hamiltonian) model introduced in \cite{Onorati17} (see \cref{sec:technical} for a precise definition). We will denote the resulting measure by $\nuslh$. Informally, the model can be visualized as follows. Divide time $t$ into small time steps of size $\Delta t$. In each time step, choose independently a random Hamiltonian consisting of {\it local} interaction terms with random coefficients and evolve the system with this Hamiltonian for time $\Delta t$. Finally, take the limit $\Delta t \to 0$ to obtain a continuous process. In a sense, the model is a continuous time version of the discrete random quantum circuit, as each time step can be thought of as analogous to applying local gates. Alternatively, one can think of a time evolution governed by a time-dependent fluctuating Hamiltonian governed by a stochastic process (white noise).

\section{Overview of results}\label{sec:overview} 

\begin{figure}[h!]
    \centering
    \definecolor{db}{rgb}{0,0,0.4}
    \definecolor{dp}{rgb}{0.28, 0.24, 0.55}
    \begin{tikzpicture}[scale=1]
    \draw[thick,dashed,color=db] (3,-0.1) -- (3,3);
    \node[align=center] at (3.3,-0.64) {$t_{sat}\approx e^{\Theta(n)}\log(1/\ep)$\\[2pt]
    {\small (\cref{res:equidistr-recurrence} + \cref{res:designs-equidistribution})}};
    \node[align=center] at (10.3,-0.58) {~~$t_{rec}\approx (1/\ep)^{e^{\Theta(n)}}$\\[2pt]
    {\small (\cref{res:equidistr-recurrence} + \cref{res:designs-equidistribution})}};
    \draw[thick,dashed,color=db] (10,-0.1) -- (10,0);
    \draw[pink,line width=1pt, decorate, decoration={random steps,segment length=5pt,amplitude=1pt}] (0,0) -- (2.3,2.3);
    \draw[dp,line width=1pt, decorate, decoration={random steps,segment length=5pt,amplitude=1pt}] (2.3,2.3) -- (3,3) -- (6.5,3);
    \draw[dp,line width=1pt, decorate, decoration={random steps,segment length=5pt,amplitude=1pt}] (7.2,3) -- (8,3) -- (10,0.02) -- (12,3) -- (12.4,3);
    \draw[thick,color=gray,opacity=0.2] (6.5,0) -- (6.5,3.8);
    \draw[thick,color=gray,opacity=0.2] (7.2,0) -- (7.2,3.8);
    \draw[thick,->] (0,0) -- (12.6,0);
    \draw[thick,->] (0,0) -- (0,4);
    \node at (12.8,0) {$t$};
    \node at (0,4.3) {$C_\ep(\h{U}_t)$};
    \node at (6.88,3) {$\cdots$};
    \node[align=center] at (-1.24,3) {$e^{\Theta(n)}\log(1/\ep)$\\[2pt]  {\small (\cref{res:maxcompl})}};
    \draw[thick,dashed,color=db] (-0.1,3) -- (0,3);
    \draw[gray,<->] (8.15,3) -- (11.85,3);
    \node[fill=white] at (10,3.1) {$e^{\Theta(n)}\log(1/\ep)$};
    \node at (10,2.65) {\small (\cref{res:recwidth})};
    \node[align=center,rotate=45] at (1.9,2.7) {$C_\ep \approx \Theta(t)$\\[2pt] {\small (\cref{res:lingrowth})}};
    \end{tikzpicture}
    \caption{Diagram depicting our results for the circuit complexity $C_\ep$ of depth-$t$ local random quantum circuits and unitaries generated by the SLH model after time $t$  acting on $n$ qudits. For typical realizations of random quantum circuits, complexity grows until an exponential time, at which point it saturates to its maximal value at time $t_{sat}\approx \exp(\Theta(n))\log(1/\ep)$. Thereafter, the complexity remains maximal until it undergoes a recurrence at doubly exponential times $t_{rec}\approx (1/\ep)^{\exp(\Theta(n))}$. The recurrence process typically takes exponential amount of time.
    }
    \label{fig:compgrowthresults}
\end{figure}
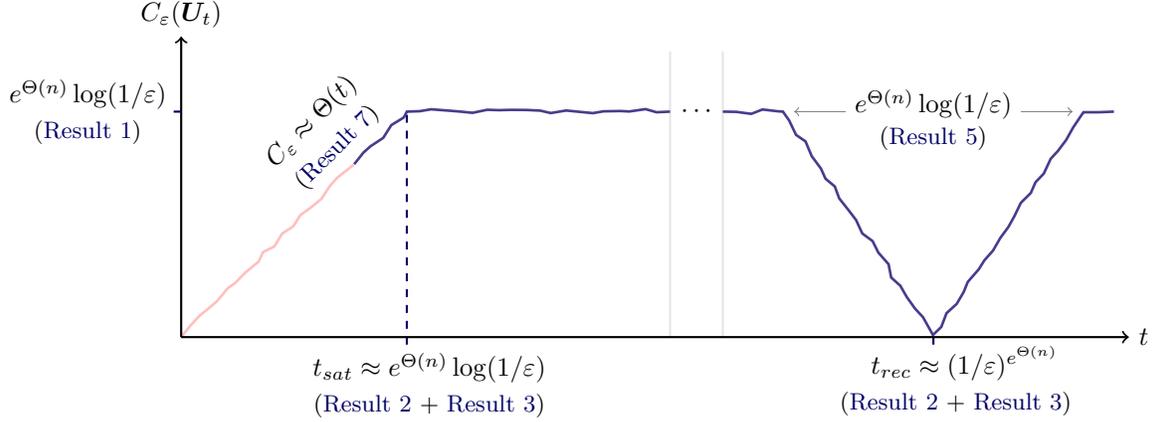
 
The summary of our contributions to late-time behaviour of complexity in random quantum circuits and unitaries generated by stochastic local Hamiltonains is presented in \cref{fig:compgrowthresults}. Our findings crucially depend on a new (to the best of our knowledge) property of probability measures on sets of unitary channels and pure quantum states, which we call the approximate equidistribution property. The approximate equidistribution property is central for establishing the saturation and recurrence of complexity for states and unitary channels.

\begin{figure}[t]
    \centering
    \begin{tikzpicture}[scale=0.42,baseline=-1mm]
    \shade[ball color = red!40, opacity = 0.2] (0,0) circle (2cm);
    \draw (0,0) circle (2cm);
    \node at (-0.4,0) {\small $V$};
    \draw (-2,0) arc (180:360:2 and 0.6);
    \draw[dashed] (2,0) arc (0:180:2 and 0.6);
    \fill[fill=black] (0,0) circle (2pt);
    \draw[dashed] (0,0) -- (0.3,0);
    \draw[dashed] (1.4,0) -- (2,0);
    \node at (0.84,0.05) {\footnotesize $\alpha R$};
    \node at (0.1,-2.64) {\small $\vol(\alpha\!\cdot\! R)$};
    \end{tikzpicture}
    ~$\leq$~
    \begin{tikzpicture}[scale=0.48,baseline=-1mm]
    \shade[ball color=blue!40,opacity=0.2] (0,0) circle (2cm);
    \draw (0,0) circle (2);
    \draw (-2,0) arc (180:360:2 and 0.6);
    \draw[dashed] (2,0) arc (0:180:2 and 0.6);
    \fill[fill=black] (0,0) circle (2pt);
    \node at (-0.4,0) {\small $V$};
    \draw[dashed] (0,0) -- (0.68,0);
    \draw[dashed] (1.4,0) -- (2,0);
    \node at (1,0.05) {{\small $R$}};
    \node at (0.1,-2.64) {{\small $\nu(B(\V,R))$}};
    \end{tikzpicture}
    ~$\leq$~
    \begin{tikzpicture}[scale=0.54,baseline=-1mm]
    \shade[ball color=black!20,opacity=0.2] (0,0) circle (2cm);
    \draw (0,0) circle (2);
    \node at (-0.4,0) {\small $V$};
    \draw (-2,0) arc (180:360:2 and 0.6);
    \draw[dashed] (2,0) arc (0:180:2 and 0.6);
    \fill[fill=black] (0,0) circle (2pt);
    \draw[dashed] (0,0) --  (0.6,0);
    \draw[dashed] (1.5,0) --  (2,0);
    \node at (1,0) {{\small $\beta R$}};
    \node at (0.1,-2.64) {{\small \rm Vol$(\beta\!\cdot\! R)$}};
    \end{tikzpicture}
    \qquad\quad
    \begin{minipage}{5cm}
    {\small
    For every ball $B(\h{V},R)$\\ with $R\geq \ep$ we have
    \begin{equation*}
    \vol(\alpha\cdot R) \leq \nu(B(\h{V},R)) \leq \vol(\beta\cdot R)
    \end{equation*}
    }
    \end{minipage}
    \caption{Diagram depicting approximate equidistribution. For every ball $B(\h{V},R)$ of radius $R$ in the space of unitaries, we have that the measure on that ball is upper and lower bounded by a comparable Haar volume.}
    \label{fig:equipart}
\end{figure}
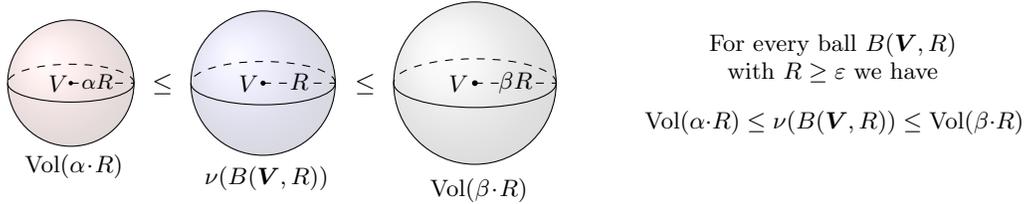

\begin{definition}[Approximate equidistribution on unitary channels and pure states]\label{def:equidistr}
Let $\alpha,\beta$ be real numbers such that $0<\alpha\leq 1 \leq \beta$. We say that a probability measure $\nu$ on $\UU(\dim)$ is $(\alpha,\beta)$-equidistributed on $\UU(\dim)$ at the scale $\ep$ if, for every ball $B(\V,R )\subset\UU(\dim)$ of radius $R\geq\ep$, we have
\begin{equation}\label{eq:equidistrUnitary}
    \vol(\alpha\cdot R)\leq \nu(B(\V,R)) \leq  \vol(\beta\cdot R)\,.
\end{equation}
Similarly, a probability measure $\nu_\S$ on pure states $\S(\dim)$ is $(\alpha,\beta)$-equidistributed on $\S(\dim)$ at the scale $\ep$ if, for every ball $B(\psi,R)\subset\S(\dim)$ of radius $R\geq\ep$, we have
\begin{equation}\label{eq:equidistrState}
    \vol_\S(\alpha\cdot R))\leq \nu_\S(B(\psi,R)) \leq  \vol_\S(\beta\cdot R)\,.
\end{equation}
\end{definition}
Informally, a measure $\nu$ satisfies this approximate equidistribution property on a given scale if the value of the measure it assigns the to every ball of size $\ep$ is comparable to the Haar volume of the of balls of similar size, e.g.\ see \cref{fig:equipart}. In the informal presentation of our results below we will use just the term {\it equidistribution}, with $\alpha,\beta$ and $\ep$ implicit.

Our first result, proven in \cref{lem:complexity-equidist}, is that the complexity of typical states and unitary channels chosen from an ensemble that equidistributes are close to maximal with overwhelming probability.

\begin{result}[Typical unitary channels and states generated by ensembles exhibiting approximate equidistribution have almost maximal complexity]\label{res:maxcompl}
Let $\nu$ and $\nu_\S$ be probability measures on $\UU(\dim)$ and $\S(\dim)$ which equidistribute on their respective spaces. Then, with high probability, unitary channels $\h{U}$ and states $\psi$ selected at random from $\nu$ and $\nu_\S$ satisfy 
\begin{equation}\label{eq:genLOWER}
    C_\ep(\h{U})\gtrsim \dim^2 \log\left(\frac{1}{\ep }\right) \quad\text{and}\quad\ C_\ep(\psi)\gtrsim \dim\log\left(\frac{1}{\ep }\right)\,.
\end{equation}
 Both lower bounds appearing in  Eq.~\eqref{eq:genLOWER}  saturate (up to terms polynomial in $n$) the known \emph{upper} bounds on complexities of unitary states and unitaries in $n$ qudit systems (see \cref{lem:ComplexityUpperBound}).  
\end{result}

Another important property of equidistributed ensembles of unitaries is that they naturally yield exponential saturation times and doubly-exponential recurrence times for complexity. This is proven in \cref{th:recurrence} in \cref{sec:CompRec} and summarized below. 

\begin{result}[From approximate equidistribution to saturation in exponential time and recurrence in doubly-exponential time]\label{res:equidistr-recurrence}
Let $\nu$ be a probability measure on $\UU(\dim)$ describing a single layer of a random quantum circuit, i.e.\ a single time step. Let $\tau$ be a circuit depth such that the distribution of depth $\tau$ random quantum circuits $\h{U}_\tau$, generated by $\nu$, is equidistributed. Then, with high probability, sequences of random unitary circuits $\left(\h{U}_t\right)_{t\in \mathbb{N}}$: 
\begin{itemize}
    \item[(i)]  Achieve complexity $C_\ep(\h{U}_t)\gtrsim d^2 \log(1/\ep)$ simultaneously for all times from $t=\tau$ to $t=\tau + T$, where $T \approx \left(\frac{1}{\ep }\right)^{\dim^2}$ 
    \item[(ii)] Have a recurrence time $t_{\mathrm{rec}}$---the first time after reaching maximal complexity as in (i) that the complexity exhibits a recurrence, i.e.\ $C_\ep(\h{U}_{t_\mathrm{rec}})=O(1)$---which satisfies
    \begin{equation}\label{eq:recurINEQUALITY}
   \left(\frac{1}{\ep }\right)^{\dim^2}  \lesssim    t_{\mathrm{rec}}\lesssim  \tau \left(\frac{1}{\ep }\right)^{\dim^2}\,
\end{equation}
\end{itemize}
Noting that  $\dim=q^n$, we observe that the complexity of random circuits undergoes recurrence at a circuit depth $e^{e^{\Theta(n)}}$ in the multiples $\tau$. An analogous result holds for sequences of states $\left(\psi_t\right)_{t\in \mathbb{N}}$ generated by $\nu$, 
the only difference being the maximal complexity and dimension of the space  proportional to $\dim$, not $\dim^2$.
\end{result}

In fact, we prove that in item (i) it is exponentially unlikely that complexity falls to a subexponential value anytime in the saturation regime, thus providing fine control over complexity fluctuations.




In order to make practical use of the above results we upper bound equidistribution times $\tau$ for local random quantum circuits by establishing a connection between spectral gaps and approximate equidistribution for unitary channels and pure quantum states. The result below is formally proved in \cref{th:equid-gap}. Here we make the equidistribution parameter $\Gamma$ explicit to highlight that a better spectral gap implies better equidistribution.


\begin{result}[Distributions with spectral gap satisfy approximate equidistribution]\label{res:designs-equidistribution}
Let $\nu$ be a measure with spectral gap $1-\delta$, where $\delta$ satisfies: 
\begin{equation}
    \delta \lesssim \left(\ep\Gamma\right)^{\dim^2 }\ 
\end{equation}
Then the distribution $\nu$ is $(1-\Gamma,1+\Gamma)$-equidistributed  in $\UU(\dim)$ on a scale $\ep$. An analogous result holds for the space of states, with $\dim^2$ replaced by $\dim$.
\end{result}

Our quantitative statements about the saturation and recurrence of complexity for general local random quantum circuits  and unitary evolutions generated by SLH follow from combining \cref{res:equidistr-recurrence} and \cref{res:designs-equidistribution} with exponential bounds on the spectral gap for random quantum circiuts \cite{BHH2016} (see \cref{prop:RQChighdegree} for details). This is formally proved in \cref{lem:eqequidistrTIME}.
 
\begin{result}[Saturation and recurrence of complexity for random quantum circuits]
\label{res:satANDrecRQC}
Let $\nu$ be the probability distribution describing a single layer of local random quantum circuit on $n$ qudit system $\H=(\C^q)^{\ot n}$ of dimension $\dim=q^n$. Then, the measure $\nu_t$, corresponding to circuits of depth $t$, becomes equidistributed when the circuit depth is 
\begin{equation}
\label{eq:res-time-uni}
    \tau \approx \dim^{4}\log\left( \frac{1}{\ep}\right) \,. 
\end{equation}
Moreover, the induced measure $\nu_{t,\mathcal{S}}$ on the space of pure states $\S(\dim)$  becomes equidistributed after a circuit depth
\begin{equation}
\label{eq:res-time-states}
    \tau_\S \approx \dim^{3}\log\left( \frac{1}{\ep} \right)\,.
\end{equation}
By inserting $\dim=q^n$ and using \cref{res:equidistr-recurrence}, 
we find that the complexity $C_\ep$ evaluated on sequences of random unitary channels $(\h{U}_t)_{t\in\mathbb{N}}$ and states $(\psi_t)_{t\in\mathbb{N}}$ generated by $\nurqc$ with high probability
\begin{itemize}
    \item[(i)] Saturates at an essentially maximal value, equal to $C_{unitary}\approx d^2 \log(1/\ep)$ and $C_{state}\approx d \log(1/\ep)$ for unitaries and states respectively, at time   $t_{sat}\lesssim e^{\Theta(n)}\log(1/\ep)$.
    \item[(ii)] The first time after saturation that complexity  reaches $C_\ep=O(1)$ is $t_{rec}\approx e^{e^{\Theta(n)}}$.
\end{itemize} 
Analogous results hold for the continuous time SLH model.
\end{result}

Interestingly, careful usage of approximate equidistribution allows us to gain insight into timescales on which recurrences  
of complexity occur. The following result shows that their typical duration scales exponentially with the system size, in accordance to what was recently conjectured in Ref.~\cite{BHexpTIME2020}. We provide a technical formulation and formal proof of this claim in  \cref{sec:recurrTIME}.

\begin{result}[Typical duration of recurrences of complexity]
\label{res:recwidth}
Let $\nu$ be a probability measure on $\UU(\dim)$ describing a single layer of  local random quantum circuit acting on an $n$ qudit system of dimension $\dim=q^n$. Assume that $C_\ep(\h{U}_t)=0$ at the time $t$. Then, \emph{conditioned on this event}, with high probability (over the distribution of trajectories $\lbrace{\h{U}_t \rbrace}_{t\in\mathbb{N}}$) we have
\begin{equation}\label{eq:qualitativeRECdip}
  C_\ep(\h{U}_{t-T}) \approx d^2 \log(1/\ep)\ \ , \  C_\ep(\h{U}_{t+T}) \approx d^2 \log(1/\ep)\ .
\end{equation}
From \cref{res:satANDrecRQC} we know that for local random quantum circuits   have $T=e^{\Theta(n)}\log(1/\ep)$. Since complexity $C_\ep$ cannot change faster than linearly with $t$, we get that with high probability recurrences of complexity have a duration exponential in the system size.  An analogous results holds for sequences of states $\left(\psi_t\right)_{t\in \mathbb{N}}$ generated by  $\nu$. 
\end{result}
A result analogous to \eqref{eq:qualitativeRECdip} holds also for unitaries and states generated by the SLH model. However, as we explain after the proof of \cref{th:compTIMESCALES}, for this model of random evolution it is difficult to bound changes of complexity $C_\epsilon (\h{U}_t)$ during a unit time step. For this reason  we can only show that in this model recurrences of complexity have at most exponential duration (but cannot exclude they might be smaller).

The following result, proven in \cref{lem:coveringNUMBERhighcomplexityEquidistribution}, ensures that in the support of an equidistributed measure the number of distinct high-complexity unitaries is comparable to the packing number of the set all unitary channels $\UU(\dim)$ (on an appropriate logarithmic scale).

\begin{result}[Approximately equidistributed ensembles of unitary channels and states have doubly-exponential distinct high-complexity elements]\label{res:paking}
Let $\nu$ be a probability measure on $\UU(\dim)$ which is equidistributed. Let $N^r_{\ep}(\nu)$ be the maximal number of pairwise $\ep$-distinct unitary channels $\h{U}$ from the support of $\nu$ on unitaries with complexity greater than $r$. Then for $r \approx \dim^2  \log(1/\ep)$ we have $N^r_{\ep}(\nu) \approx \left( \frac{1}{\ep}  \right)^{\dim^2}$. As $\dim=q^n$, we get that both $N^r_{\ep}(\nu)$ exhibits a \emph{doubly exponential} scaling with the number of qudits. The same result holds for states, with $\dim^2$ replaced by $\dim$. 
\end{result}

Taken together, \cref{res:maxcompl} and \cref{res:paking} show that probability measures which approximately equidistribute describe ensembles of states or unitary channels with essentially maximal complexity. Furthermore, the number of distinct high complexity unitaries and states in the support of such measures scales doubly-exponentially with the number of qudits $n$. Crucially the dependence on $\ep$ in Eq.~\eqref{eq:genLOWER} is optimal.

\begin{figure}
    \definecolor{db}{rgb}{0,0,0.6}
    \definecolor{dr}{rgb}{0.86,0,0}
    \centering
    \begin{tikzpicture}[scale=1]
    \draw[dr,thick,decorate, decoration={random steps,segment length=5pt,amplitude=0.5pt}] (0,0) -- (3,3) -- (8,3);
    \draw[db,thick, decorate, decoration={random steps,segment length=5pt,amplitude=0.5pt}] (0,0) -- (3.8,1.8) -- (5,3) -- (8,3);
    \draw[thick,->] (0,0) -- (8.1,0);
    \draw[thick,->] (0,0) -- (0,4);
    \node at (8.3,0) {$t$};
    \node at (0,4.3) {$C_\ep(\U_t)$};
    \draw[thick,dashed] (3,-0.1) -- (3,3);
    \node at (3,-0.4) {$d^2\log(1/\ep)$};
    \draw[thick,dashed,color=db] (5,-0.1) -- (5,3);
    \node at (5.3,-0.4) {\textcolor{db}{$d^4\log(1/\ep)$}};
    \node at (-1,3) {$d^2\log(1/\ep)$};
    \draw[thick,dashed] (-0.1,3) -- (0,3);
    \node[align=center,rotate=45] at 
    (4.42,2.12)
    {\footnotesize (\cref{res:lingrowth})};
    \node[align=center,rotate=25] at 
    (1.85,0.65)
    {\footnotesize (previous results)};
    \end{tikzpicture}
    \caption{A plot depicting the conjectured random circuit complexity growth (upper curve \textcolor{dr}{in red}), a long-time linear growth saturating when the circuit depth is $d^2$, alongside the best established lower bounds (lower curve \textcolor{db}{in purple}), a long-time sublinear growth (previously known bounds, see \cref{sec:lineargrowth}), followed by a linear regime  and saturation at circuit depth $d^4$.}
    \label{fig:conjgrowth}
\end{figure}
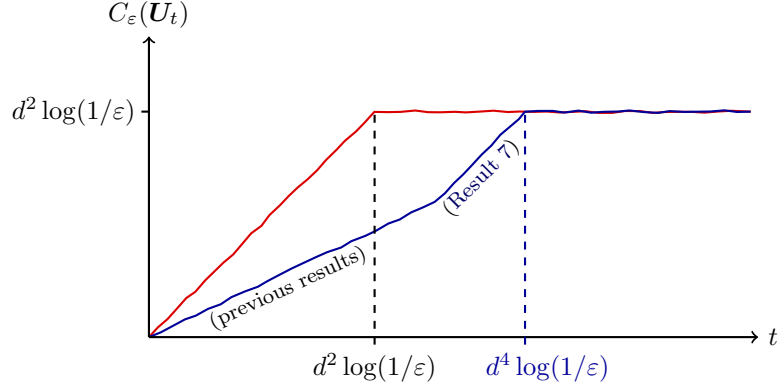

Lastly, we combine some previous results to show that an exponentially-small but moment-independent spectral gap for random quantum circuits implies a linear lower bound  complexity growth for deep circuits. See Figure \ref{fig:conjgrowth} and \cref{prop:linear-growth} for a detailed statement. We emphasize that in contrast to the results involving approximate equidistribution, these complexity lower bounds have a trivial $\ep$ dependence since instead of equidistribution they rely on bounds using moments and approximate $k$-designs. Such techniques can be used to obtain similar results as ours \cite{complexitygrowth2019}, but they yield weaker results not capturing the correct scaling in $\ep$.


\begin{result}[Linear lower bound for complexity  at exponential depths]\label{res:lingrowth}
Local random quantum circuits of depth $t$, when the circuit depth is $t\gtrsim d^2 \log(1/\delta)$, undergo a linear complexity growth up until the saturation time $t\approx d^4$. Specifically, with very high probability, a depth $t$ random quantum circuit has complexity lower bounded as $C_\ep (\U_t) \gtrsim (t/d^2)$. Similarly, the states generated by RQCs of the same depth have complexity $C_\ep (\psi) \gtrsim (t/d^2)$, such that a linear growth of state complexity begins at $t\approx d^2$ and proceeds until saturation. Analogous results hold for the continuous time SLH model.
\end{result}

In the formulation of above results we were not specific about the notion of locality of gates used to define random quantum circuits. The first three results and Result \ref{res:paking} hold for abstract random circuits and do not involve any locality. Results \ref{res:satANDrecRQC},\ref{res:recwidth} and 
\ref{res:lingrowth} hold for circuits acting on graphs $G$ containing a Hamiltonian path (see \cref{def:glocRQC}). In particular they hold e.g. for a $1D$ chain or a $2D$ lattice. 

Importantly, we can also choose the gates of the circuit randomly from a finite universal gateset $\G$ rather than from the Haar measure (see \cref{def:glocRQC-gates}). If the gateset $\G$ has the property called the spectral gap (e.g. if all the entries of gates are algebraic, by a deep result of Bourgain and Gamburd \cite{Bourgain2011}), the same proof techniques as for the Haar measure apply. If the gateset $\G$ is not known to have a spectral gap, we instead have to rely on $k$-designs properties for high $k$ in order to obtain equidistribution. This is achieved in \cref{th:equid} and is more technically involved than the approach based on spectral gaps. In such a case the leading dependence on $\ep$ of the equidistribution times becomes $\sim \log(1/\ep)^3$ rather than $\log(1/\ep)$. We remark that there is a (difficult) conjecture that every universal gateset has a spectral gap, which, if true, would allow us to dispense with the $k$-design approach entirely.

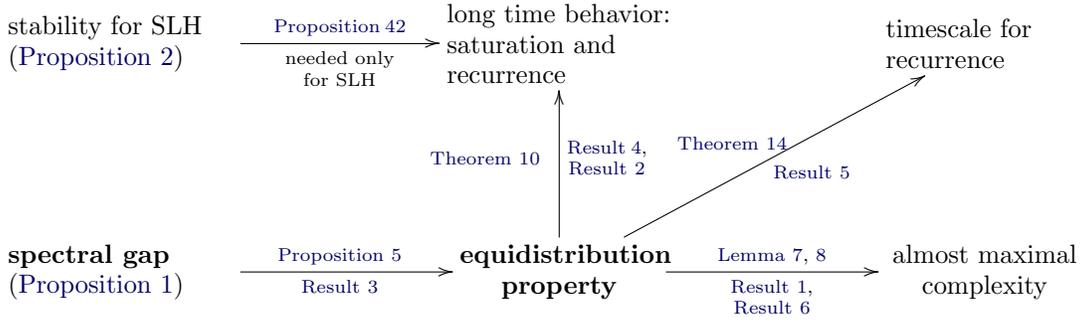
\begin{figure}
\begin{equation*}
 \xymatrix@C=75pt@R=55pt{
  \parbox{85pt}{\raggedright stability for SLH (\cref{thm:continuity})} 
 \ar@{->}[r]^(0.5){\parbox{50pt}{ \scriptsize \cref{prop:rec-SLH} } }_(0.5){\parbox{50pt}{\scriptsize needed only for SLH}}
 &  \parbox{85pt}{\raggedright long time behavior: saturation and recurrence}
   & \parbox{75pt}{\raggedright timescale for recurrence} \\
 %
 %
 \parbox{85pt}{\raggedright {\bf spectral gap} (\cref{prop:RQChighdegree}) }
  \ar@{->}[r]^(0.5){\parbox{50pt}{ \scriptsize \cref{lem:eqequidistrTIME} } }_{\parbox{50pt}{\scriptsize \cref{res:designs-equidistribution}}}
 %
  & \parbox{75pt}{{\bf equidistribution property}}  
  \ar@{->}[r]^(0.5){\parbox{50pt}{ \scriptsize \cref{lem:complexity-equidist}, \ref{lem:coveringNUMBERhighcomplexityEquidistribution} } }_{\parbox{50pt}{\scriptsize \cref{res:maxcompl}, \cref{res:paking}}}
    \ar@/_0.0pc/@{->}[ur]^(0.5){\parbox{50pt}{\scriptsize \cref{th:compTIMESCALES}} }_{\parbox{50pt}{\scriptsize \cref{res:recwidth}}}
    \ar@/_0.0pc/@{->}[u]^(0.5){\parbox{50pt}{\scriptsize \cref{th:recurrence}} }_{\parbox{30pt}{\scriptsize \cref{res:satANDrecRQC}, \cref{res:equidistr-recurrence}}}
  %
  & \parbox{75pt}{almost maximal complexity}
  %
 %
 }
 \end{equation*}
 \caption{Structure of the main results of the paper. Arrows are labelled by one of the informal results presented above together with the relevant rigorous statement. }
    \label{fig:diagram}
\end{figure}

To help the reader navigate the paper, we present in \cref{fig:diagram} the logical structure of technical statements and proofs of the results listed above.

\subsection*{Open problems}

\ni Our work leaves open a number of interesting directions for future work:


The results presented in this work, along with those in Ref.~\cite{complexitygrowth2019}, provide a consistent framework for understanding the growth and saturation of quantum complexity. The techniques directly apply to {\it time-dependent} unitary evolutions that form approximate designs. Nevertheless, in physics we are often interested in the energy-conserving evolution of {\it time-independent} Hamiltonians, which will not (in general) form unitary designs \cite{ChaosRMT}. Proving statements about the complexity growth of the ensemble $\{e^{-iHt}, \ H\in \CE_H\}$, for an ensemble of Hamiltonians, demands a different proof technique. Some progress in this direction, for specific random matrix models, has been made in \cite{gue-complexity}.

An exponentially small lower bound on the spectral gap for random circuits, $\Delta\geq e^{-\Omega(n)}$ and independent of $k$, gave that random quantum circuits form designs of exponential degree. Combined with the conditions for distributions to equidistribute over the space of unitaries, this gives the circuit depth at which complexity saturates and recurs. Can one substantially improve the $n$-dependence of the spectral gap bound for random circuits? An even more challenging question: can one prove an unconditional constant lower bound on the spectral gap? This would prove the optimal design depth and, via the results in Ref.~\cite{complexitygrowth2019} and in this work, would directly imply the conjectured linear complexity growth for random circuits.

Recently, Ref.~\cite{entflucs20} studied another long-time property of random circuit evolution---the fluctuations of subsystem entropies of states evolved by random circuits, which become increasingly suppressed for an exponentially long time. The results here should give the timescale at which the subsystem entropy recurs and the evolved-state becomes unentangled. More generally, it would be intriguing to rigorously explore the connection between quantum complexity and other long-time properties of many-body systems.

Furthermore, in our analysis we rely heavily on the approximate equidistribution property to estimate the measure on balls $\nu(B(\h{V},\ep))$, $\nu_\S(B(\psi,\ep))$ in the set of unitary channels and in the space of pure quantum states. The use of traditional techniques to estimate these quantities---e.g.\ based on (approximate) designs, moment bounds, or Levy's lemma---does not give the correct scaling with $\ep$. This suggests that in the regime of high-degree designs it should be possible to develop stronger concentration inequalities, interpolating between the known design bounds \cite{LowDeviation09} and the properties of the Haar measure.  

More speculatively, for holographic complexity there are various expectations of the behavior of circuit complexity coming from explicit gravity calculations in the bulk \cite{BHexpTIME2020}. It would be interesting to see if our methods were able to identify any particular features in the long-time behavior of complexity relevant to either the Complexity/Action or Complexity/Volume conjectures, in particular subtleties that arise at exponential time scales.

\section{Technical tools}\label{sec:technical}

\subsubsection*{Haar measure, volumes in the space of unitary channels and pure states}

The manifold of unitary channels $\UU(\dim)$ inherits the unique invariant probability measure from the Haar measure on the unitary group $\Ug(\dim)$ according to the following prescription: for $\mathcal{A}\subset\UU(\dim)$ we set $\mu_P(\mathcal{A})=\mu\left(\mathcal{A}'\right)$, where $\mu_P,\mu$ are Haar measures on $\Ug(\dim)$ and $\UU(\dim)$ respectively, and $\mathcal{A}'$ denotes collection of all unitary operators defining  quantum channels from  $\mathcal{A}$. 
In what follows we will not differentiate between unitary channels and unitary operators, as well as Haar measures defined on these sets, unless it leads to ambiguity. In particular, we let $\vol(\mathcal{A})$ denote the Haar measure of a subset $\mathcal{A}$ of unitary channels $\U(\dim)$ or unitary group $\Ug(\dim)$, depending on the context. Finally, the Haar measure $\mu$ on $\U(\dim)$ induces the unique invariant normalized measure $\mu_\S$ on the set of pure quantum states by $\S(\dim)$. This measure can be defined by $\mu_\S(\mathcal{B})=\mu(\varphi^{-1}(\mathcal{B}))$, where $\mathcal{B}\subset \S(\dim)$ and $\varphi:\UU(\dim)\rightarrow\S(\dim)$ is defined by $\varphi(\U)=\U(\phi_0)$, for arbitrarily chosen pure state $\phi_0\in\S(\dim)$. In analogous way, for a fixed choice of a referential state $\phi_0\in\S(\dim)$,  any  measure $\nu$ on $\UU(\dim)$ defines a measure on the set of pure states via 
\begin{equation}\label{eq:stateTRANSPORT}
    \nu_\S(\mathcal{B})=\nu(\varphi^{-1}(\mathcal{B}))\,.
\end{equation}

We will be frequently working with balls of radius $\ep$ in the space of unitary channels $\UU(\dim)$, induced by the diamond norm
\begin{equation}
    B(\U,\ep)=\{\V: \d(\V,\U)\leq\ep\}\,.
\end{equation}
We denote the Haar measure of such a ball by $  \vol(B(\U,\ep))=\vol(\ep)$. 
From the results in \cite{Szarek98,Szarek82}, we have
\begin{align}
\label{eq:szarek}
(\clower\, \ep)^{\dim^2-1}  \leq   \vol(\ep) \leq (\cupper\, \ep)^{\dim^2-1}\,,
\end{align}
where $\clower$ and $\cupper$ are absolute constants. 
In Ref.~\cite{OSH2020} one finds that $\clower= \frac{1}{9 \pi}$ and $\cupper= 87$. 
Similarly for pure states, we have balls of radius $\ep$ in the space of states $B(\psi,\ep)=\{\phi: \d(\phi,\psi)\leq \ep\}$. We denote the volume of this set according to the uniform measure $\mu_\S$ in $\S$ induced from the Haar measure on $\UU(\dim)$ by $\vol_\S(B(\psi,\ep))=\vol_\S(\ep)$. In this case it is possible to find an exact expression for the volume
\begin{equation}\label{eq:pureSTATESvol}
    \vol_\S(\ep)=\ep^{2(\dim-1)} \,.
\end{equation}
The above equation arises from the distribution of the overlaps of pure states $x=\tr(\psi\phi)$.
The explicit expression given by Ref.~\cite{Zyczkowski_2000} for this distribution is
\begin{equation}\label{eq:overlapdistribution}
    p(x)=(\dim-1)(1-x)^{\dim-2}\ .
\end{equation}

\subsubsection*{Approximate unitary designs and spectral gaps}
The pseudo-random properties of probability distributions over the unitary group e elegantly captured by the notions of (approximate) unitary $k$-designs and spectral gaps. Here we provide basic concepts and terminology regarding unitary designs. For a more comprehensive treatment of this subject see, for instance, Ref.~\cite{LowThesis}. 

The central concept used in the proofs is the spectral gap of the averaging operator associated to a measure $\nu$. Let $\nu$ be a probability measure on $\UU(\dim)$ and let $\elltwo$ be the space of square integrable functions on $\UU(\dim)$ equipped with the usual inner product. We can associate with $\nu$ its averaging operator $M_{\nu}: \elltwo\to\elltwo$ acting as:
\begin{equation}\label{eq:averaging-operator}
    (M_{\nu}f)(\U) := \int_{\UU(\dim)}\dt \nu(\V) f(\V^{-1}\U)
\end{equation}
Note that if $\mu$ is the Haar measure on $\UU(\dim)$, its averaging operator $M_{\mu}$ is simply the projection onto the subspace of constant functions. 

\begin{definition}[Spectral gap]\label{def:expander}
Let $\delta>0$ be a positive number. A probability distribution $\nu$ on $\UU(\dim)$ is said to have spectral gap $\delta$ if: 
\begin{equation}\label{eq:expinfDEF}
    \|M_{\nu}-M_{\mu}\|_\infty \leq 1 - \delta
\end{equation}
where $\mu$ is the Haar measure, where $\|\cdot \|_\infty$ denotes the operator norm in the space of bounded operators acting on $\elltwo$. 
\end{definition}

We now decribe the notion of $k$-designs. To any measure $\nu$ on $\UU(\dim)$ we associate the $k$-fold channel, defined as
\begin{equation}
    \Phi^{(k)}_\nu (\op) \coloneqq \int_{\UU(\dim)} \dt\nu(\h{U})\, \U^{\otimes k}(\op)\ ,
\end{equation}
where $\op$ is an operator acting on $\left(\C^{\dim}\right)^{\ot k}$. A measure $\nu$ is said to form an exact $k$-design if $\Phi^{(k)}_\nu (\op)=\Phi^{(k)}_\mu (\op)$, where $\mu$ denotes the Haar measure on $\UU(\dim)$. Alternatively, for any measure $\nu$ we can define the $k$-th moment operator as
\begin{equation}
    M_{\nu,k} \coloneqq \int_{\UU(\dim)} \dt \nu(\U) U^{\otimes k}\otimes \bar{U}^{\otimes k}\ .
\end{equation}
The property of $\nu$ being an exact $k$-design can be equivalently expressed by demanding $M_{\nu,k}=M_{\mu,k}$ (this is because  a moment operator $M_{k,\nu}$ is a vectorization of a $k$-fold channel  $\Phi^{(k)}_\nu$). Approximate $k$-designs are measures for which the aforementioned equalities are satisfied approximately. It is common to define approximate designs such that the channels are close in diamond norm $\| \Phi^{(k)}_\nu -  \Phi^{(k)}_\mu \|_\diamond$, or even stronger definitions involving small relative error. In this work it will be sufficient to work with a weaker definition of approximate design in terms of the operator norm.

\begin{definition}[Approximate unitary designs/expanders]\label{def:unitarydesign}
Let $\delta>0$ be a positive number. A probability distribution $\nu$ on $\UU(\dim)$ is called a  $\delta$-approximate unitary $k$-design (or approximate $k$-expander) if 
\begin{equation}\label{eq:expDEF}
    \|M_{\nu,k}-M_{\mu,k}\|_\infty \leq \delta\,.
\end{equation}
The quantity $g(\nu,k):=   \|M_{\nu,k}-M_{\mu,k}\|_\infty$ is called the expander norm of $\nu$.
\end{definition}

Given our definition of design in terms of the expander norm, we will use the terms approximate $k$-design and approximate $k$-expander interchangeably.

Approximate $k$-expanders yield approximate designs defined in terms of stronger norms when the error is exponentially small \cite{BHH2016}. Nevertheless, approximate $k$-expanders, despite being a weaker notion, are often more convenient to work with. This is because $g(\nu,k)$ can be connected to the \emph{spectral gap} of $\nu$, $\Delta(\nu,k) = 1-g(\nu,k)$, defined as the highest nontrivial eigenvalue of the moment operator $M_{\nu,k}$.
This allows one to estimate $g(\nu,k)$ for local random circuits via the application of tools from many-body physics, which, after iterating the property $g(\nu^{\ast t},k)=g(\nu,k)^t$ gives practical estimates on the circuit depth needed to ensure a strong design-like properties. 

The spectral gap of $M_{\nu}$ is related to the gaps of the moment operators $M_{\nu,k}$ in the following way (proved in Section 2 of \cite{SlowikSawicki2023}): 
\begin{equation}\label{eq:spectral-gap}
    \|M_{\nu}-M_{\mu}\|_\infty = 
    \sup_k \|M_{\nu,k}-M_{\mu,k}\|_\infty
\end{equation}
This means that a bound on the spectral gap for the averaging operator $M_{\nu}$ corresponds to a bound on expander norms $g(\nu, k)$ which is uniform regardless of $k$. One can understand the expander norms $g(\nu,k)$ as corresponding to spectral gaps restricted to the subspaces of $\elltwo$ consisting of polynomials of degree at most $k$.

\subsubsection*{Discrete and continuous time local quantum evolutions: definitions}

The first class of models we consider are discrete time models, i.e. random quantum circuits. For simplicity, we will first consider the random quantum circuit architecture defined on a 1D chain of $n$ qudits with local dimension $q$. We will then introduce a more general circuit architecture.

\begin{definition}[Local random quantum circuits]
\label{def:localRQCs}
Let $\nurqc$ denote the probability distribution on $\UU(\dim)$ for a single step of the random circuit, defined by choosing a site $i$ uniformly at random from $[1,\ldots,n-1]$ and applying a Haar random 2-site unitary $U_{i,i+1}$, drawn from $\Ug(q^2)$, to an adjacent pair of qudits. We are assuming open boundary conditions. Depth $t$ random quantum circuits are then given by the probability distribution $\nurqct := (\nurqc)^{*t}$. The measure $\nurqct$ induces the measure on states $\nurqcts$ as defined previously in \cref{sec:intro}.
\end{definition}

The 1D RQCs with local Haar-random gates introduced above are very concrete in that they allow for an explicit calculation of the scaling and constants involved. To demonstrate the applicability of our results, we also introduce a more general random circuit architecture, with notably few restrictions on the graph on which the qudits live. We consider $n$ qudits on a graph $\Grqc$ which contains a Hamiltonian path, i.e.\ there is no restriction on the geometric locality of the qudit interaction graph but the graph is connected and there exists a path that visits each qudit exactly once. We can also replace Haar-random gates with gates drawn from a finite universal gateset, see \cref{app:rqcproofs} for details.

\begin{definition}[$\Grqc$-local random quantum circuits]\label{def:glocRQC}
Consider $n$ qudits defined on a graph $\Grqc$ which contains a Hamiltonian path.
Let $\nu_\Grqc$ be the distribution for a single step of the random circuit, obtained by choosing a pair $(i,j)$ uniformly at random from the set of edges of $\Grqc$, and applying a Haar random 2-site unitary $U_{i,j}$, drawn from $\Ug(q^2)$, to qubits $(i,j)$.  Depth $t$ $\Grqc$-local random quantum circuits are then defined by the probability distribution  $\nugrqct  := (\nu_\Grqc)^{*t}$. 
\end{definition}

The second type of models we consider is a continuous stochastic process called the Stochastic Local Hamiltonian model (SLH) that can be thought of as a continuous time version of the random quantum circuit model. The process, defined in \cite{Onorati17} (also in \cite{FastScrambling} under the name of Brownian quantum circuits), will be a Brownian motion on the unitary group generated by a stochastic local Hamiltonian. The process $U_t$ is constructed using a limit of finite time increments, namely:
\begin{equation}\label{eq:slh-definition}
U_t = \lim_{\Delta t \to 0}\prod^{1}_{l=\frac{t}{\Delta t}}\exp\{i H_{l ,\Delta t} \Delta t\} U_0
\end{equation}
where $H_{l ,\Delta t}$ are finite increments defined below and $U_0$ is the initial point which we always take to be the identity. 

Fix an interaction graph $G=(V,E)$, where vertices correspond to $q$-level subsystems and only systems connected by an edge $e \in E$ can interact. As in the case of discrete circuits we assume that $G$ contains a Hamiltonian path. We define the increment as:
\begin{equation}\label{eq:theta}
H_{l, \Delta t} = \sum_{e \in E}\theta^{(e)}_{l, \Delta t}
\end{equation}
where each local term $\theta^{(e)}_{l, \Delta t}$ is given by:
\begin{equation}\label{eq:small-theta}
\theta^{(e)}_{l, \Delta t} = \sum_{\mu} A_{\mu}^{(e)}\xi^{(e, \mu)}_{l, \Delta t}
\end{equation}
with $A_{\mu}^{(e)}$ being the local operators corresponding to an edge $e$ and $\xi^{(e, \mu)}_{l, \Delta t}$ the random variables representing the noise. Let $\{A_{\mu}\}_{\mu}$ be a basis for the real Lie algebra of Hermitian matrices denoted as:
\begin{equation}
i\cdot \mathfrak{u}(q^2) = \{ X \in \C^{q^2 \times q^2} : X = X^{\dagger}\} 
\end{equation}
For each edge $e \in G$, let $A^{(e)}_{\mu}$ be the operator that acts as $A_{\mu}$ on vertices connected by $e$ and as the identity elsewhere. For each pair $(e, \mu)$ we let $\xi^{(e, \mu)}_{l, \Delta t}$ to be a Gaussian random variable with mean $0$ and covariance given by:
\begin{equation}\label{eq:covariance}
 \BE\left( \xi^{(e, \mu)}_{l, \Delta t}\xi^{(e', \mu')}_{l', \Delta t}\right) =
\frac{1}{\Delta t}\delta_{l,l'}\delta_{e,e'}(\kappa^{-1})_{\mu,\mu'}
\end{equation}
where the matrix $\kappa$ is the Killing metric tensor given by:
\begin{equation}\label{eq:killing}
    \kappa_{\mu,\mu'} =  2q^2 \tr(A_{\mu}A^{\dagger}_{\mu'})
\end{equation}
In the simplest case where $A_{\mu}$ form an orthonormal basis with respect to the inner product \eqref{eq:killing}, the variables $\xi^{(e, \mu)}_{l, \Delta t}$, are simply uncorrelated for different values of $l, e, \mu$.

\begin{definition}[Stochastic Local Hamiltonian (SLH) model]
\label{def:SLH}
The Stochastic Local Hamiltonian model (SLH) at time $t$ is defined as the distribution of $U_t$ on the unitary group, where $U_t$ is the continuous time Markov process constructed via \eqref{eq:slh-definition} with the increments defined as above (see \cite{Onorati17} and [\cite{McKean}, Chapter 4.8] for a proof). We will denote this distribution by $\nuslh$, where the interaction graph $G$ is implicit.
\end{definition}

We remark that just as the Brownian motion on the real line can be thought of as a continuous version of the random walk with discrete steps, the SLH process can be thought of as a continuous version of the random quantum circuit model. Formally, the SLH process is an example of a Brownian motion on the unitary group. Any such process $U_t$ is characterized by the following properties (see \cite{Onorati17}):
\begin{enumerate}
    \item For any $0 < t_1 < t_2 < \dots < t_n$, the (left) increments $U_{t_1}U_{0}^{\dagger}, U_{t_2}U_{1}^{\dagger}, \dots, U_{t_n}U_{t_{n-1}}^{\dagger}$ are independent.
    \item For any time $t \geq 0$, the increments are stationary, i.e., for any $\Delta t >0$ the increment $U_{\Delta t}U_{0}^{\dagger}$ is equal in distribution to $U_{t+\Delta t}U_{t}^{\dagger}$.
    \item The paths $t \mapsto U_t$ are continuous almost surely.
\end{enumerate}

\subsubsection*{Discrete and continuous time local quantum evolutions: properties}


In the paper, we will make extensive use of the following property of local random quantum evolutions introduced in the previous section, relating their depth (time for which the evolution is run) and the spectral gap (see \cref{def:expander}). In the statement $G$ is any interaction graph that contains a Hamiltonian path (see \cref{def:glocRQC}).

\begin{proposition}[Spectral gap for quantum evolutions]
\label{prop:RQChighdegree}
Fix any $\delta>0$ and let $\nu=\nu_t$ be the distribution corresponding either to the $G$-local random quantum circuit of depth $t$ or the SLH model on $n$ qudits at time $t$. Then the distribution $\nu$ has spectral gap equal to $1-\delta$ (i.e. $\norm{M_{\nu} - M_{\mu}}_{\infty}<\delta$, see Definition \ref{def:expander}) for:
\begin{equation}\label{eq:time-for-gap-rqc}
    t \geq c_2 n^6 d^2 \log\left(\frac{1}{\delta}\right)
\end{equation}
if $\nu_t$ corresponds to the $G$-local random quantum circuit or:
\begin{equation}\label{eq:time-for-gap-slh}
    t \geq 4c_2 n^4 d^2 \log\left(\frac{1}{\delta}\right)
\end{equation}
if $\nu_t$ corresponds to the SLH model, where $c_2 = 10^5$.
\end{proposition}

The proof of \eqref{eq:time-for-gap-rqc} follows from a number of Lemmas in Ref.~\cite{BHH2016}, which we present for completeness in \cref{app:rqcproofs} (\cref{th:gapGlocalCIRC}). The proof of \eqref{eq:time-for-gap-slh} follows along similar lines, but also uses some arguments from \cite{Onorati17} and we present it in \cref{app:slh} (\cref{th:gap-SLH}). The right hand side of both \eqref{eq:time-for-gap-rqc} can be reduced by a factor of $n$ if the underlying graph is a 1D chain. Note that the difference between timescales in equations \eqref{eq:time-for-gap-rqc} and \eqref{eq:time-for-gap-slh} by a factor of $n^2$ is a consequence of normalization of local interactions defining SLH model (cf. \eqref{eq:theta}, where the Hamiltonian is {\it not} divided by the number of edges in the graph $G$, so the energy per unit time scales with the number of edges).

A result analogous to \cref{prop:RQChighdegree}, guaranteeing that $\nu_t$ is a approximate $k$-expander (for large $k$), can be shown with different technical tools, and holds if instead of Haar random gates we pick gates chosen randomly from a finite universal gateset (see \cref{app:rqcproofs} for formal statements and proofs, specifically \cref{prop:GRQCdesigns}). 


In the proofs for the SLH model, we will also need a technical stability result for the SLH process $U_t$ which states that the process cannot stray too far from the identity in a short time. This is captured by the following Proposition, proved in Appendix \ref{app:slh}, which is an interesting concentration result in its own right:

\begin{proposition}\label{thm:continuity}
For any $s >0$ and $x>0$, the channel corresponding to the SLH process $U_t$ satisfies:
\begin{equation}\label{eq:maximal-inequality}
\Pr\left( \max_{t \in [0,s]} \d(\h{U}_t, \h{I}) > \frac{m t}{2} + x \right) \leq 2d  \exp\left( -\frac{x^2}{2ms} \right)
\end{equation}
where $m = \abs{E}$ is the number of edges in the interaction graph, $d=q^n$ is the total dimension and $\d$ is the distance on the unitary group defined by \eqref{eq:distancesDEF}.
\end{proposition}

\section{Approximate equidistribution property for ensembles of unitary channels and states}
\label{sec:approxEQUIfromDESINGS}

Recall that we are interested in the complexity properties of sequences of random circuits and states:
\begin{equation}\label{eq:randomWALKdef}
    \h{U}_t \ ,\ \psi_{t}\ ,\ t=0,1,2,\ldots
\end{equation}
with initial conditions $\h{U}_0=\I$ and  $\psi_0=\ketbra{0}^{\ot n}$ (a computationally trivial state). The
 distributions of $\h{U}_t$  and $\psi_t$ are controlled by the $t$-fold convolution of $\nu^{\ast t}$:
\begin{equation}
   \nu_t = \nu^{\ast t} \ ,\  \nu_{t,\S}=\nu^{\ast t}(\psi_0)\,.
\end{equation}
 We also consider the continuous walk generated by stochastic local Hamiltonians (SLH) as defined in \cref{def:SLH}. It generates a measure $\nuslh$ which is not a convolution of some initial measure $\nu$ on unitaries, but is generated by a random process on Hamiltonians.

\begin{definition}[Equidistribution time]\label{def:equidistibTIME}
Let $\nu$ be a probability distribution on $\UU(\dim)$. We define unitary and state equidistribution times as follows:
\begin{equation}
    \tau(\nu,\ep)=\min\{t: \nu_t\ (\alpha,\beta)\text{-equidistributes in}\ \UU(\dim)\ \text{on a scale}\ \ep\}\,,
\end{equation}
\begin{equation}
    \tau_{\S}(\nu,\ep)=\min\{t: \nu_{t,\S}\ (\alpha,\beta)\text{-equidistributes in}\ \S(\dim)\ \text{on a scale}\ \ep\}\,.
\end{equation}

For the SLH model  we define: 
\begin{equation}
    \tau(SLH,\ep)=\min\{t: \nuslh\ (\alpha,\beta)\text{-equidistributes in}\ \UU(\dim)\ \text{on a scale}\ \ep\}\,.
\end{equation}
The notation above omits mentioning $\alpha, \beta$ explicitly, as these will be clear from the context.
\end{definition}

\begin{fact}
\label{fact:equi}
If a probability measure $\nu$ on $\UU(\dim)$ is $(\alpha,\beta)$-equidistributed on $\UU(\dim)$ on the scale $\ep$ then so does 
measure the $\nu  \ast \vartheta$ for any measure $\vartheta$ on $\UU(\dim)$. Similarly, if $\nu_\S$ is $(\alpha,\beta)$-equidistributed in $\S(\dim)$, then so is $(\nu\ast\vartheta)_\S$ for any measure $\vartheta$ on $\UU(\dim)$.
\end{fact}
\begin{proof}
Assume first that $\vartheta=\delta_{\h{U}}$, i.e.\ a Dirac delta concentrated at $\h{U}\in\UU(\dim)$. For arbitrary $\h{V}\in\UU(\dim)$ we have
\begin{equation}
    \nu \ast \delta_{\h{U}}\left(B(\h{V},\ep)\right)=
    \nu \left(\h{U}^{-1}\left[B(\h{V},\ep)\right]\right)= 
     \nu \left(B(\h{U}^{-1}\h{V},\ep)\right)\,,
\end{equation}
where the first equality follows from definition of convolution while in the second equality we used $\h{V}^{-1}\left[B(\h{U},\ep)\right]=B(\h{V}^{-1}\h{U},\ep)$, which follows from the unitary invariance of the distance measure $\d(\cdot,\cdot)$. By invoking the approximate equidistribution property (as in  Eq.~\eqref{eq:equidistrUnitary}) of $\nu$ to a ball $B(\h{V}^{-1}\h{U},\ep)$, we get that the measure $\nu\ast\delta_{\h{U}}$ also satisfies this property. For general measures $\vartheta$ the claim follows because an arbitrary measure $\vartheta$ can be written as a (possibly infinite) convex combination of Dirac deltas: $\vartheta=\sum_\alpha p_\alpha \delta_{\h{U}_\alpha}$. We then have
\begin{equation}
    \nu \ast \vartheta \left(B(\h{V},\ep)\right)=
   \sum_\alpha  p_\alpha
     \nu \left(B(\h{U}_{\alpha}^{-1}\h{V},\ep)\right)\,,
\end{equation}
and approximate equidistribution follows by employing approximate equidistribution of $\nu$ for every summand and using $\sum_\alpha p_\alpha =1$. The proof for pure states is completely analogous.
\end{proof}
The above fact shows that \cref{def:equidistibTIME} is well-justified since equidistribution for time (circuit depth) $t$ implies equidistribution for time $t+1$ (for both states and unitaries). The same can be easily proved also for the continuous time SLH model because it is Markovian.

We now state and prove the main technical theorem establishing a connection between approximate equidistribution and spectral gaps. It was stated informally in \cref{res:designs-equidistribution}. 

\begin{theorem}[Equidistribution of unitaries and states from spectral gap]\label{th:equid-gap}
Fix $\ep > 0$. Let $\mu$ be the Haar measure and let $\nu$ be a probability measure on $\UU(\dim)$. Let $M_{\mu}, M_{\nu}$ be the associated averaging operators as defined in Eq.~\eqref{eq:averaging-operator}. Fix $\beta \geq 1 \geq \alpha >0$ and suppose that:
\begin{equation}\label{eq:condition-lambda}
\norm{M_{\nu} - M_{\mu}}_{\infty} \leq \left(  \sqrt{\frac{3}{2}} C_1 \ep \cdot \min\left\{\left( \frac{(\beta-1)^3}{\beta+1} \right)^{1/2} , \left(\frac{(1-\alpha)^3}{1+\alpha}\right)^{1/2} \right\}\right)^{d^2-1}
    \end{equation}
where $C_1 =\frac{\clower^{3/2}}{4(\cupper)^{1/2}}\cdot \sqrt{\frac{2}{3}}$ and $\cupper, \clower$ are absolute constants from \eqref{eq:szarek}. Then $\nu$ is $(\alpha, \beta)$-equidistributed on scale $\ep$. Likewise, if $\nu$ satisfies:
\begin{equation}\label{eq:condition-lambda-states}
    \norm{M_{\nu} - M_{\mu}}_{\infty} \leq \left(  \sqrt{\frac{3}{2}} C_2 \ep \cdot \min\left\{\left( \frac{(\beta-1)^3}{\beta+1} \right)^{1/2} , \left(\frac{(1-\alpha)^3}{1+\alpha}\right)^{1/2} \right\}\right)^{2(d-1)}
    \end{equation}
where $C_2 = 1/4 \cdot \sqrt{\frac{2}{3}}$, then the induced measure on the set of states $\nu_{\S}$ is $(\alpha, \beta)$-equidistributed on scale $\ep$. Specializing to $\beta=1+\Gamma, \alpha=1-\Gamma$ for $\Gamma \in (0,1)$ yields simpler conditions for $(1-\Gamma, 1+\Gamma)$-equidistribution:
\begin{equation}\label{eq:condition-lambda-gamma}
    \norm{M_{\nu} - M_{\mu}}_{\infty} \leq \left(C_1 \ep \cdot\Gamma^{3/2} \right)^{d^2-1}
\end{equation}
and
\begin{equation}\label{eq:condition-lambda-states-gamma}
    \norm{M_{\nu} - M_{\mu}}_{\infty} \leq \left( C_2 \ep \cdot\Gamma^{3/2} \right)^{2(d-1)}
    \end{equation}
    respectively.
\end{theorem}

\begin{proof} 
We first prove equidistribution for unitaries, the proof for states is analogous and is sketched at the end. The argument below is based on a similar reasoning as in \cite{Harrow2002}. In what follows, for $f\in \elltwo$,  we will use $\|f \|$ to denote the standard norm in $\elltwo$ i.e. $\|f\|^2=\int_{\UU(d)}\dt\mu(\U)  |f(\U)|^2 $. Let $\lambda := \norm{M_{\nu} - M_{\mu}}_{\infty}$. Fix some $\kappa \in (0,1)$ to be chosen later and recall that $\vol(r)$ denotes the Haar volume of a ball of radius $r$ in the space of unitary channels $\UU(d)$. Fix any $\V \in \UU(\dim)$ and let $\indc_{B(\V,r)}$ denote the indicator function of the ball $B(\V,r)$. Since $M_{\mu}$ is the orthogonal projection onto constant functions, we have $M_{\mu}\indc_{B(\V, r)} = \vol(r)\cdot \indc$, where $\indc$ is the constant function in $\UU(d)$ equal to one. From this we easily get:
\begin{align}
\norm{\indc_{B(\V, r)}}^2 = \norm{M_{\mu} \indc_{B(\V, r)}} = \vol(r)
\end{align}
and
\begin{align}
\scalar{\indc_{B(\V, r)}}{M_{\mu} \indc_{B(\h{I}, \kappa r)}} = \vol(r) \vol(\kappa r) \ .
\end{align}
We can now write:
\begin{align}
    &\scalar{\indc_{B(\V, r)}}{M_{\nu} \indc_{B(\h{I}, \kappa r)}} =
    \scalar{\indc_{B(\V, r)}}{(M_{\nu}-M_{\mu}) \indc_{B(\h{I}, \kappa r)}} +
    \scalar{\indc_{B(\V, r)}}{M_{\mu} \indc_{B(\h{I}, \kappa r)}} = \\ \nonumber
    &\scalar{\indc_{B(\V, r)}}{(M_{\nu}-M_{\mu}) \indc_{B(\h{I}, \kappa r)}} + \vol(r)\  \vol(\kappa r)\ .
\end{align}
Since $\lambda = \norm{M_{\nu} - M_{\nu}}_{\infty}$ and $M_{\mu}$ is the projection onto constant functions, it follows that for every function $f \in \elltwo$ such that $\BE f = 0$, with expectation taken over the Haar measure on $\UU(\dim)$, we have:
\begin{equation}\label{eq:spectral-aux}
    \norm{M_{\nu}f} \leq \lambda \norm{f}\ .
\end{equation}
It is easy to see that for any function $f$ we have $M_{\nu} M_{\mu}f=M_{\mu}f$, which by \eqref{eq:spectral-aux} implies that:
\begin{equation}
\norm{(M_{\nu}-M_{\mu})f} = \norm{M_{\nu}(\iden-M_{\mu})f} \leq \lambda \norm{(\iden - M_{\mu})f} \leq \lambda\norm{f} \ .
\end{equation}
Thus by applying Cauchy-Schwartz we obtain:
\begin{equation}
    \abs{\scalar{\indc_{B(\V, r)}}{(M_{\nu}-M_{\mu}) \indc_{B(\h{I}, \kappa r)}}} \leq \lambda \norm{\indc_{B(\V,r)}}\norm{\indc_{B(\h{I}, \kappa r)}} = \lambda \sqrt{\vol(r) \vol(\kappa r)}
\end{equation}
In this way we obtain
\begin{equation}\label{eq:spectral}
    \vol(r) \vol(\kappa r) - \lambda \sqrt{\vol(r) \vol(\kappa r)} \leq \scalar{\indc_{B(\V, r)}}{M_{\nu} \indc_{B(\h{I}, \kappa r)}} \leq \vol(r) \vol(\kappa r) + \lambda \sqrt{\vol(r) \vol(\kappa r)}\ .
\end{equation}
Now, since $\scalar{f}{g} = \int_{\UU(d)}\dt\mu(\U)f(\U)\overline{g(\U)}$, we observe that $\scalar{\indc_{B(\V, r)}}{M_{\nu} \indc_{B(\h{I}, \kappa r)}}$ can be written as below:
\begin{align}\label{eq:intersection}
    &\scalar{\indc_{B(\V, r)}}{M_{\nu} \indc_{B(\h{I}, \kappa r)}} = 
    \int_{\UU(d)}\dt\mu(\U) \left[ \indc_{B(\V, r)}(\U) \cdot \int_{\UU(d)}\dt\nu(\h{W}) \indc_{B(\h{I}, \kappa r)}(\h{W}^{-1}\U)  \right] =\\ \nonumber
    & \int_{\UU(d)}\dt\nu(\h{W}) \left[ \int_{\UU(d)}\dt\mu(\U)\left( \indc_{B(\V, r)}(\U) \cdot \indc_{B(\h{W}, \kappa r)}(\U) \right) \right] = 
    \int_{\UU(d)}\dt\nu(\h{W}) \left[ \mu\left( B(\V, r) \cap B(\h{W}, \kappa r) \right)  \right]\ .
\end{align}
We first lower bound the right hand side of \eqref{eq:intersection}. Note that if $\h{W} \in B(\V, (1-\kappa)r)$, then by the triangle inequality we have $B(\h{W}, \kappa r) \subseteq B(\V, r)$. Thus, we can lower bound \eqref{eq:intersection} by restricting only to $\h{W} \in B(\V, (1-\kappa)r)$ and on that condition use $\mu\left( B(\V, r) \cap B(\h{W}, \kappa r)\right) = \mu\left( B(\h{W}, \kappa r)\right)$. This implies:
\begin{equation}
    \int_{\UU(d)}\dt\nu(\h{W}) \left[ \mu\left( B(\V, r) \cap B(\h{W}, \kappa r) \right)  \right] \geq
    \nu\left( B(\V, (1-\kappa)r)\right) \cdot \vol(\kappa r)\ .
\end{equation}
Combining this with the RHS of \eqref{eq:spectral} we get:
\begin{equation}
    \nu\left( B(\V, (1-\kappa)r)\right) \cdot  \leq \vol(r)  + \lambda \sqrt{\frac{\vol(r)}{ \vol(\kappa r)}}\ .
\end{equation}
Let us put $\kappa = \frac{\beta-1}{\beta+1}$ and replace $r$ with $r/(1-\kappa)$ to obtain:
\begin{equation}
    \nu\left( B(\V, r)\right) \cdot  \leq \vol\left(\frac{\beta+1}{2}\cdot r\right)  + \lambda \sqrt{\frac{\vol(\frac{\beta+1}{2} r)}{ \vol(\frac{\beta-1}{2} r)}}\ . 
\end{equation}
It is easy to see that $\vol(\frac{\beta+1}{2}\cdot r) + \vol(\frac{\beta-1}{4} r) \leq \vol(\beta r)$, since a ball of radius $\beta r$ contains two disjoint balls of radii, respectively, $\frac{\beta+1}{2}r$ and $\frac{\beta-1}{4} r$. To prove the upper half of the approximate equidistribution condition it thus suffices to have:
\begin{equation}\label{eq:gamma-condition}
    \lambda\cdot \sqrt{\frac{\vol\left(\frac{\beta+1}{2}\cdot r\right) }{ \vol\left(\frac{\beta-1}{2}\cdot r\right) }} \leq \vol\left(\frac{\beta-1}{4}\cdot r\right) \ .
\end{equation}
This follows after a short computation from condition \eqref{eq:condition-lambda} by invoking inequalities \eqref{eq:szarek} and using the assumption $r \geq \varepsilon$.

The proof of the lower half proceeds in the same fashion. This time we upper bound the right hand side of \eqref{eq:intersection} by noting that if $B(\V, r) \cap B(\h{W}, \kappa r)$ is nonempty, then $\h{W} \in B(\V ,(1+\kappa)r)$. We can thus restrict only to such $\h{W}$ and use the upper bound $\mu\left( B(\V, r) \cap B(\h{W}, \kappa r)\right) \leq \mu\left(  B(\h{W}, \kappa r)\right)$ to get:
\begin{equation}
    \int_{\UU(d)}\dt\nu(\h{W}) \left[ \mu\left( B(\V, r) \cap B(\h{W}, \kappa r) \right)  \right] \leq
    \nu\left( B(\V, (1+\kappa)r)\right) \cdot \vol(\kappa r)\ .
\end{equation}
Combining this with the LHS of \eqref{eq:spectral} we obtain:
\begin{equation}
  \vol(r)  - \lambda \sqrt{\frac{\vol(r)}{ \vol(\kappa r)}} \leq   \nu\left( B(V, (1+\kappa)r)\right) \ .
\end{equation}
By putting $\kappa = \frac{1-\alpha}{1+\alpha}, r \to r/(1+\kappa)$ and employing an analogous reasoning as in the upper bound, we arrive at the condition:
\begin{equation}
     \lambda\cdot \sqrt{\frac{\vol(\frac{1+\alpha}{2}r)}{ \vol(\frac{1-\alpha}{2}r)}} \leq \vol\left(\frac{1-\alpha}{4} r\right) \ .
\end{equation}
which again follows from \eqref{eq:condition-lambda} by invoking \eqref{eq:szarek} and $r \geq \varepsilon$.

To obtain condition \eqref{eq:condition-lambda-gamma} it suffices to plug $\beta=1+\Gamma, \alpha=1-\Gamma$ and use  the inequality $\frac{\Gamma^3}{1+\Gamma/2} \geq \frac{2}{3}\Gamma^3$, which is valid since $\Gamma \leq 1$.

To use the above argument to prove equidistribution for states, we employ the same argument, only replacing $M_{\nu}$ and $M_{\mu}$ acting on $\elltwo$ by $\avgstate{\nu}$ and $\avgstate{\mu}$ (defined in \eqref{eq:averaging-for-states}) acting on $\L^2(\S(\dim))$. By \eqref{eq:T-states} and \eqref{eq:spectral-gap}, we have $\norm{\avgstate{\nu} - \avgstate{\mu}} \leq \norm{M_{\nu} - M_{\nu}}$, so condition \eqref{eq:condition-lambda-states} holds for $\norm{\avgstate{\nu} - \avgstate{\mu}}$ as well. We then follow the above proof verbatim and estimate the volumes of balls using \eqref{eq:pureSTATESvol} instead of \eqref{eq:szarek} as for unitaries.
\end{proof}

We are now ready to formulate the proposition which constitutes a technical version of \cref{res:satANDrecRQC}.

\begin{proposition}[Equidistribution time for random quantum circuits and SLH]\label{lem:eqequidistrTIME}
Let $\alpha,\beta$ be fixed parameters entering the definition of approximate equidistribution, $\Gamma=\min\lbrace{1-\alpha,\beta-1\rbrace}$, and let $\ep\in(0,\frac{1}{4})$. Let $\nurqc$ be a single step of a local random  quantum circuit (see \cref{def:localRQCs}). Then we have:
\begin{equation}\label{eq:localTIMES}
    \tau (\nurqc,\ep) \leq \crqc n^6d^4 \log\frac{1}{C_1\ep \Gamma^{3/2}},\quad 
    \tau_{\S} (\nurqc,\ep)
    \leq 
    \crqc n^6d^3 \log\frac{1}{C_2\ep \Gamma^{3/2}}
    \,.
\end{equation}
where $c_2$ is the absolute constant from \cref{prop:RQChighdegree} and $C_1, C_2$ are constants from \cref{th:equid-gap}. 
Likewise, for $\nu_t$ corresponding to SLH model we have: 
\begin{align}
\label{eq:equi-time-for-slh}
    \tau(SLH,\ep)\leq
    4 c_2n^4  d^4\log\frac{1}{C_1\ep\Gamma^{3/2}}.
\end{align}
 
Finally, let $\nugrqc$ be a single step of $(\Grqc, \G)$-local random quantum circuits (i.e. with general gateset and underlying graph, see \cref{def:glocRQC-gates}).  Then we have 
\begin{align}
\label{eq:grqc-times}
 \tau (\nugrqc,\ep) \lesssim
  c(\G) n^{10} d^4  
 \log^3\frac{1}{\ep\Gamma},\quad
 \tau_\S (\nugrqc,\ep) \lesssim
  c(\G) n^{9} d^3 
 \log^3\frac{1}{\ep\Gamma}\,.
\end{align}
where $c(\G)$ is the constant from \cref{prop:GRQCdesigns}.
\end{proposition}
\begin{proof}

First we prove \eqref{eq:localTIMES} and \eqref{eq:equi-time-for-slh} (RQC with Haar random gates and SLH). To this end, we simply invoke the bounds on spectral gaps contained in \cref{prop:RQChighdegree} and combine them with \cref{th:equid-gap} (specifically conditions \eqref{eq:condition-lambda-gamma} and \eqref{eq:condition-lambda-states-gamma}), which states that spectral gap implies equidistribution.

To prove \eqref{eq:grqc-times}, we combine \cref{th:equid} 
(saying that $\delta$-approximate $k$-expanders equidistribute for high enough $k$ and low enough $\delta$) and \cref{prop:GRQCdesigns} relating high-degree designs with very deep random quantum circuits. We simply insert the values $\delta$ and $k$ from \cref{th:equid} (Eq.~\eqref{eq:equi-delta-k-unitary} for unitaries and Eq.~\eqref{eq:equi-delta-k-states} for states) into the expression bounding lengths of circuits from \cref{prop:GRQCdesigns}, specifically ~\eqref{eq:time-crqc}.
\end{proof}

\begin{remark}
Qualitatively \cref{lem:eqequidistrTIME} states that for local random quantum circuits acting on an $n$-qudit system, the timescale needed for equidistribution on scale $\ep$ with parameters $\alpha,\beta$ satisfying $\Gamma=\min\lbrace{1-\alpha,\beta-1\rbrace}$ is
\begin{equation}
    \tau_{RQC}= \exp(\Theta(n))\log\left(\frac{1}{\Gamma\ep}\right)
\end{equation}
for $\nu=\nurqc$ and 
\begin{equation}
     \tau_G= \exp(\Theta(n))\log\left(\frac{1}{\Gamma\ep}\right)^3
\end{equation}
for $\nu=\nugrqc$, i.e. general $(G,\G)$-local circuits for which global (i.e. moment independent) spectral gap is not necessarily established. 
\end{remark}

\section{Maximal complexity and approximate equidistribution}
\label{sec:MaxcompApproxequi}

In this section we prove that ensembles of quantum states and unitaries which exhibit the approximate equidistribution property describe high-complexity unitary states and channels. Specifically, we prove that 
\begin{itemize}
    \item Typical states and unitaries from these ensembles have essentially maximal complexity (\cref{lem:complexity-equidist}).
    \item The ensembles themselves contain doubly-exponentially many (in the number of qudits $n$) high-complexity unitaries and states 
    (\cref{lem:coveringNUMBERhighcomplexityEquidistribution}).
\end{itemize}

\subsection{Typical values of complexity for approximately equidistributing ensembles}

We begin by stating the following upper bounds on the complexity of arbitrary states and unitaries with respect to a general $\ell$-local gate set $\gset$, with $\ell=O(1)$.

\begin{lemma}\label{lem:ComplexityUpperBound}
Let $\G$ be an arbitrary $\ell$-local universal gate set on a Hilbert space $\H=(\C^q)^{\ot n}$ of dimension $\dim=q^n$. Then for an arbitrary unitary $\U\in\UU(\dim)$ and a state $\psi\in\S(\dim)$ we have 
\begin{equation}\label{eq:upper_bound_compl}
    C_\ep(\U)  \leq  A \dim^2  \log\left(\dim^2/\ep\right)^\gamma    \ \ ,\ \  C_\ep(\psi) \leq A (2\dim-2)  \log\left(\dim/\ep\right)^\gamma\,,
\end{equation}
where $A$ is a constant depending on $\G$ and $\gamma<3$ is the constant appearing in the Solovay-Kitaev theorem \cite{Kitaev2002}.
\end{lemma}

\begin{proof}
The proof of this result can be found in \cite{NielsenBook} but we reproduce it here for completeness.  
The idea is to use an \emph{exact} procedure for compilation of unitary gates given in Ref.~\cite{QuditCompilation2005} and then apply Solovay-Kitaev theorem. We explicitly consider the case of unitaries as the argument for states is virtually identical. Ref.~\cite{QuditCompilation2005} gives an algorithm that allows one to write every unitary $\U\in\UU(\dim)$ as a product $\U=\prod_{i=1}^{N_{\mathrm{gates}}}\U_i $, where the number of gates $N_{\mathrm{gates}}=O(\dim^2)$ and the gates $\U_i$ act non-trivially on at most two qubits. Every such two qudit gate can approximated to an accuracy of $\tilde{\ep}$ by a sequence of gates from $\gset$ using the Solovay-Kitaev theorem (which also holds for gate sets that are not symmetric, i.e.\ do not necessitate the inclusion of inverses). Specifically, it suffices to use $L_{\mathrm{SK}}(\tilde{\ep})=O(\log(1/\tilde{\ep})^\gamma)$ gates from $\gset$, where the constant hidden in $O(\cdot)$ depends on the locality $\ell$, to exactly realize a gate $\tilde{\U}_i$ that approximates $\U_i$ to accuracy $\tilde{\ep}$ in distance $\d$. Using the standard telescopic argument, we obtain that the gate $\tilde{\U}=\prod_{i=1}^{N_{\mathrm{gates}}}\tilde{\U}_i$ approximates the target unitary $\U$ to accuracy $\delta=N_{\mathrm{gates}}\tilde{\ep}$. By setting $\tilde{\ep}=\ep/\dim^2$ we see that it is possible to approximate the target gate to an accuracy $\ep$ using a total of $N_{\mathrm{gates}}\L_{SK}(\dim^2/\ep)$ gates. Inserting bounds for these quantities establishes first inequality in Eq.~\eqref{eq:upper_bound_compl}. The argument for pure quantum states is analogous. The only difference is that in this case $N_{\mathrm{states}}=O(\dim)$ two qudit gates suffice to obtain any pure quantum state from $\ket{0}^{\ot n}$ \cite{QuditCompilation2005}. 
\end{proof}

\begin{remark}
We note that in light of the results in Ref.~\cite{Varju2013} (see also \cite{OSH2020}) the gate set $\gset$ does not have to be symmetric, i.e.\ it is not required to contain inverses. Moreover, for the so-called \emph{magic gate sets} $\gset$ (for which the corresponding transition operator has a gap) one can chose $\gamma=1$ \cite{Harrow2002}. 
\end{remark}

\begin{lemma}\label{lem:complexity-equidist}
Let $\G$ be an arbitrary $\ell$-local universal gate set acting on a Hilbert space $\H=(\C^q)^{\ot n}$ of dimension $\dim=q^n$. Let $\U$ be a random unitary channel chosen according to a measure $\nu$, which $(\alpha,\beta)$-equidistributes on scale $\ep$. Then for any $\Delta \in (0,1)$:
\begin{equation}\label{eq:equidistrCOMPLbound}
    \underset{\U\sim \nu}{\Prob}\left( C_\ep(\U) \geq  \frac{(\dim^2-1) \log\frac{1}{\cupper\,\beta\ep} - \log\frac{1}{\Delta} }{\log|\gset|}\right)\geq 1-\Delta\,.
\end{equation}
Moreover, let  $\psi$ be a random pure state chosen according to $\nu_\S$ which $(\alpha,\beta)$-equidistributes on scale $\ep$. We then have
\begin{equation}
    \underset{\psi\sim \nu_\S}{\Prob}\left( C_\ep(\psi) \geq  \frac{ (2 \dim-2) \log\frac{1}{\beta\ep} - \log\frac{1}{\Delta} }{\log|\gset|}\right)\geq 1-\Delta\,.
\end{equation}
\end{lemma}

Note that up to a factor polynomial in $n$ and the factor of $1/\log|\G|$, the lower bounds matches upper bound from \cref{lem:ComplexityUpperBound}. The latter does not play a significant role as the gate sets used in quantum computing have $|\gset|=\poly(n)$. 
Thus, \cref{lem:complexity-equidist} shows that most unitaries and states coming from measures that exhibit approximate equidistribution are, in essence, maximally complex. Finally, let us remark that setting $\beta=1$ in the above lemma allows us to derive statements for the Haar measure $\mu$ on $\UU(\dim)$ as well as the induced uniform measure $\mu_\S$ on the set of pure states $\S(\dim)$.  

\begin{proof}[Proof of \cref{lem:complexity-equidist}]
We first tackle the case of unitary complexity. We shall bound the volume of all unitaries that have complexity at most $r$: $\CC^{r}_\ep=\{\U:  \exists \V\in\gset^l,\ l\leq r\ ~\text{s.t.}~ \d(\U,\V) \leq \ep\}$. By definition, the set of unitaries with $\ep$-complexity smaller than $r$ is a union of $\ep$-balls centered on circuits from $\gset^l$ for $l\leq r$,
\begin{equation}
    \CC^{r}_\ep =\bigcup_{l\leq r} \bigcup_{\V\in\gset^l} B(\V,\ep)\,.
\end{equation}
Therefore, we have 
\begin{equation}\label{eq:smallComplexitySet}
    \nu\left(\CC^{ r}_\ep\right)\leq
    \sum_{l\leq r} \nu\left(\bigcup_{\V\in\gset^l} B(\V,\ep)\right) \leq 
    \sum_{l\leq r} |\gset|^l \nu(B(\h{I,}\ep))
    \leq |\gset|^{r+1} (\beta \cupper\, \ep )^{\dim^2-1} \,,
\end{equation}
where we use the fact that for $a\geq 2$ we have 
$1 + a + ... + a^r \leq a^{r+1}$, and the last inequality comes from the equidistribution property and  Eq.~\eqref{eq:szarek}. Thus, if 
\begin{align}
|\gset|^{r+1} (\beta \cupper\, \ep )^{\dim^2-1} \leq \Delta \,,
\end{align}
then with probability at least $1-\Delta$, a unitary  $\h{U}\in \UU(\dim)$  has complexity larger than $r$.  We obtain the lemma by solving the above inequality for $r$. The case of state complexity is proven in a completely analogous manner along with the use of Eq.~\eqref{eq:pureSTATESvol}. 
\end{proof}

\subsection{Number of high complexity unitaries and states}

Let us draw our attention to another feature that justifies the claim that probability distributions which approximately equidistribute are, in a sense, maximally complex. The following lemma gives a lower bound on the packing number of high complexity unitaries and states in such ensembles. Interestingly, unless $r$ is very large this number is comparable to the packing number of unitary channels $\UU(\dim)$ and quantum states $\S(\dim)$, respectively.

\begin{lemma}[The number of distinct high-complexity unitaries and states is large in equidistributed ensembles]\label{lem:coveringNUMBERhighcomplexityEquidistribution}
Let $\nu$ be a measure on $\UU(\dim)$ that $(\alpha,\beta)$-equidistributes on a scale $\ep$. Let $\ucre=\{\U\in\supp(\nu): C_\ep(\U)> r\}$ be the set of unitaries from the support of measure $\nu$ with complexity larger than $r$. Then, provided
\begin{equation}\label{eq:rcondUNIT}
    r< (\dim^2-1)\frac{\log(A/\ep)}{\log(|\gset|)}-2\,, \quad\text{where}\quad A=\frac{\clower}{4\beta (\cupper)^2}\,, 
\end{equation}
the packing number of $\ucre$, i.e.\ the number of high-complexity unitaries from the support of $\nu$, is lower bounded by
\begin{equation}\label{eq:COMPLEXpackUNI}
    \npack(\ucre,\ep)\geq\frac{1}{2} \npack(\supp(\nu),\ep)\,,
\end{equation}
where $\npack(\supp(\nu),\ep)\geq\vol(2  \ep)^{-1}\geq(\frac{1}{2\ep \cupper})^{\dim^2-1}$.

Moreover, let $\nu_\S$ be a measure on $\S(\dim)$ that $(\alpha,\beta)$-equidistributes on scale $\ep$. Furthermore, let $\scre=\{\psi\in\supp(\nu_\S): C_\ep(\psi)> r\}$ be the set of states from $\supp(\nu_\S)$ with complexity larger than $r$. Then, provided 
\begin{equation}\label{eq:rcondSTATE}
    r< (2\dim-2)\frac{\log(1/(4\beta \ep))}{\log(|\gset|)}\  -2\,,
\end{equation}
the packing number of $\scr$, i.e.\ the number of high-complexity states from the support of $\nu_\S$, is lower bounded by
\begin{equation}\label{eq:COMPLEXpackSTATE}
    \npack(\scre,\ep)\geq \frac{1}{2} \npack(\supp(\nu_S),\ep)\,,
\end{equation}
where $\npack(\supp(\nu_\S),\ep)\geq\vol_\S(2 \beta \ep)^{-1}\geq(\frac{1}{2\ep \beta})^{2\dim-2}$.
\end{lemma}

\begin{proof}
We describe in detail the case of unitary channels. The arguments for quantum states are analogous.
We have $\supp(\nu)=\ucre\cup \CC^r_\ep(\nu)$, where $\CC^r_\ep(\nu)=\CC^r_\ep \cap\supp(\nu)$ are unitaries from the support of $\nu$ with complexity at most $r$. Using the inequality in Eq.~\eqref{eq:npack-sum-rule} we have 
\begin{align}
    \npack(\supp(\nu),\ep)\leq \npack(\ucre,\ep) +\npack\bigl(\CC^r_\ep(\nu),\ep\bigr)\,.
\end{align}
Then using the characterization of $\CC^r_\ep$ given in Eq.~\eqref{eq:smallComplexitySet} and again employing the bound Eq.~\eqref{eq:npack-sum-rule}, we find
\begin{equation}
    \npack\bigl(\CC^r_\ep(\nu),\ep\bigr) \leq \sum_{l\leq r}\sum_{\U\in \gset^l}
        \npack(B(\U,\ep)\cap\supp(\nu),\ep)\leq |\gset|^{r+1} \npack(B(\U,\ep),\ep)\,,  
\end{equation}
where in the last inequality we used the monotonicity of the packing number subject to inclusion of sets.
Furthermore, since any disjoint balls of radius $\ep$ that have centers in $B(\U,\ep)$ are contained in $B(\U,2\ep)$, a simple volume comparison gives that $\npack(B(\U,\ep),\ep) \vol(\ep) \leq \vol(2\ep)$, which in turn yields
\begin{equation}\label{eq:packBALLuni}
   \npack(B(\U,\ep),\ep) \leq \frac{\vol(2\ep)}{\vol(\ep)}\leq   \left(\frac{2\cupper}{\clower} \right)^{\dim^2-1} \,,
\end{equation}
where in the second inequality we used Eq.~\eqref{eq:szarek}. Altogether, we obtain 
\begin{align}
    \npack(\ucre,\ep) \geq \npack(\supp(\nu),\ep)-|\gset|^{r+1}  \left(\frac{2\cupper}{\clower} \right)^{\dim^2-1} \,.
\end{align}
Now, choosing $r$ in such a way that $|\gset|^{r+1}  \left(\frac{2\cupper}{\clower} \right)^{\dim^2-1} \leq \frac{1}{2} \npack(\supp(\nu),\ep)$ 
subsequently ensures that $\npack(\ucre,\ep) \geq \frac{1}{2}\npack(\supp,\ep)$. To estimate $r$ we use the following bound lower on $\npack(\supp(\nu),\ep)$:
\begin{equation}\label{eq:loverCOVsupp}
    \npack(\supp(\nu),\ep) \geq  \frac{1}{\vol(\beta\cdot 2 \ep)}\,.
\end{equation}
The above inequality follows from the relation $\npack(\supp(\nu),\ep) \geq \ncov(\supp(\nu),2\ep)$, which in turn follows from the fact that the number of balls needed to fully cover the space cannot increase if we increase their radii, as well as the bound
\begin{equation}
    1\leq  \ncov(\supp(\nu),2\ep)\vol(\beta\cdot 2\ep)\,,
\end{equation}
which follows from the definition of the covering number and the equidistribution property. Finally, from Eq.~\eqref{eq:loverCOVsupp} and Eq.~\eqref{eq:szarek-pack}, we find that
\begin{equation}\label{eq:compar}
|\gset|^{r+1}  \left(\frac{2\cupper}{\clower} \right)^{\dim^2-1} 
    \leq \frac12
    \left(\frac{1}{ \cupper \beta 2\ep}\right)^{\dim^2-1}\ .  
\end{equation}
It can be easily seen that Eq.~\eqref{eq:compar} is implied by Eq.~\eqref{eq:rcondUNIT}. This concludes the proof of the first part of the lemma. 

The proof for pure states proceeds identically. Just like for unitaries we separate the support $\supp(\nu_\S)$ into states of high complexity, $\scre$, and low complexity, the complement of $\scre$ denoted by $(\scre)^c$. For these states we apply Eq.~\eqref{eq:npack-sum-rule} and proceed as before by (i) upper bounding of $\npack((\scre)^c,\ep)$ and (ii) lower bounding $\npack(\supp(\nu_S),\ep)$. Upon implementing (i) we use the estimate  
\begin{equation}
    \npack(B(\psi,\ep),\ep) \leq \frac{\vol_\S(2\ep)}{\vol_S(\ep)}=2^{2\dim-2} \,,
\end{equation}
which is proven analogously to Eq.~\eqref{eq:packBALLuni}.
\end{proof}

\section{Recurrence of complexity from probability distributions exhibiting approximate equidistribution}
\label{sec:CompRec}

Here we show that the notion of equidistribution allows us to derive strong \emph{quantitative statements} concerning the recurrence of circuit complexity in random quantum circuits. Before moving on to formal results and proofs, we first present informal reasoning that justifies, on an intuitive level, why approximate equidistribution can be relevant for studying the recurrence of complexity. 

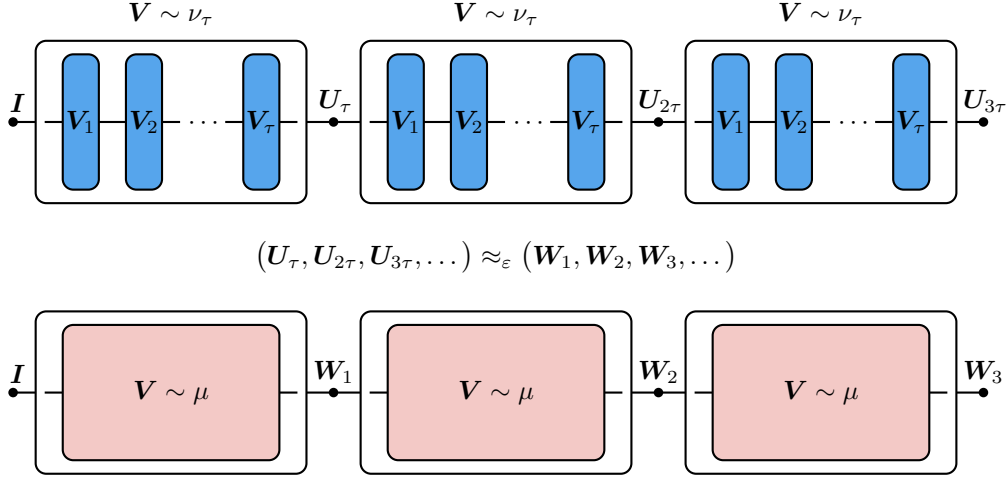
\begin{figure}
    \centering
    \definecolor{bb}{rgb}{0.337,0.647,0.925}
    \definecolor{lr}{rgb}{0.88,0.412,0.38}
    \begin{tikzpicture}[scale=0.6,thick]
    \foreach[count=\i] \p in {-7.2,0,7.2}
    {\foreach \y in {0,-6}
    {\draw[rounded corners] (\p+0.4,5.8+\y) rectangle (\p+6.4,2.2+\y);
    \draw (\p+0.4,4+\y) -- (\p-0.1,4+\y); \draw (\p+6.4,4+\y) -- (\p+7,4+\y);
    \fill[radius=3pt] (\p+7,4+\y) circle;}
    \draw (\p+0.6,4) -- (\p+3.6,4); \draw (\p+4.6,4) -- (\p+6.2,4);
    \node at (\p+4.17,3.98) {$\cdots$};
    \foreach \x in {1,2.4,5}
    {\draw[fill=bb,rounded corners] (\p+\x,5.5) -- ++(0,-3) -- ++(0.8,0) -- ++(0,3) -- cycle;}
    \foreach[count=\j] \x in {1,2.4} {\node at (\p+\x+0.4,4) {$\V_\j$};}
    \node at (\p+5+0.4,4) {$\V_\tau$};
    \node at (\p+3.4,6.4) {$\V\sim \nu_\tau$};
    \node[anchor=south] at (\p+7.05,4) {\ifnum\i=1 {$\U_\tau$} \else {$\U_{\i\tau}$}\fi};
    \draw (\p+0.6,4-6) -- (\p+1,4-6); \draw (\p+5.8,4-6) -- (\p+6.2,4-6);
    \draw[fill=lr,fill opacity=0.36,rounded corners] (\p+1,5.5-6) rectangle (\p+1+0.8+4,5.5-6-3);
    \node at (\p+3.4,4-6) {$\V\sim \mu$};
    \node[anchor=south] at (\p+7,4-6) {$\W_\i$};
    }
    \foreach \y in {0,-6} {\fill[radius=3pt] (-7.3,\y+4) circle; \node[anchor=south] at (-7.2,\y+4) {$\I$};}
    \node at (3.4,1) {$\big(\U_{\tau},\U_{2\tau},\U_{3\tau},\ldots\big) \approx_\ep \big(\W_1, \W_2, \W_3, \ldots\big)$};
    \end{tikzpicture}
    \caption{Illustration of the main idea allowing us to bound recurrence times for a random walk $\U_t$ generated by a probability measure $\nu$. Assume that $\U_t$ satisfies approximate equidistribution on a scale $\ep$ after $\tau$ time steps. Then the statistical properties of the sequence $(\U_\tau,\U_{2\tau},\ldots,\U_{j\cdot\tau})$ on the scale $\ep$ are similar to that of $(\W_1,\W_2,\ldots,\W_{j})$, where the unitaries $\W_i$ are independently chosen according to Haar measure $\mu$ on $\UU(\dim)$.
    }
    \label{fig:eqdistwalk}
\end{figure}

The main idea is that the space of quantum states or unitaries gets divided into objects of low and high complexity (for unitaries we denoted them as $\CC^r_\ep$ and $\ucr$, respectively). These sets have well-defined volumes, computed according to the Haar measure on $\UU(\dim)$ and the uniform measure on $\S(\dim)$. Consider now the ensemble of unitaries $\h{U}_t$ induced by a measure $\nu_t=\nu^{\ast t}$ resulting from $t$ steps of a random walk $\nu$ (specified by the random quantum circuit model) and initialized at identity $\h{I}$. Assume that at time $\tau$ the measure $\nu_\tau$ approximately equidistributes on the scale $\ep$. Then the probability of a unitary $\h{U}_\tau$ having complexity smaller than $r$ roughly scales like the Haar measure of low-complexity unitaries, i.e.\ $\vol(\CC^r_\ep)$. Importantly, the approximate equidistribution property does not depend on the initial state of a random walk and therefore the probability that $\h{U}_{2\tau}$ has complexity smaller than $r$ again exhibits the same behaviour, \emph{regardless} of the state of the walker at time $\tau$. The same holds for the subsequent multiples of $\tau$. It follows that, when observed every $\tau$ times, the walker effectively behaves as if at every interrogation time $t=j\cdot \tau$ the unitaries $\h{U}_t$ are independent random variables distributed according to the Haar measure on $\UU(\dim)$, as illustrated in \cref{fig:eqdistwalk}. Therefore, the expected time of return to a low complexity region goes as $\vol(\CC^r_\ep)^{-1}$, which scales doubly-exponentially in the number of qudits $n$. Importantly, our findings do not depend on the details of the measure $\nu$ on $\UU(\dim)$, which is used to specify the walk.

We start with the following crucial lemma that gives insight into the typical long-time behaviour of the complexity of the random sequences of unitaries   $(\h{U}_{t})_{t\in\mathbb{N}}$ (resp.\ states $(\psi_{t})_{t\in\mathbb{N}}$) generated by a random circuit architecture $\nu$. The lemma will allow us to prove the main result of this section contained in \cref{th:recurrence}.

\begin{lemma}[Bounds on recurrence events for random quantum circuits]
\label{lem:bouns-rec-events}
Consider a system of $n$ qudits described by a Hilbert space $\H=(\C^q)^{\ot n}$ of dimension $\dim=q^n$. Let $\nu$ be a measure (random quantum circuit architecture) on $\UU(\dim)$ that defines walks $\U_{t}$, $\psi_t$ in their respective spaces, as in Eq.~\eqref{eq:randomWALKdef}. Let $\tau$ and $\tau_\S$ be the corresponding equidistribution times of $\nu_t$ and $\nu_{\S,t}$ on the scale of $\ep$ (with parameters $\alpha,\beta$, as in \cref{def:equidistr}). We then have the following inequalities
\begin{equation}\label{eq:lowerBOUNDrecurUnitary}
\Prob\left(\text{There exists}~ k= 0 ,\ldots,  K-1 \ ~\text{s.t.}~  C_{\ep}(\U_{\tau+k})\leq r \right)  \leq K |\gset|^{r+1} \vol(\beta \ep)\,,
\end{equation}
\begin{equation}\label{eq:lowerBOUNDrecurState}
\Prob\left(\text{There exists}~ k= 0 ,\ldots,  K-1 \ ~\text{s.t.}~ C_{\ep}(\psi_{\tau_\S+k})\leq r \right)  \leq K |\gset|^{r+1}  \vol_\S(\beta \ep)\,.
\end{equation}
Moreover, for $K\geq \tau$ (resp.  $K\geq \tau_\S$ for states)  we also have bounds in the opposite direction
\begin{equation}\label{eq:upperBOUNDrecurUnitary}
\Prob\left(\text{There exists}~ k=1,\ldots, K\ ~\text{s.t.}~ C_{\ep}(\U_{\tau+k}) \leq r \right)  \geq 1- \bigl(1 -\npack(\gset^r,\ep)\vol(\alpha\ep)\bigr)^{\lfloor\frac{K}{\tau}\rfloor}\,,
\end{equation}
\begin{equation}\label{eq:upperBOUNDrecurStates}
\Prob\left(\text{There exist}~  k=1,\ldots, K\ ~\text{s.t.}~  C_{\ep}\left(\psi_{\tau_\S+k}\leq r\right) \right)  \geq 1-  \bigl(1 -\npack(\S^r,\ep)\vol_\S(\alpha\ep)\bigr)^{\lfloor\frac{K}{\tau}\rfloor}\,,
\end{equation}
where $\npack(\gset^r,\ep)$ ($\npack(\S^r,\ep)$) denotes the  packing number of the set of unitaries (states) that can be realized as depth $r$ circuits formed by gates in  $\gset$.
\end{lemma}
\begin{proof}
We start with the proof for unitary channels. Consider first the inequality Eq.~\eqref{eq:lowerBOUNDrecurUnitary}.
We  apply the union bound:
\begin{align}
    \Prob\left(\text{There exists } k= 0 ,\ldots,  K-1 \ \text{s.t.}\  C_{\ep}(\U_{\tau+k})\leq r \right)
    \leq 
     \sum_{k= 0}^{K-1}\Prob(C_{\ep}(\U_{\tau+k})\leq r)
    = \sum_{k= 0}^{K-1}  \nu^{\ast (\tau+k)} \left(\CC^r_\ep\right)\,.
\end{align}

Now, by definition  we have $ \CC^r_\ep =\bigcup_{l\leq r} \bigcup_{\V\in\gset^l} B(\V,\ep)$ and as a result
\begin{equation}
    \nu^{\ast (\tau+k)}\left(\CC^r_\ep\right)\leq
    \sum_{l\leq r}\sum_{\V\in\gset^l} \nu^{\ast (\tau+k)}\left( B(\V,\ep)\right) \leq 
    \sum_{l\leq r} |\gset|^l \vol(\beta \ep)
    \ .
\end{equation}
In the first inequality we again used the union bound, while in the second we applied \cref{fact:equi}, which implies that since $\nu^{\ast \tau}$ equidistributes, so does $\nu^{\ast (\tau+k)}$. 

Consider now the inequality in Eq.~\eqref{eq:upperBOUNDrecurUnitary}.
To prove it, we will upper bound the probability of the opposite event
\begin{align}
    \Prob\left(\text{For all } k=1,\ldots,K \, \,    \CC_{\ep}(\U^{(\tau+k)}) > r \right)\,.
\end{align}
Denoting $\CA=\{\U: \CC_\ep(\U)>r\} $ and using the Markov property we obtain
\begin{align}
&\Prob\left(\text{For all}~ k=1,\ldots,K \, \,    \CC_{\ep}(\U_{\tau+k}) > r \right) = 
\Prob\left(\text{For all}~ k=1,\ldots,K \, \,    \U_{\tau+k} \in \CA \right) \nn
&\leq\Prob \left(
\U_{2\tau}\in \CA,
\U_{3\tau}\in \CA,\ldots,
\U_{\tau+K/\tau}\in \CA
\right) \nn
&= \Prob\left(\U_{2\tau} \in \CA)\, \Prob(\U_{3\tau} \in \CA | \U_{2\tau}\in \CA) \ldots  
\Prob(\U_{\tau+K/\tau} \in \CA | \U_{K/\tau}\in \CA\right)\,,
\end{align}
where for clarity, we have assumed that  $K/\tau$ is an integer. 
We now have
\begin{align}
\label{eq:l-to-VU}
    \Prob\left(\U_{(l+1)\tau}\in \CA|
    \U_{l\tau}\in \CA\right)
    =  
    \Prob\left(\V \U\in \CA|
    \U \in \CA\right)\,,
\end{align}
where $\V$ is distributed according to $\nu^{\ast \tau}$ and $\U$ according to $\nu^{\ast l \tau}$
To avoid complicated notation, we shall now assume that $\nu^{\ast l \tau}$ is a discrete measure (the proof for the case of a continuous measure is analogous) and denote it by $\lbrace{p_\alpha,\U_\alpha\rbrace}$. We then write 
\begin{align}
\label{eq:VU-to-nutau}
    \Prob\left(\V \U\in \CA|
    \U \in \CA\right) = \sum_{\U_\alpha\in \CA}p_\alpha \nu^{\ast \tau}_{\U_\alpha} (\CA)\,,
\end{align}
where $\nu^{\ast \tau}_{\U_\alpha}= \nu^{\ast \tau} \ast \delta_{\U_\alpha}$ with $\delta_{\U}$ being a Dirac delta measure concentrated on $\U$. Since, by definition, $\CA = \UU(\dim) \setminus \CC_\ep^r$ and 
the set $\CC_\ep^r$ contains an $\ep$-ball centered around some $\U_0$, we have 
\begin{align}
    \nu^{\ast \tau}_{\U_\alpha} (\CA) 
    = 1 -  \nu^{\ast \tau}_{\U_\alpha}(\CC_\ep^r)
     \leq 
    1 -  \nu^{\ast \tau}_{\U_\alpha}(B(\U_0,\ep) ) \leq 
    1 - \vol(\alpha \ep)\,,
\end{align}
where we used the equidistribution property of $\nu^{\ast \tau}$. From Eqs.~\eqref{eq:l-to-VU} and \eqref{eq:VU-to-nutau} we get that for integer $l\geq 1$
\begin{align}
    \Prob\left(\U_{(l+1)\tau}\in \CA|
    \U_{l\tau}\in \CA\right) \leq 1- \vol(\alpha \ep)\,.
\end{align}
Since $\nu^{\ast 2 \tau}$ equidistributes, we have the same estimate for $\Prob\left(\U_{2\tau}\in \CA\right)$, and we obtain
\begin{align}
    \Prob\left(\text{For all } k=1,\ldots,K \, \,    \CC_{\ep}(\U_{\tau+k}) > r \right) \leq (1-\vol(\alpha \ep))^{K/\tau}\,,
\end{align}
which gives the sought-after lower bound  
\begin{equation}
   \Prob\left(\text{There exists}~ k=1,\ldots, K\ ~\text{s.t.}~ C_{\ep}(\U_{\tau+k}) \leq r \right)  \geq 1- (1 -\vol(\alpha \ep))^{\frac{K}{\tau}}\,. 
\end{equation}
If $K$ is not a multiple of $\tau$, the formula remains valid if we replace $K/\tau$ with $\lfloor K/\tau \rfloor$. But we can obtain a better bound on 
$\nu^{\ast \tau}_{\U} (\CC_\ep^r) $ by considering the set $X\subset \CC_\ep^r$ of disjoint $\ep$ balls centered in $\gset^r$, i.e.\ words of length $r$ formed from unitaries in $\gset$. The maximal number of such balls is the packing number $\npack(\gset^r,\ep)$.
Then we have that
\begin{align}
     \nu^{\ast \tau}_{\U}(\CC_\ep^r)
     \geq \sum_{\W  \in X} 
     \nu^{\ast \tau}_{\U}(B(\W,\ep/2))
     \geq \npack(\gset^r,\ep) \vol(\alpha \ep)\,,
\end{align}
which gives us that
\begin{align}
\Prob\left(\text{There exists}~ k=1,\ldots, K\ ~\text{s.t.}~ C_{\ep}(\U_{\tau+k}) \leq r \right)  \geq 1- \left(1 -\npack(\gset^r,\ep)\vol\left(\alpha \ep/2\right)\right)^{\frac{K}{\tau}}\,.
\end{align}
The proof for the case of pure states follows very similar reasoning, with the only notable difference being that it requires repeated usage of the equidistribution property of $(\nu^{\ast \tau})_\S$ and the measures related to it.
\end{proof}

We are now ready to present the main result of this section, which is a technical formulation of items (i) and (ii) in  \cref{res:satANDrecRQC}. To establish timescales claimed there it suffices to substitute the timescales $\tau,\tau_\S$ in the bounds on the equidistribution time given in \cref{lem:eqequidistrTIME}.

\begin{theorem}[Long time behavior for random walks]
\label{th:recurrence}
Consider a system of $n$ qudits described by a Hilbert space $\H=(\C^q)^{\ot n}$ of dimension $\dim=q^n$. Let $\nu$ be a measure (random quantum circuit architecture) on $\U(\dim)$ that defines walks $\h{U}_t$, $\psi_t$ in the space of unitary channels (as in Eq.~\eqref{eq:randomWALKdef}). Let $ \tau,  \tau_\S$ be the corresponding $(\alpha,\beta)$-equidistribution times of these walks on the scale of $\ep$. 
 Let $\nuslh$ define the SLH walk $\h{U}_t^{SLH}$, with the corresponding equidistribution time $\tau_{SLH}$. 
Pick $\Delta,\Delta_1,\Delta_2 \in (0,1)$. Let $r_1>r_2$ and 
\begin{align} 
\label{eq:T1T2U}
    &T_1= \frac{\Delta_1}{|\gset|^{r_1+1}}\left(\frac{1}{\beta \cupper \ep}\right)^{\dim^2-1}\ ,\quad T_2 = \tau\log(1/\Delta_2) \left(\frac{2  \aalpha(r_2)}{\dim^2 (1-\ep^2)}\right)^{a(r_2)} 
    \left(\frac{1}{\alpha \clower \ep}\right)^{\dim^2-1}\,, \\
    \label{eq:T1T2SLH}
    & T_{1,SLH}= \frac{\Delta_1}{|\gset|^{r_1+2}}\frac{1}{64d^2 m}\left(\frac{1}{2\beta \cupper \ep}\right)^{\dim^2-2}\ ,\quad \text{$T_{2,SLH}$ \rm as $T_2$, with $\tau$ replaced by $\tau_{SLH}$} 
    \\
    \label{eq:T1T2S}
    &T_{1,\S}=\frac{\Delta_1}{|\gset|^{r_1+1}}\left(\frac{1}{\beta \ep}\right)^{2\dim-2},
    \quad T_{2,\S} = \tau_\S\log(1/\Delta_2) \left(\frac{  \aalpha(r_2)}{\dim (1-\ep^2)}\right)^{a(r_2)} \left(\frac{1}{\alpha  \ep}\right)^{2\dim-2}\,,
\end{align}
 with $\aalpha(r) = \lfloor(r/n^2c(\gset))^{1/11}\rfloor $
 
where $c(\gset)$ is a constant that depends only on gate set $\gset$ and $\clower, \cupper$ are as in \eqref{eq:szarek}.  Assume further that $r_1$ is such that $T_1, T_{1,SLH}, T_{1,\S} \geq 1$.

Then we have $T_2 \geq T_1, T_{2,SLH} \geq T_{1,SLH}, T_{2,\S} \geq T_{1,\S}$ and with probability greater than $1- \Delta- \Delta_1-\Delta_2$:
\begin{itemize}
    \item[(a)] Saturation of complexity at time $\tau$ ($\tau_\S$)
    \begin{align}
       C_\ep(\h{U}_\tau), \,C_\ep(\h{U}_{\tau^{SLH}}^{SLH})\geq \frac{(\dim^2-1) \log\frac{1}{\cupper\,\beta\ep} - \log\frac{1}{\Delta} }{\log|\gset|},\quad  C_\ep(\psi_{ \tau_{\S}})\geq\frac{(2\dim-2) \log\frac{1}{\beta\ep} - \log\frac{1}{\Delta} }{\log|\gset|}\, ,
    \end{align}
     \item[(b)] 
    Large complexity for {\bf all times} $t$ satisfying $\tau \leq t\leq  \tau  + T_1 $ ( $\tau_{SLH} \leq t\leq \tau_{SLH}+ T_{1,SLH}$,  $\tau_\S \leq t\leq \tau_{\S}+ T_{1,\S}$)
    \begin{align}
        C_\ep(\h{U}_t) > r_1\, , \quad  C_\ep(\psi_t)> r_1 \ .
    \end{align}   
    \item[c)] Small complexity for {\bf some time} $t$ satisfying $ \tau+ T_1< 
    t\leq  \tau +  T_2$ 
    ( $ \tau_{SLH}+ T_{1,SLH}< 
    t\leq  \tau_{\S} + T_{2,SLH}$,   $
     \tau_{\S}+ T_{1,\S}< 
     t\leq  \tau_{\S} + T_{2,\S}$)
    \begin{align}
         C_\ep(\h{U}_t) \leq r_2\, , \quad  C_\ep(\psi_t)\leq r_2 \ .
    \end{align}
\end{itemize}
\end{theorem}

\begin{remark}
For $r_2=0$ (recurrence to trivial complexity), the expressions for the above time-scales simplify significantly: 
\begin{align} 
\label{eq:T1T2U-simple}
    &T_1=\frac{\Delta_1}{|\gset|^{r_1+1}}\left(\frac{1}{\beta \cupper \ep}\right)^{\dim^2-1}\ ,\quad T_2= \tau\log(1/\Delta_2) 
    \left(\frac{1}{\alpha \clower \ep}\right)^{\dim^2-1}, \\
    &T_{1,S}= \frac{\Delta_1}{|\gset|^{r_1+1}}\left(\frac{1}{\beta \ep}\right)^{2\dim-2},
    \quad T_{2,S} = \tau_\S\log(1/\Delta_2)  \left(\frac{1}{\alpha  \ep}\right)^{2\dim-2}. 
\end{align}
It is clear that both lower bounds ($T_1,T_{1,\S}$) as well as upper bounds ($T_2,T_{2,\S}$) for the recurrence time exhibit the same exponential dependence on the inverse of the accuracy parameter $\ep$ (ignoring the dependence on the equidistribution times $\tau,\tau_S$). 
\end{remark}

\begin{proof}
We  present the proof just for  unitaries within the random circuit model as the reasoning for states is proceeds analogously.
We then point out changes needed for the proof in the case of the unitary SLH model.
Let us define the following three events in the space of realizations of the random walk $(\U_t)_{t\in\mathbb{N}}$:
\begin{equation}
\begin{split}
    I &= \left\lbrace C_\ep(\h{U}_\tau)\geq \frac{(\dim^2-1) \log\frac{1}{\cupper\,\beta\ep} - \log\frac{1}{\Delta} }{\log|\gset|} \right\rbrace\,,\\
    II &= \{C_\ep(\U_t)\geq r_1 ~\text{for all}~ t ~\text{s.t.}~ \tau \leq t \leq  \tau+T_1 \}\,,\\
    III &= \{C_\ep(\U_t)<r_2 ~\text{for some}~ t ~\text{s.t.}~ 
     \tau \leq t \leq \tau+ T_2 \}\,, \\ 
    IV &= \{C_\ep(\U_t)<r_2 ~\text{for some}~ t ~\text{s.t.}~ 
     \tau+T_1\leq  t \leq \tau+ T_2 \}\ .
\end{split}
\end{equation}
As for event $I$, by the definition of the equidistribution time $\tau$, \cref{lem:complexity-equidist} ensures that $\Pr(I)\geq 1-\Delta$. 

To control the probability of event $II$, we use \cref{lem:bouns-rec-events}. In conjunction with 
lower and upper bounds for the volume of balls given in  Eq.~\eqref{eq:szarek},
Eq.~\eqref{eq:lowerBOUNDrecurUnitary} implies that, for $T_1$ as given by first inequality of Eq.~\eqref{eq:T1T2U}, we have $\Prob(II)\geq 1- \Delta_1$.  

Next, to deal with event $III$ we use \cref{lem:bouns-rec-events} and the lower bounds on the packing number $\npack(\gset^r,\ep)$ from \cref{lem:lowerboundLOWcomplexity} (the same lemma contains the lower bound of the packing number $\npack(\S^r,\ep)$, relevant for states). Namely, the inequalities Eq.~\eqref{eq:upperBOUNDrecurUnitary} and Eq.~\eqref{eq:lem-pack-U} along with Eq.~\eqref{eq:szarek} imply that, for $T_2$ as given by the second inequality in Eq.~\eqref{eq:T1T2U}, $\Prob(III)\geq 1- \Delta_2$.
By union bound we get $\Prob(I\cap II\cap III) \geq 1- \Delta-\Delta_1-\Delta_2$.

Finally, we note that the event $IV$ which we are interested in, rather than in $III$, satisfies $II\cap III\subset IV$, because $r_2\leq r_1$. Thus also 
$\Prob(I\cap II\cap IV) \geq 1- \Delta-\Delta_1-\Delta_2$. Note that this also proves that $T_1$ and $T_2$ in fact satisfy $T_1\leq T_2$.

To prove the results for the SLH model, the following changes are needed. Regarding the event $I$, it depends on the walk only via the value of the equidistribution time, since \cref{lem:complexity-equidist} used to bound $I$ uses only the equidistribution property. Thus, it suffices to just write $\tau_{SLH}$ instead of $\tau$.  
Regarding the event $II$, we proceed analogously as above -- the only change is that we  use \cref{prop:rec-SLH} in place of \cref{lem:bouns-rec-events} (\cref{prop:rec-SLH} is the analogue of \cref{lem:bouns-rec-events} in the case of a continuous walk). Specifically, we set $t_1=\tau_{SLH}$ and $t_2 = t_1 + T_{1,SLH}$ so that the probability of $II$ is bounded from below by $1-\Delta_1$. 
Regarding item $III$, no modification is needed, it suffices to replace $\tau$ with $\tau_{SLH}$. Indeed, 
the event $III$ in the case of the SLH walk contains as a subset an event constructed from discrete times:
\begin{align}
    III' &= \left\{C_\ep(\U_t)<r_2 ~\text{for some}~ t ~\text{s.t.}~ 
     t= 2\tau_{SLH}, 3 \tau_{SLH}\ldots, \left\lfloor \frac{T_2}{\tau_{SLH}}  \right\rfloor \tau_{SLH}\right\}\,.
\end{align}
Thus in SLH case, $\Prob(III)\geq \Prob(III')$. 
Then, as in the case of event $I$, the bound for probability of analogous event in \cref{lem:bouns-rec-events}
given by \eqref{eq:upperBOUNDrecurUnitary}
is obtained solely using equidistribution. 
Hence it is enough to replace $\tau$ with $\tau_{SLH}$ as done in \eqref{eq:T1T2S}.

\end{proof}

Here we present a lower bound on the packing number 
that we have used in the proof of \cref{th:recurrence}.

\begin{lemma}[Lower bounds for packing number of low complexity states and unitaries]\label{lem:lowerboundLOWcomplexity}
The packing number  of the set $\gset^r$ (circuits of depth $r$ generated by gates in $\gset)$ satisfies
\begin{align}
\label{eq:lem-pack-U}
\npack(\gset^r,\ep) \geq \left(\frac{\dim^2 (1-4\ep^2)}{ \aalpha(r)}\right)^{\aalpha(r)}\ ,
\end{align}
where $ \aalpha(r)  := \lfloor(r/n^2c(\gset))^{1/11}\rfloor $ and $c(\gset)$ is a constant depending on the gate set $\G$. 

Similarly, the packing number of the set $\S^r$ (states generated by depth $r$ circuits constructed from $\gset$ and applied to computationally trivial state $\psi_0$) satisfies
\begin{align}\label{eq:lem-pack-S}
\npack(\S^r,\ep) \geq \left(\frac{\dim (1-4\ep^2)}{  \aalpha(r)}\right)^{\aalpha(r)}\ .
\end{align}
\end{lemma}
\begin{remark}
For small depth $r$, Eqs.~\eqref{eq:lem-pack-U} and \eqref{eq:lem-pack-S} appear singular. However, in this regime one can easily lower bound $\npack(\gset^r,\ep)$ and $\npack(\S^r,\ep)$ by $1$, by considering a single $\ep$-ball centered around $\h{I}$ (resp.\ $\psi_0$).  
\end{remark}

\begin{proof}[Proof of \cref{lem:lowerboundLOWcomplexity}]
For the gate set $\gset$, the set $\gset^r$ contains the support of the measure given by random quantum circuits generated from gate set $\gset$, denoted as $\supp(\nu_{\rm RQC})$. To proceed with the proof, we first analyze unitary expanders, and then use the fact that random quantum circuits are approximate $k$-expanders. From \cref{lem:volUnitaryDesignsMoments} we get that for $\delta$-approximate $k$-expanders and arbitrary $\V\in\UU(\dim)$ we have $\nu(B(\V,\ep)\leq  \frac{k! + \dim^{2k}\delta}{\dim^{2k}(1-\ep^2)^{k}}$.
For local random quantum circuits, we now employ \cref{prop:GRQCdesigns} to conclude that circuits of size $T\geq c(\gset)n^2k^{11}$ form $\delta$-approximate $k$-expanders with $\delta=1/d^{2k}$. For random quantum circuits $\nu_{\rm RQC}$ of size $r$, it follows that
\begin{equation}
    \nu_{\rm RQC}\big(B(\V,\ep)\big)
    \leq\frac{k! +1}{\dim^{2k}(1-\ep^2)^{k}} \leq \left(\frac{ \lfloor (r/n^2c(\gset))^{1/11}\rfloor}{\dim^2 (1-\ep^2)}\right)^{\lfloor(r/n^2c(\gset))^{1/11}\rfloor} =: f(\ep,r)\ .
\end{equation}
We now construct a subset of elements $\supp(\nu_{\rm RQC})=\gset^{r}$ that are pairwise far apart by more than a distance $\ep$. First, take an arbitrary element $\V_1$ of $\supp(\nu_{\rm RQC})$. Since $\nu_{\rm RQC}(B(\V_1,2\ep))\leq f(2\ep,r)<1$ is strictly less than one, a condition which can easily be satisfied provided $r$ is not too large, then there must be some $\V_2$ from $\supp(\nu_{RQC})$ which does not belong to the ball, i.e.\ $\d(\V_1,\V_2)> 2 \ep$. By a union bound we then have 
\begin{align}
    \nu_{\rm RQC}(B(\V_1,2 \ep)\cup B(\V_2,2\ep)) \leq 
    2 f(2 \ep,r)<1
\end{align}
There must then exist a $\V_3$ that does not belong to both balls, $\d(\V_3,\V_1)> 2\ep$ and $\d(\V_3,\V_2)> 2\ep$.
Continuing this reasoning, 
we obtain $N$ unitaries $\{\V_i\}_{i=1,\ldots,N}$ such that the distance  $\d(\V_i,\V_j)>2\ep$ for all pairs $(i,j)$, provided that $N f(2\ep,r)<1$.
Thus the support of $\nu_{\rm RQC}$ contains at least $1/f(2\ep,r)-1$ unitaries that are pairwise far apart by $2\ep$. If we consider balls of radius $\ep$ with centers placed at those unitaries, the balls will be pairwise disjoint and so the packing number of $\gset^r$ is bounded as follows
\begin{equation}
    \npack(\gset^r,\ep) \geq \frac{1}{f(2\ep,r)}\,,
\end{equation}
which gives the sought after estimate.

The proof of the lemma for the set of states is proceeds identically to the unitary case. We construct, using a probabilistic argument, a set of pairwise disjoint $\ep$ balls  centered in states generated by depth $r$ quantum circuits acting on $\psi_0$. The only difference is a different bound on the measure of an $\ep$-ball according to probability distribution $\nu_\S$ induced on $\S(d)$ from $\delta$-approximate unitary $k$-designs on $\UU(\dim)$: $\nu_\S(B(\phi,\ep))\leq \frac{k!+\dim^k \delta}{\dim^k(1-\ep^2)^k}$, which we prove in \cref{lem:volStatesDesignsMoments}.

\end{proof}

\section{Recurrences of complexity have exponential size} \label{sec:recurrTIME}
In this section we quantify timescales on which recurrences of complexity take place for local random quantum circuits. We approach this problem by showing that under certain conditions 
\begin{equation}\label{eq:conditionPROBo}
    \Prob\Big(C_\ep(\U_{t-T})>r\, ,\, C_\ep(\U_{t+T})>r\ |\  C_\ep(\U_t)=0   \Big) \approx 1\,.
\end{equation} 
The quantity appearing above is the conditional probability in the space of realizations of a random walk $\lbrace\U_t\rbrace_{t\in\mathbb{N}}$ (see \cref{def:localRQCs}). We want to show that this probability is close to $1 $ for a large enough time $t$, and for $r$ comparable to the maximal complexity $C_{\max}$ and times $T$ comparable to the equidistribution time $\tau$. Establishing this together with the ``temporal stability'' of complexity, which is proved in \cref{lem:stabCompl} below, shows that recurrences of complexity in random quantum circuits indeed take place on exponential timescales, as conjectured in recent work by Susskind \cite{BHexpTIME2020}.

\begin{lemma}[Stability of complexity under action of local gates]\label{lem:stabCompl}
Consider a Hilbert space of $n$ qudits with local dimension $q$, $\H=(\C^q)^{\otimes n}$ of dimension $\dim=q^n$. Let $\gset$ be a universal gate set used to define complexity, as in \cref{def:complexityUnitary} and \cref{def:complexityState}. Let  $\U\in\UU(\dim)$ and $\psi\in\S(\dim)$. Furthermore, let $\h{W}_1,\h{W}_2, \ldots, \h{W}_m$ be a sequence of $l$-local gates acting in $\H$ and let $\ep>0$. Then we have
\begin{equation}\label{eq:stabilityComplexity}
  C_{2\ep}(\h{W}_m \h{W}_{m-1} \ldots \h{W}_1 \U) \leq  C_\ep(\U) +  A m \log(m/\ep)^\gamma\,\ ,\   C_{2\ep}(\h{W}_m \h{W}_{m-1} \ldots \h{W}_1 (\psi)) \leq C_\ep(\psi) + A m \log(m/\ep)^\gamma\,, 
\end{equation}
where $A>0$ depends on $\gset$ and $\gamma<3$ is an absolute constant ($\gamma=1$ provided $\gset$ has a spectral gap). 
\end{lemma}
Note that an analogous bound can be derived, with the same proof, to bound $C_\ep(\U)$ also from above.
\begin{proof}
We present the proof for unitaries as the reasoning for states is analogous. By definition of the complexity $C_\ep$, a unitary channel $\U$ can be approximated to accuracy $\ep$ by sequence of gates from $\gset$ of length $r=C_\ep(\U)$. Furthermore, each of the gates $\h{W}_i$ ($i=1,\ldots,m$) can be approximated to accuracy $\ep'=\ep/m$ via sequences of gates from $\gset$ of length $r_i = A \log(1/\ep')^\gamma$ (as in the proof of \cref{lem:ComplexityUpperBound}). Putting together this gives an $\ep + m \cdot \ep'= 2\ep$ approximation of $\h{W}_m \h{W}_{m-1} \ldots \h{W}_1 \U$ via a sequence of length $r+r_1+\ldots+ r_m = C_\ep(\U) + A m \log(m/\ep)^\gamma$.
\end{proof}

\begin{remark}
In the case when unitaries $\h{W}_1,\h{W}_2, \ldots, \h{W}_m$ in the above lemma can be \emph{exactly} obtained from gates in $\gset$, then the bounds in Eq.~\eqref{eq:stabilityComplexity} can be strengthened to 
\begin{equation}\label{eq:exactSTAB}
     C_{\ep}(\h{W}_m \h{W}_{m-1} \ldots \h{W}_1 \U) \leq  C_\ep(\U) + A m\,, \quad C_{\ep}(\h{W}_m \h{W}_{m-1} \ldots \h{W}_1 (\psi)) \leq  C_\ep(\psi) + A m \,.
\end{equation}
Note that in the above inequalities complexities are computed on the same scale $\ep$ and no $\ep$-dependence is present in the upper bounds. 
\end{remark}
The following proposition gives a sufficient condition on the time $T$ such that the complexity of the `past' unitary $\U_{t-T}$ is essentially maximal, conditioned on $\U_t$ having the trivial (i.e.\ zero) complexity. 

\begin{proposition}\label{prop:abstractRECtimescales}
Let $\gset$ be a fixed universal gate set used in the definition of unitary and state complexity on $n$ qudit system $\CH=(\mathbb{C}^q)^{\ot n}$ of dimension $\dim=q^n$. Let $\nu$ be a probability distribution on $\UU(\dim)$ generating random walks $\U_t$ and $\psi_t$ on the space of unitary channels and quantum states. Let $\ep\in(0,1/4)$ and $\Delta\in(0,1)$. Assume that $T$ is a time such that the measure $\nu^{\ast T}$ is $(\alpha,\beta)$-equidistributed on a scale $\ep$ for $\Gamma=\min\lbrace{1-\alpha,\beta-1\rbrace}\leq \frac{\Delta}{6}(\clower/3)^{d^2-1}$. Then for $t>2T$ and 
\begin{equation}\label{eq:rconUNIT}
    r= \frac{(\dim^2-1) \log\frac{1}{\cupper\,\beta\ep} - \log\frac{2}{\Delta} }{\log|\gset|}\,,
\end{equation}
we have 
\begin{equation}\label{eq:condLOWboundUni}
\Prob\Big(C_\ep(\U_{t-T})>r\, |\  C_\ep(\U_t)=0   \Big)\geq 1-\Delta\ .
\end{equation}
Similarly for pure quantum states, we let $T_\S$ be a time such that the measure $(\nu^{\ast T_\S})_\S$ is $(\alpha,\beta)$-equidistributed on a scale $\ep$ for $\Gamma_\S=\min\lbrace{1-\alpha,\beta-1\rbrace}\leq\frac{\Delta\log(2)}{16d}$, where $\Delta\in(0,1)$. Then for $t>2T_\S$ and 
\begin{equation}\label{eq:rconSTATE}
    r_\S=\frac{(2\dim-2) \log\frac{1}{\beta\ep} - \log\frac{2}{\Delta} }{\log|\gset|}
\end{equation}
we have 
\begin{equation}
\Prob\Big(C_\ep(\psi_{t-T_\S})>r_\S\, |\  C_\ep(\psi_t)=0   \Big)\geq 1-\Delta .
\end{equation}
\end{proposition}
\begin{proof}
In order to simplify the notation let us denote $\CA=\lbrace{\U\in\UU(\dim)| C_\ep(\U)>r\rbrace}$, $\CB=\lbrace{\U\in\UU(\dim)| C_\ep(\U)=0\rbrace}$. We then write
\begin{equation}\label{eq:bayesUNIT}
    \Prob\Big(C_\ep(\U_{t-T})>r\, |\  C_\ep(\U_t)=0   \Big) = \Prob(\U_{t-T}\in\CA\ |\  \U_t\in\CB)= \frac{\Prob(\U_{t-T}\in\CA\ ,\ \U_t\in\CB)}{\Prob(\U_t\in\CB)}\,,
\end{equation}
where in the second equality we used Bayes theorem. Let us assume for simplicity that $\nu$ is a discrete measure (the proof in the general case is analogous). It then follows that the distribution of $\U_{t-T}$ is also discrete. For the measure $\nu_{t-T}=\lbrace{p_i, \V_i\rbrace}$, we use the definition of $\U_t$ to obtain
\begin{equation}
    \Prob(\U_{t-T}\in\CA\ ,\ \U_t\in\CB)= \sum_{i:\V_i\in\CA} p_i \Prob(\U_t\in\CB |\U_{t-T}=\V_i)\,.
\end{equation}
The distribution of $\U_t$ conditioned on $\U_{t-T}=\V_i$ is simply $\nu_T \ast\delta_{\V_i}$ and therefore 
\begin{equation}
     \Prob(\U_{t-T}\in\CA\ ,\ \U_t\in\CB)= \sum_{i:\V_i\in\CA} p_i  \nu_T \ast\delta_{\V_i}(\CB)\,.
\end{equation}
Using $\CB=B(\I,\ep)$ and the fact that for every $i$ the measure $\nu_T \ast\delta_{\V_i}$ approximately equidistributes on scale $\ep$ (this follows from the assumption about $\nu_T$ and \cref{fact:equi}), we obtain
\begin{equation}
     \Prob(\U_{t-T}\in\CA\ ,\ \U_t\in\CB)\geq \Prob(\U_{t-T}\in \CA) \vol(\alpha \ep)\,. 
\end{equation}
Since $\nu_t$ is also $(\alpha,\beta)$-equidistributed we can upper bound the denominator of Eq.~\eqref{eq:bayesUNIT} by $\vol(\beta \ep)$, which gives 
\begin{equation}\label{eq:unitRECsemiBOUND}
     \Prob\Big(C_\ep(\U_{t-T})>r\, |\  C_\ep(\U_t)=0\Big)\geq \Prob\Big(C_\ep(\U_{t-T})>r\Big) \frac{\vol(\alpha \ep)}{\vol(\beta\ep)}\,. 
\end{equation}
Furthermore, we have 
\begin{equation}
    \frac{\vol(\beta \ep)}{\vol(\alpha\ep)}=1 + \frac{\vol(A(\alpha\ep,\beta\ep))}{\vol(\alpha \ep)}\,,
\end{equation}
where $A(\alpha\ep,\beta\ep)=\lbrace{\U\in\UU(\dim)| \alpha\ep \leq \d(\U,\I)\leq \beta \ep\rbrace}$ is an annulus in the space of unitaries. Using \cref{prop:ringVOL} for $\kappa=\alpha \ep$ and $\lambda=\beta/\alpha$ we find 
\begin{equation}
    \vol(A(\alpha\ep,\beta\ep))\leq (2\beta \ep)^{d^2-1}(\lambda-1)\,.
\end{equation}
Combining this bound with $\vol(\alpha\ep)\geq (\clower \beta \ep)^{d^2-1}$ we obtain
\begin{equation}
    \frac{\vol(\beta \ep)}{\vol(\alpha\ep)}\leq 1 + \left(\frac{2\beta}{\alpha \clower} \right)^{d^2-1} (\lambda-1)\,.
\end{equation}
It is straightforward to see that $\Gamma\leq \frac{\Delta}{6}(\clower/3)^{d^2-1}$ implies
\begin{equation}
      \lambda  =\frac{\beta}{\alpha}\leq\frac{1+\Gamma}{1-\Gamma}\leq (1+\Gamma)^2\leq \frac{3}{2}\ ,\  1-\Gamma\geq \frac{2}{3}\,,
\end{equation}
\begin{equation}\label{eq:trivUNIT}
    (\lambda-1)\left(\frac{2\beta}{\clower\alpha}\right)^{d^2-1} \leq \left(\frac{3}{\clower}\right)^{d^2-1}\frac{2\Gamma}{1-\Gamma}\leq 3\Gamma \left(\frac{3}{\clower}\right)^{d^2-1}\leq\frac{\Delta}{2}\,.
\end{equation}
On the other hand, from \cref{lem:complexity-equidist} and the fact that $\nu_{t-T}$ equidistributes it follows that for $r$ as in Eq.~$\eqref{eq:rconUNIT}$ we have $ \Prob\Big(C_\ep(\U_{t-T})>r\Big)\geq 1-\Delta/2 $. Combining these estimates with Eq.~\eqref{eq:unitRECsemiBOUND} gives the desired result:
\begin{equation}
\Prob\Big(C_\ep(\U_{t-T})>r\, |\  C_\ep(\U_t)=0\Big)\geq \frac{1-\Delta/2}{1+\Delta/2}\geq 1-\Delta\,.
\end{equation}

The considerations for quantum states are similar. By repeating the same steps we obtain
\begin{equation}
    \Prob\Big(C_\ep(\psi_{t-T_\S})>r_\S\, |\  C_\ep(\psi_t)=0  \Big)\geq \Prob\Big(C_\ep(\psi_{t-T_\S})>r_\S\Big) \frac{\vol_\S(\alpha \ep)}{\vol_\S(\beta\ep)}\ .
\end{equation}
In this case volumes of balls in the space of pure states are explicitly known: $\vol_\S(\ep)=\ep^{2d-2}$. We can then bound the ratio of volumes as
\begin{equation}
    \frac{\vol_\S(\beta \ep)}{\vol_\S(\alpha\ep)}\leq\Big(\frac{1+\Gamma_\S}{1-\Gamma_\S} \Big)^{2(d-1)}\leq (1+2\Gamma_\S)^{4(d-1)}\leq \exp(8d \Gamma_\S)\,,
\end{equation}
where in the second inequality we used $\Gamma_\S\leq1$, which is implied by the assumptions. Using the inequality $\log(1+x)\geq \log(2) x$, valid for $x\in[0,1]$, it is easy to show that requiring
\begin{equation}
    8d\Gamma_\S \leq \frac{\log(2)\Delta}{2}\ ,
\end{equation}
which is equivalent to $\Gamma_\S\leq \frac{\Delta\log(2)}{16d}$, implies that $\exp(8d \Gamma_\S)\leq 1+\Delta/2$. Similarly as for the case of unitaries, using  \cref{lem:complexity-equidist} and the fact that $\nu_{t-T,\S}$ equidistributes it follows that for $r_\S$ as in Eq.~\eqref{eq:rconSTATE} we have $\Prob(C_\ep(\psi_{t-T})>r_\S)\geq 1-\Delta/2$. Proceeding as before we get
\begin{equation}
\Prob\Big(C_\ep(\psi_{t-T})>r_\S\, |\  C_\ep(\psi_t)=0\Big)\geq \frac{1-\Delta/2}{1+\Delta/2}\geq 1-\Delta\,.
\end{equation}

\end{proof}

\begin{remark}
We note that the assumptions on $\Gamma$ in the above proposition are much more stringent for unitary channels, compared to pure states. We believe that this is solely an artifact of our proof and that the assumptions in the former case can be greatly weakened. In order to achieve this one would likely need much stronger control over the ratio of volumes $\vol(\beta \ep)/\vol(\alpha \ep)$ in the space of unitary channels.
\end{remark}

We are ready to formulate and prove the main result of this section, which constitutes technical formulation of \cref{res:recwidth}. To establish the timescales claimed there for local random quantum circuits, it suffices to substitute as timescales $T,T_\S$ the bounds on equidistribution time given in \cref{lem:eqequidistrTIME} (for suitable parameters $\Gamma$ required by the Theorem).

\begin{theorem}[Timescales of complexity recurrences]\label{th:compTIMESCALES}
Let $\gset$ be a fixed universal gate set used in the definition of unitary and state complexity on $n$ qudit system $\CH=(\mathbb{C}^q)^{\ot n}$ of dimension $\dim=q^n$. Let $\nu=\lbrace{\nu_i,\V_i\rbrace}$ be a probability distribution on $\UU(\dim)$ generating random walks $\U_t$ and $\psi_t$ on the space of unitary channels and quantum states. Let $\ep\in(0,1/4)$ and $\Delta\in(0,1)$ and assume that $T$ is a time such that the measure $\nu^{\ast T}$ is $(\alpha,\beta)$-equidistributed on a scale $\ep$ for $\Gamma=\min\lbrace{1-\alpha,\beta-1\rbrace}\leq \frac{\Delta}{6}(\clower/3)^{d^2-1}$. Then for $t>2T$ and 
\begin{equation}\label{eq:rconUNITfin}
    r= \frac{(\dim^2-1) \log\frac{1}{\cupper\,\beta\ep} - \log\frac{2}{\Delta} }{\log|\gset|}\,,
\end{equation}
we then have
\begin{equation}\label{eq:conditionPROB}
    \Prob\Big(C_\ep(\U_{t-T})>r\, ,\, C_\ep(\U_{t+T})>r\ |\  C_\ep(\U_t)=0   \Big) \geq 1-\frac{3}{2}\Delta\,.
\end{equation} 

Furthermore, assuming that each of the gates $\lbrace{\V_i\rbrace}$, $\lbrace{\V^{-1}_i\rbrace}$ are implemented exactly using at most $k$ gates from $\gset$, then conditioned on $C_\ep(\U_t)=0$ we have
\begin{equation}\label{eq:unitarylightcone}
    \max\lbrace{C_\ep(\U_{t-m}), C_\ep(\U_{t+m})\rbrace} \leq A\cdot k\cdot m \,,
\end{equation}
where the constant $A$ depends only on $\gset$. Taken together, Eq.~\eqref{eq:conditionPROB} and Eq.~\eqref{eq:unitarylightcone} show that for typical realizations of a random walk $\U_t$ complexity recurrences occur for $m_\ast$ satisfying
\begin{equation}
    \exp(\Theta(n))\log(1/\ep)\leq m_\ast \leq T\,.
\end{equation}

Analogous results holds for recurrences of complexity in pure quantum states $\psi_t$. The only difference is that in this case the role of timescale $T$ is replaced by $T_\S$  required for measure $\nu_{t,\S}$ to distribute on scale $\ep$ with parameters satisfying $\Gamma_\S=\min\lbrace{1-\alpha,\beta-1\rbrace}\leq \frac{\Delta}{24d}$  (see \cref{prop:abstractRECtimescales}). Furthermore, the value of $r$ is replaced by $r_\S$ from Eq.~\eqref{eq:rconSTATE}.
\end{theorem}

\begin{proof}
Using a union bound we obtain
\begin{align}
\begin{split}
    &\Prob\Big(C_\ep(\U_{t-T})>r\, ,\, C_\ep(\U_{t+T})>r\ |\  C_\ep(\U_t)=0   \Big)\\ &\quad \geq \Prob\Big(C_\ep(\U_{t-T})>r |\  C_\ep(\U_t)=0   \Big)+\Prob\Big(C_\ep(\U_{t+T})>r |\  C_\ep(\U_t)=0   \Big)-1\,.
\end{split}
\end{align}
For $r$ as in Eq.~\eqref{eq:rconUNITfin} we then obtain Eq.~\eqref{eq:conditionPROB} by lower bounding both conditional probabilities appearing on the right hand side of above inequality using \cref{prop:abstractRECtimescales} and \cref{lem:complexity-equidist}. On the other hand, the inequality Eq.~\eqref{eq:unitarylightcone} follows directly from the assumptions and considerations following the proof \cref{lem:stabCompl}, particularly Eq.~\eqref{eq:exactSTAB}.
\end{proof}

\begin{remark}
 We are not able to prove such a result for the SLH model. Indeed, in the discrete case of random circuits, we know that after $n$ steps the complexity can roughly increase by at most $n$ (Eq.~\eqref{eq:stabilityComplexity}).  In the case of continuous walks (such as SLH) we do not have an analogous bound that would tell us that complexity increases at most proportionally to time. 
\end{remark}

\section{Linear lower bound for complexity at late times}
\label{sec:lineargrowth}

Thus far we have primarily focused on the saturation and recurrence of complexity in random quantum circuits, but there has been significant interest in proving the growth of quantum complexity, specifically establishing a long-time strictly linear growth of complexity. Such a long-time linear growth has been conjectured by Brown and Susskind for random quantum circuits and chaotic quantum many-body systems more generally. 
Ref.~\cite{complexitygrowth2019} proved a sublinear algebraic growth and Ref.~\cite{exactcomplexity2021} proved linear growth until exponential times for the restrictive notion of exact complexity. But it is worth emphasizing that RQC spectral gaps are sufficient to prove a linear lower bound which begins at exponential times, from $t\approx d^2$ until $t\approx d^4$. This linear lower bound follows from the exponentially small gaps proved in Refs.~\cite{BHH2016,Haferkamp22}.

\begin{proposition}[Linear complexity lower bound]\label{prop:linear-growth}
For depth $t$ local random quantum circuits on $n$ qudits, $\nu_t^{(n)}$, if the circuit depth obeys $t\leq 2\crqc n^6 d^4$ then with probability greater than $1-\Delta$ a given random circuit instance has circuit complexity lower bounded as
\begin{equation}
    C_\ep(\U_t) \geq \frac{t}{d^2} \frac{\log(2(1-\ep^2))}{\crqc n^6 \log(|\gset|)} - \frac{\log(1/\Delta)}{\log(|\gset|)}\,,
\end{equation}
where the constant $\crqc=10^5$. This implies a linear complexity lower bound from circuit depth $t=\tilde O(d^2)$ until $t=\tilde O(d^4)$.

Similarly, consider the states generated by depth $t$ local random quantum circuits $\nu_t^{(n)}$, if the circuit depth is $t\leq \crqc n^6 d^3$ then with probability greater than $1-\Delta$ a state has complexity lower bounded as
\begin{equation}
    C_\ep(\psi_t) \geq \frac{t}{d^2} \frac{\log(2(1-\ep^2))}{\crqc n^6 \log(|\gset|)} - \frac{\log(1/\Delta)}{\log(|\gset|)}\,,
\end{equation}
which implies a linear state complexity lower bound from $t=\tilde O(d)$ until $t=\tilde O(d^3)$. 
\end{proposition} 
Note that in the above proposition, as we are interested in regimes where $t\geq d^2$ for unitaries and $t\geq d$ for states, we may take $\Delta$ to be exponentially small in the total dimension $d$.

Similar complexity lower bounds hold for $(\Grqc, \G)$-local random quantum circuits (\cref{def:glocRQC-gates}), albeit with a slightly sublinear scaling in $t$ -- the complexity instead grows as $C_\ep(\U) \gtrsim t/{\rm polylog}(t)$, where the polylog factor is due to the $\log(k)$ factors in the design depth for $(\Grqc, \G)$-local random circuits. We remark that in order to have a single proof for all considered models, we work solely on the level of $k$-expanders, not spectral gaps as elsewhere in the paper.

\begin{proof}
We start by computing complexity lower bounds for elements of high degree designs. Let $\nu$ be a $\delta$-approximate unitary $k$-expander. The probability that a unitary $\U$ drawn from $\nu$ has complexity at most $r$ can be union bounded, similar to the proof of \cref{lem:complexity-equidist} and Eq.~\eqref{eq:smallComplexitySet}, as
\begin{equation}
    \Pr\big( C_\ep(\U)\leq r\big) = \nu\bigg( \bigcup_{\ell\leq r} \bigcup_{\V\in \G^\ell} B(\V,\ep)\bigg) \leq \sum_{\ell\leq r}\sum_{\V\in \G^\ell} \nu\big(B(\V,\ep)\big)\,.
\end{equation}
Using the upper bound on the volumes of balls according to $\nu$ in \cref{lem:volUnitaryDesignsMoments}, we have that for unitaries drawn from a $\delta$-approximate unitary $k$-expander
\begin{equation}
    \Pr(C_\ep(\U)\leq r) \leq |\gset|^{r+1} \frac{k!+d^{2k}\delta}{d^{2k}(1-\ep^2)^k}\,.
\end{equation}
First, we take $\delta = 1/d^{2k}$ and then for any $k\geq 6$ we have that $k!+1\leq (k/2)^k$, thus
\begin{equation}
    \frac{k!+1}{d^{2k}(1-\ep^2)^k}\leq \left( \frac{k}{2 d^2(1-\ep^2)}\right)^k.
\end{equation}
Restricting to $k\leq d^2$, we find
\begin{equation}
    \Pr(C_\ep(\U)\leq r) \leq |\gset|^{r+1} \frac{1}{(2(1-\ep^2))^k}\,.
\end{equation}
Computing the value of $r$ for which $ \Pr(C_\ep(\U)\leq r)\leq \Delta$ and taking the negation,
we find that the circuit complexity of a unitary $k$-design element is 
\begin{equation}
    C_\ep(\U) \geq \frac{1}{\log(|\gset|)} \big(k\log(2(1-\ep^2)) - \log(1/\Delta)\big)
\end{equation}
with probability greater than $1-\Delta$. Furthermore, \cref{prop:RQChighdegree} gives that local RQCs of circuit depth $t=2\crqc d^2 n^6k$ form $\delta$-approximate $k$-expanders with $\delta=1/d^{2k}$, which establishes the first statement of the proposition with the restriction that $6\leq k\leq d^2$. Note the irrelevance of the lower restriction on $k$ as we wish to apply the proposition to exponentially deep circuits $t=\Omega(d^2)$, but the upper bound restricts to circuits of depth $t\leq \tilde O(d^4)$.

The proof for states proceeds similarly. Let $\nu_\S$ be a $\delta$-approximate state $k$-expander. Again using a union bound and the upper bound on state design volumes in \cref{lem:volStatesDesignsMoments}, the probability that a state $\psi$ drawn from $\nu$ has complexity at most $r$ can be upper bounded as
\begin{equation}
    \Pr(C_\ep(\psi)\leq r) \leq |\gset|^{r+1} \frac{k!+d^k\delta}{d^k(1-\ep^2)^k}\leq |\gset|^{r+1} \frac{1}{(2(1-\ep^2))^k}\,,
\end{equation}
where we take $\delta=1/d^k$ and restrict to $6\leq k\leq d$. Proceeding, we get that the circuit complexity of a state drawn from an approximate state $k$-design is lower bounded as
\begin{equation}
        C_\ep(\psi) \geq \frac{1}{\log(|\gset|)} \big(k\log(2(1-\ep^2)) - \log(1/\Delta)\big)\,.
\end{equation}
\cref{prop:RQChighdegree} establishes that local random circuits of depth $t=\crqc d^2n^6 k$ form $\delta$-approximate state $k$-expanders with $\delta=1/d^k$, which yields the second statement of the proposition for circuits of depth $t\leq \tilde O(d^3)$.

Now we turn to complexity lower bounds for $(\Grqc, \G)$-local random quantum circuits. \cref{prop:GRQCdesigns} gives a circuit depth of $t=2c(\gset) n^8 k \log^2(k) d^2$ for approximate $k$-expanders with $\delta=1/d^{2k}$. Inverting this equation to express $k$ as a function of circuit depth $t$ can be done using the Lambert $W$-function: $x = W(x)e^{W(x)}$, where we take the principal branch for $x\geq 0$. Using this defining relation, we can show that the solution to $x=f(x)\log^2(f(x))$ is $f(x)=x/(2W(\sqrt{x}/2))^2$. Noting that $W(x)\leq \log(x)$ for $x\geq e$, we have $f(x)\geq x/\log^2(x)$. Thus, $G$-local random quantum circuits of circuit depth $t$ are $k=\lfloor t/(\gamma \log^2(t/\gamma))\rfloor$ degree approximate expanders, where $\gamma=2c(\gset)n^8 d^2$ and for $\delta=1/d^{2k}$, which in turn gives the desired complexity lower bounds.

\end{proof}

We conclude by noting that bounding the volumes of $\ep$-balls, as prescribed by the distribution $\nu$, using moments of exponential degree allows one to see some aspects of complexity saturation. Specifically, \cref{res:maxcompl} and \cref{res:designs-equidistribution} establish that unitaries drawn from high degree designs have nearly maximal complexity, which similarly follows from \cref{lem:volUnitaryDesignsMoments} for $k=O(d^2)$ and \cref{lem:volStatesDesignsMoments} for $k=O(d)$, albeit with trivial $\ep$ dependence. The volume bounds increase negligibly as we take $\ep\ra 0$, in stark contrast to the behavior which follows from approximate equidistribution.

\subsection*{Acknowledgments}

We thank Adam Bouland, Richard Kueng, Dan Ranard, Adam Sawicki, and Zhenbin Yang for helpful discussions.
MO acknowledges financial support from the National Science Centre, Poland under the grant OPUS: UMO2020/37/B/ST2/02478.
MK acknowledges financial support from the TEAM-NET project (contract no.\ POIR.04.04.00-00-17C1/18-00).
MH acknowledges support from the Foundation for Polish Science through an IRAP project co-financed by the EU within the Smart Growth Operational Programme (contract no.\ 2018/MAB/5). 
NHJ is supported in part by the Stanford Q-FARM Bloch Fellowship in Quantum Science and Engineering as well as the US Department of Energy under grant {DE}-{SC0020360}. NHJ would like to thank the Aspen Center for Physics and the Simons Institute for the Theory of Computing for their hospitality during the completion of part of this work.
Research at Perimeter Institute is supported in part by the Government of Canada through the Department of Innovation, Science and Economic Development Canada and by the Province of Ontario through the Ministry of Colleges and Universities.

\appendix

\section{Equidistribution from unitary designs}
\label{app:equidist}

  In this section we prove equidistribution using $\delta$-approximate $k$-expanders for unitaries and states. For unitaries, to improve readability, we first prove equidistribution for exact $k$-expanders and subsequently extend to approximate expanders, as in \cref{prop:equi-exact}. For states, we prove the property directly for approximate expanders in \cref{lem:equidistribution-appr-des-states}. 
For reader's convenience we collect the two propositions into \cref{th:equid} stated below.

\begin{theorem}[Equidistribution of states and unitaries from approximate designs]\label{th:equid}
Consider $\ep$, $\alpha$, and $\beta$ satisfying $0<\ep\leq 1/4$ and $1/2\leq \alpha\leq 1\leq \beta\leq 3/2 $, and let $\Gamma=\min\{1-\alpha,\beta-1\}$. Let $\nu$ be a $\delta$-approximate unitary $k$-expander in $\UU(\dim)$ for 
\begin{equation}\label{eq:equiUNI}
    k=  30 \frac{\dim^{5/2}}{\ep\Gamma}
   \eta(\ep\Gamma,\dim) \ \quad\text{and }\quad 
   \delta\leq \left( \frac{\clower \Gamma^2 \ep^2}{ 
    36
    \dim \sqrt{\log\frac{6}{\clower \ep\Gamma}}}\right)^{d^2-1}\,,
\end{equation}
where
\begin{align}
\label{eq:eta}
\eta(\ep,\dim)=
\sqrt{\log\frac{6}{\clower \ep}}
 \sqrt{\frac{1}{ 2  }\log\frac{6}{\clower \ep} + \log\biggl(\frac{6\dim}{\ep}\log\frac{6}{\clower \ep}\biggr)}\,,
\end{align}
i.e.\ for 
\begin{align}
\label{eq:equi-delta-k-unitary}
k\approx
\frac{d^{5/2}}{\ep\Gamma},\quad \delta \approx \left( \frac{\ep^2\Gamma^2}{d}\right)^{d^2}.
\end{align}
Then $\nu$ is $(\alpha,\beta)$-equidistributed on $\UU(\dim)$ on a scale $\ep$. 

Moreover, let $\nu$ be a $\delta$-approximate unitary $k$-expander for 
\begin{equation}\label{eq:equiPROJ}
    k= 1296\,\frac{\dim}{(\ep\Gamma)^4}\log\left(
    \frac{3}{\ep\Gamma}\right) \ \quad\text{and }\quad \delta = \left(\frac{\ep\Gamma}{9 \log\left(
    \frac{3}{\ep\Gamma}\right)}\right)^{6d},
\end{equation}
i.e.\ for 
\begin{align}
\label{eq:equi-delta-k-states}
    k\approx \frac{d}{(\ep \Gamma)^4},\quad \delta\approx (\ep \Gamma)^{6d},
\end{align}
Then the induced measure $\nu_\S$ on $\S(\dim) $ is $(\alpha,\beta)$-equidistributed on $\S(\dim)$ on a scale $\ep$. 
\end{theorem}

 The techniques used there are inspired by the ones used in Ref.~\cite{OSH2020}. The essential ingredient of this approach is the construction of the suitable polynomial approximation of Dirac delta on appropriate manifold of interest. 

\begin{remark}
While we prove equidistribution for state designs that are induced by unitary designs, one could consider a more direct definition of state designs. Namely, one can say that measure $\xi$ is state $k$-design if 
\begin{align}
\int_{\S(\dim)} \dt \xi(\psi)\, \ketbra{\psi}^{\otimes k}
=
\int_{\S(\dim)} \dt \mu_\S(\psi)\, \ketbra{\psi}^{\otimes k}\,,
\end{align}
where $\mu_\S$ is uniform measure on states (induced by the Haar measure). One can prove equidistribution for such designs (both exact and approximate) as well. However, we are primarily interested in distributions over states that come from quantum circuits, and therefore we do not consider such directly defined state designs. 
\end{remark}

\subsection{Equidistribution on \ensuremath{\UU(\dim)} from exact designs}

We quote the following theorem from Ref.~\cite{OSH2020}.
\begin{theorem}[Efficient polynomial approximation of the Dirac delta on unitary channels \cite{OSH2020}]\label{th:FUNC}
Let $\ep\in(0,2/3]$ and $\sigma\leq \frac{\ep}{6 \sqrt{d}}$.  There exists a function $\F^{\sigma}_k:\U(d)\rightarrow \R$ with the following properties. 
\begin{enumerate}
\item Normalization: $\int_{\UU(d)} \dt \mu (\h{U}) \F^{\sigma}_k(\h{U})  =1$.
\item Vanishing integral of modulus outside of the ball $B(\h{I},\ep)$:
for 
\begin{equation}\label{eq:degreeTAIL}
k\geq 5 \frac{d^{\frac{3}{2}}}{\sigma}  \sqrt{\frac{1}{8}\frac{\ep^2}{d^2\sigma^2} + \log\Big(\frac{1}{\sigma}\Big)}
\end{equation}
we have 
\begin{equation}
\label{eq:bound-for-g}
\int_{ B(\h{I},\ep)^c } \dt \mu (\h{U}) |\F^\sigma_k (\h{U})| \leq 9 \exp\left(-\frac{ \ep^2 }{4\sigma^2}\right) \left(\frac{\pi}{2}\right)^{d(d-1)} \,.
\end{equation}
\item Low degree polynomial: $\F^{\sigma}_k (\h{U})$ can be represented as a balanced polynomial in $U$ and $\bar{U}$  of degree $k$.
\item Bounded $L^2$-norm:  for $k\geq d/\sigma$ we have 
\begin{equation}\label{eq:L2norm}
\left\|\F^\sigma_k\right\|_{2}=\sqrt{\int_{\UU(d)} \dt  \mu (\h{U})  \left|\F^{\sigma}_k(\h{U}) \right|^2} \leq 
8 \times 2^{d^2} \, \sigma^{-d(d-\frac12)}
\,.
\end{equation}
\item $L^1$-norm  close to $1$: for $k$ satisfying Eq.~\eqref{eq:degreeTAIL}, we have 
\begin{align}
\label{eq:bound-for-h}
    1\leq \|\F_k^\sigma\|_1\leq  1 + 
6 \exp\left(-\frac{ \ep^2 }{4\sigma^2}\right) \left(\frac{\pi}{2}\right)^{d(d-1)} \,.
\end{align}
\end{enumerate}

\end{theorem}

In order to simplify the notation let us denote 
\begin{align}
    \gdirac(\ep)
    =\int_{B(\h{I},\ep)^c} 
     \dt \mu (\h{U})\, |\F_k^\sigma(\h{U})|\,\ ,\quad h_k^\sigma=\|\F_k^\sigma\|_1-1\,.
\end{align}
Under the assumptions of the theorem we then have
\begin{equation}
\label{eq:hkgk-bounds}
    \gdirac(\ep) \leq 
    9 \exp\left(-\frac{ \ep^2 }{4\sigma^2}\right) \left(\frac{\pi}{2}\right)^{d(d-1)}, \quad h_k^\sigma
    \leq 
    6 \exp\left(-\frac{ \ep^2 }{4\sigma^2}\right) \left(\frac{\pi}{2}\right)^{d(d-1)}\,.
\end{equation}
Further note that $h$ does not depend on $\ep$ explicitly, but under the assumptions of the theorem $k$ depends on $\ep$.
Hence the bound on $h^\sigma_k$ depends on $\ep$ as well. The following lemma will be crucial for establishing the connection between equidistribution and unitary designs.

\begin{lemma}
\label{lem:equidistribution-exact-des}
Let $\nu=\{\nu_\alpha,\h{U}_\alpha\}$ be exact $k$-design. 
Then for any  $\h{U} \in \UU(d)$, $\sigma>0 $, and for any $\kapls,\kapgr$ and $\ep$ satisfying $0<\kapls<\ep<\kapgr$ we have
\begin{align}
    \vol\left (\kapls\right) - \gdirac\left(\ep-\kapls\right) -h_k^\sigma \leq 
    \nu(B(\h{U},\ep)) \leq  
    \vol(\kapgr) + \gdirac(\kapgr-\ep)+h_k^\sigma\,.
\end{align}
\end{lemma}

\begin{proof}
We first prove the result for $\h{U}=\h{I}$ and then will argue that for general $\h{U}$ the proof is analogous. 
For unitary channels $\h{V}$ and $\h{W}$, let us denote $ \F_k^\sigma (\h{V}^{-1} \h{W}) = \F^\sigma_{k,\h{V}}(\h{W})$, so that $\F_{k,\h{V}}^\sigma$ is centered around $\h{V}$.
To prove a lower bound, we consider two balls around $\h{I}$, with radii $\kapls$ and $\ep$ (recall that  $\ep> \kapls$). Using the exact $k$-design property, we have that for every $\U\in\UU(\dim)$, $\sum_\alpha \nu_\alpha\, \F^\sigma_{k,\h{U}_\alpha}(\h{U})=1$, and therefore
\begin{align}
\label{eq:vol-vs-des-1}
\vol(\kapls) &=\int_{B(\h{I},\kapls)} \dt \mu(\h{U}) \sum_\alpha \nu_\alpha\, \F^\sigma_{k,\h{U}_\alpha}(\h{U})\nn
&=\sum_{\alpha: \d(\h{U}_\alpha,\h{I})> \ep} \nu_\alpha
\int_{B(\h{I},\kapls)} \dt \mu(\h{U}) \, \F^\sigma_{k,\h{U}_\alpha}(\h{U})+
\sum_{\alpha: \d(\h{U}_\alpha,\h{I})\leq \ep} \nu_\alpha
\int_{B(\h{I},\kapls)} \dt \mu(\h{U}) \, \F^\sigma_{k,\h{U}_\alpha}(\h{U})\,.
\end{align}
For $\h{U}_\alpha$ satisfying $\d(\h{U}_\alpha,\h{I})> \ep$, we estimate
\begin{equation}
\label{eq:sum-p-alpha}
    \sum_{\alpha: \d(\h{U}_\alpha,\h{I})> \ep} \nu_\alpha\leq 1\,,
\end{equation}
and 
\begin{align}
\int_{B(\h{I},\kapls)} \dt \mu(\h{U})  \,\F^\sigma_{k,\h{U}_\alpha}(\h{U})  \leq 
\int_{B(\h{I},\kapls)} \dt \mu(\h{U})  |\F^\sigma_{k,\h{U}_\alpha}|
\leq \int_{B(\h{I},\ep- \kapls)^c} \dt \mu(\h{U})
\,|\F^\sigma_{k}|= \gdirac(\ep-\kapls)\,.
\end{align}
Furthermore, for $\h{U}_\alpha$ with $\d(\h{U}_\alpha,\h{I})\leq  \ep$ we have
\begin{align}
    \sum_{\alpha: \d(\h{U}_\alpha,\h{I})\leq \ep} \nu_\alpha
    =\nu(B(\h{I},\ep))\,.
\end{align}
We then estimate
\begin{align}
    \int_{B(\h{I},\kapls)} \dt \mu(\h{U})  \F^\sigma_{k,\h{U}_\alpha}(\h{U})  &\leq 
    \int_{B(\h{I},\kapls)} \dt \mu(\h{U})  \left| \F^\sigma_{k,\h{U}_\alpha}(\h{U}) \right|
    \leq \int_{\UU(d)} \dt \mu(\h{U})  \left| \F^\sigma_{k,\h{U}_\alpha}(\h{U})\right|
    =\| \F^\sigma_k \|_1\,,
\end{align}
where in the last equality we used the identity $\| \F^\sigma_k \|_1= \| \F^\sigma_{k,\U_\alpha} \|_1$. Inserting the above estimates into Eq.~\eqref{eq:vol-vs-des-1}, we obtain
\begin{align}
    \vol(\kapls) \leq  \|\F^\sigma_{k}\|_1 \,  \nu(B(\h{I},\ep)) \, +
    \gdirac(\ep-\kapls)
    \leq 
    \nu(B(\h{I},\ep)) \, +
    \gdirac(\ep-\kapls) + h_k^\sigma\,,
\end{align}
where in the last inequality we have used Eq.~\eqref{eq:bound-for-h} and definition of $h^{\sigma}_k$. 
This gives the required lower bound.

The upper bound is proven analogously. 
We consider balls with radii $\ep$ and $\kapgr$ (recall that this time $\ep\leq\kapgr$). 
We have
\begin{align}
\label{eq:vol-vs-des-2}
1-\vol(\kapgr) &= \int_{B(\h{I},\kapgr)^c} \dt \mu(\h{U}) \sum_\alpha \nu_\alpha \F^\sigma_{k,\h{U}_\alpha}(\h{U})\\
&=\sum_{\alpha: \d(\h{U}_\alpha,\h{I})> \ep} \nu_\alpha
\int_{B(\h{I},\kapgr)^c} \dt \mu(\h{U})  \,\F^\sigma_{k,\h{U}_\alpha}(\h{U})+
\sum_{\alpha: \d(\h{U}_\alpha,\h{I})\leq \ep} \nu_\alpha
\int_{B(\h{I},\kapgr)^c} \dt \mu(\h{U})  \F^\sigma_{k,\h{U}_\alpha}(\h{U})\,.\nonumber
\end{align}
As before, consider the first term. We have 
\begin{align}
    \sum_{\alpha: \d(\h{U}_\alpha,\h{I})> \ep} \nu_\alpha= 1-\nu(B(\h{I},\ep))
\end{align}
and for $\h{U}_\alpha$ with  $\d(\h{U}_\alpha,\h{I})> \ep$ we  estimate
\begin{align}
\int_{B(\h{I},\kapgr)^c} \dt \mu(\h{U})  \F^\sigma_{k,\h{U}_\alpha}(\h{U})  \leq\|\F^\sigma_{k}\|_1\,.
\end{align}
For the second term use
\begin{equation}
    \sum_{\alpha: \d(\h{U}_\alpha,\h{I})\leq \ep} \nu_\alpha
    \leq 1\,,
\end{equation}
and  for $\h{U}_\alpha$ with  $\d(\h{U}_\alpha,\h{I})\leq \ep$ we estimate
\begin{align}
\int_{B(\h{I},\kapgr)^c} \dt \mu(\h{U})  \F^\sigma_{k,\h{U}_\alpha}(\h{U})  \leq
\int_{B(\h{I},\kapgr)^c} \dt \mu(\h{U})  |\F^\sigma_{k,\h{U}_\alpha}(\h{U})|
\leq \int_{B(\h{I},\kapgr- \ep)^c} \dt \mu(\h{U})
|\F^\sigma_k(\h{U})|= \gdirac(\kapgr-\ep)\,.
\end{align}
Putting the estimates together we obtain 
\begin{align}
    1-\vol(\kapgr)\leq \left(1 -  \nu(B(\h{I},\ep)) \right)\|\F_k^\sigma\|_1 + 
    \gdirac(\kapgr-\ep)\,.
\end{align}
Subsequently, from the definition of $h^\sigma_k$ and its positivity, we arrive at
\begin{align}
\nu(B(\h{I},\ep)) \leq  \vol(\kapgr)
    + \gdirac(\kapgr-\ep)+h_k^\sigma\,,
\end{align}
which gives the required upper bound.
To finish the proof, we have to argue 
that the same holds for $\nu(B(\h{U},\ep))$
for all $\h{U}$. One obtains this by taking the original function not just $\F^\sigma_k$
but $\F^\sigma_{k,U}$, so that it is centered around $\h{U}$, and considering balls around $\h{U}$. 
\end{proof}

Now we proceed to proving equidistribution. To this end we first need to establish the following technical lemma.

\begin{lemma}
\label{lem:bound-for-h-g}
For any  $\ep\in(0,2/3)$, $d\geq 16$, and
\begin{equation}
\label{eq:ktilde}
    k\geq 
     k^{\ast}(\ep) = 5 \frac{d^{\frac{3}{2}}}{\sigma}  \sqrt{\frac{1}{8}\frac{\ep^2}{d^2\sigma^2} + \log\Big(\frac{1}{\sigma}\Big)}
\end{equation}
with 
\begin{equation}
\label{eq:sigma-star}
\sigma\leq \sigma_\ast(d,\ep)=\frac{\ep}{ 2  d}\frac{1}{\log(2/(\clower \ep))^{\frac12}}\ ,
\end{equation}
it holds that $\sigma\leq \frac{\ep}{6\sqrt{\dim}}$
as well as that
\begin{equation}
\label{eq:g-vs-vol}
     g_k^\sigma(\ep) 
     \leq \frac12 \vol(\ep),
     \quad h_k^\sigma \leq \frac13\vol(\ep)\,.
\end{equation}
\end{lemma}
\begin{remark}
The restriction $d \geq 16$ is needed to make sure that $\sigma\leq \ep/(6\sqrt{\dim})$, which in turn is needed later when we discuss approximate designs. However, in the original paper the bound $\ep/(6\sqrt{\dim})$ 
is a strict joint lower bound on $4/(\pi \sqrt{\dim})$ and $1/4$. 
\end{remark}
\begin{proof}
First, one directly verifies that  $\sigma$ of Eq.~\eqref{eq:sigma-star} with $\ep\leq1$ 
satisfies $\sigma\leq \ep/(6\sqrt{d})$. 
Let us now prove the first bound of Eq.~\eqref{eq:g-vs-vol} using Eq.~\eqref{eq:hkgk-bounds}. Due to Eq.~\eqref{eq:szarek}, we need to show that for $\sigma$ satisfying Eq.~\eqref{eq:sigma-star} we have 
\begin{equation}
\label{eq:lipa}
    9 \exp\left(-\frac{\ep^2}{4\sigma^2}\right) \left(\frac{\pi}{2}\right)^{d(d-1)}\leq \frac{1}{2}\left(\clower \ep\right)^{d^2-1}\,.
\end{equation}
To prove the above estimate, we drop $-1$ from exponent on the right hand side, insert $\sigma^*$, and get that the following estimate is to be proven:
\begin{align}
    18\frac{1}{2^{d^2}} \left( \frac{\pi}{2}\right)^{d^2-d}\leq 1\,,
\end{align}
which indeed holds for $d\geq 3$. To prove the second estimate of Eq.~\eqref{eq:g-vs-vol} it is enough to see that,
according to bound on $h$ in Eq.~\eqref{eq:hkgk-bounds}, changing $9$ into $6$ in Eq.~\eqref{eq:lipa} we get $1/3$ on right hand side.
\end{proof}

Note that we have a fixed constant $\clower=\frac{1}{9\pi}$, but the above lemma and following proposition would hold for any such constant in the volume lower bound so long as it is less than one.

\begin{proposition}
\label{prop:equi-exact}
Let $\nu$ be exact $k$-design. Then for $0<\ep< 1/4$, $1/2\leq\alpha<1<\beta\leq 3/2$, $d\geq 16$ and 
\begin{equation}\label{eq:settingK}
   k\geq  30 \frac{\dim^{5/2}}{\ep\Gamma}
   \eta(\ep\Gamma,\dim)
\end{equation}
with  
\begin{equation}
\eta(\ep,\dim)=
\sqrt{\log\frac{6}{\clower \ep}}
 \sqrt{\frac{1}{2}\log\frac{6}{\clower \ep} + \log\biggl(\frac{6\dim}{\ep}\log\frac{6}{\clower \ep}\biggr)}\,,
 \quad \Gamma=\min\left\{1-\alpha,\beta-1\right\}\,,
 \end{equation}
we have 
\begin{equation}
    \vol(\alpha\ep)\leq \nu(B(\h{U},\ep))\leq \vol(\beta\ep)\,.
\end{equation}
\end{proposition}
\begin{proof}
{\bf Upper bound}. 
\cref{lem:equidistribution-exact-des} tells us that
 \begin{align}
 \label{eq:kappagr}
\nu(B(\h{U},\ep)) \leq  
\vol(\kapgr) + \gdirac(\kapgr-\ep) +h_k^\sigma\,.
 \end{align}
From \cref{lem:bound-for-h-g}
it follows that for any $k\geq k_\ast(\kapgr-\ep) $ with $k_\ast$ given by Eq.~\eqref{eq:ktilde}, and $\sigma=\sigma_\ast(\kapgr-\ep)$, with $\sigma_\ast$ given by Eq.~\eqref{eq:sigma-star}, we have that
\begin{equation}
    g_k^\sigma(\kapgr-\ep)+h^\sigma_k 
    \leq  
    \vol(\kapgr-\ep)\,,
\end{equation}
where we have dropped the factor of $5/6$. From this we obtain 
\begin{align}
\label{eq:nu_kappa_kappa_ep}
    \nu(B(\h{U},\ep)) \leq  
\vol(\kapgr) + \vol(\kapgr-\ep)\,.
\end{align}
Combining this statement with \cref{fact:vol-2}, we find that
\begin{equation}
\label{eq:vol-g-h-intermediate-bound1}
     \nu(B(\h{U},\ep)) \leq  \vol(\kapgr+2(\kapgr-\ep))\,.
\end{equation}
Choosing $\kapgr=\ep(\beta+2)/3$ so that $\kapgr>\ep$ as required, implies 
\begin{align}
    \vol(\kapgr) + \gdirac(\kapgr-\ep) +h_k^\sigma \leq 
    \vol(\beta \ep).
\end{align}
Together with Eq.~\eqref{eq:kappagr} this gives the desired bound for $k\geq k^*\left((\beta-1)\ep/3\right)$, with $\sigma=\sigma^*(\left((\beta-1)\ep/3\right))$, and where we used that for $\kapgr$ as above we have $\kapgr-\ep=(\beta-1)\ep/3$.

\noindent {\bf Lower bound.} Again from \cref{lem:equidistribution-exact-des}, we know that
\begin{align}
\label{eq:kapls}
    \vol\left (\kapls\right) - \gdirac\left(\ep-\kapls\right) -h_k^\sigma \leq 
    \nu(B(\h{U},\ep))\,.
\end{align}
As before, from \cref{lem:bound-for-h-g} we get 
\begin{align}
    g_k^\sigma(\ep-\kapls)+h^\sigma_k 
    \leq  \vol(\ep-\kapls)\,,
\end{align}
but this time for $k\geq k_\ast(\ep-\kapls)$ with $\sigma=\sigma_\ast(\ep-\kapls)$, which gives 
\begin{align}
\label{eq:vol-kappa-vol-alpha}
    \vol(\kapls)-\vol(\ep-\kapls)\leq\nu(B(\h{U},\ep))\,.
\end{align}
We want the left hand side to be bounded from below by $\vol(\alpha\ep)$, which would complete the proof. Equivalently, we want 
\begin{align}
\label{eq:kapls2}
    \vol(\kapls)\geq \vol(\ep-\kapls)+\vol(\alpha\ep)\,.
\end{align}
To show this, we set $\kapls=\ep (\alpha+2)/3$,
so that $\kapls=\alpha\ep + 2(\ep-\kapls)$ and $\ep-\kapls=(1-\alpha)\ep/3$. For such a $\kapls$ we can then use \cref{fact:vol-2} to get the required bound Eq.~\eqref{eq:kapls2}, which works for $k\geq k^*\left((1-\alpha)\ep/3\right)$ with $\sigma=\sigma^*\left((1-\alpha)\ep/3\right)$. 

Since $k^*$, upon inserting $\sigma=\sigma^*(\ep)$, is a decreasing function of $\ep$, we get that both the upper and lower bounds hold when $k\geq k^*(\ep\Gamma/3)$ with $\sigma=\sigma^*(\ep\Gamma/3)$, 
where 
\begin{align}
    \Gamma=\min\left\{1-\alpha,\beta-1\right\}\,.
\end{align}
The formula Eq.~\eqref{eq:settingK} is then obtained by inserting $\sigma^*(\ep \Gamma/3)$ into $k^*(\ep \Gamma/3)$ and neglecting one of square roots.

Note that using \cref{fact:vol-2} requires $\kapls\leq 1/2$, 
which, since $\kapls< \ep$, only holds when $\ep\leq 1/2$ as the diameter of $\UU$ with our distance $D$ is greater than $1$. It also requires that $\kapgr+2(\kapgr-\ep)\leq1/2$, \ie $\ep\beta\leq1/2$. 
Since $\beta\leq 3/2$, we need to take $\ep\leq 1/3$. We have also used \cref{lem:bound-for-h-g}, which requires $\ep -\kapls\leq 2/3$
and $\kapgr-\ep\leq 2/3$. Since $\beta\leq 3/2$ and $\alpha\geq 1/2$, we thus require $\ep\leq 1/4$. 
\end{proof}

\begin{fact}
\label{fact:vol-2}
Let $\vol(\ep)$ be volume of an $\ep$-ball in 
the set of unitary channels $\UU(\dim)$ equipped with distance $\d(\U,\V)$, as in Eq.~\eqref{eq:distancesDEF}. Let $\ep_1,\ep_2>0$ be such that $\ep_1+\ep_2\leq 2=\mathrm{diam}(\UU(\dim))$.  Then 
we have
\begin{equation}
\label{eq:vol-2}
    \vol(\ep_1+2\ep_2)\geq  \vol(\ep_1)+\vol(\ep_2)\,.
\end{equation}
\end{fact}
\begin{proof}
Let $\U_1\in \UU(\dim)$ be a fixed unitary channel. By assumption there exist $\U_2$ such that $\d(\U_1,\U_2)=\ep_1+\ep_2$. Using the triangle inequality for distance $\d$ it is straightforward show that balls $B(\U_1,\ep_1), B(\U_2,\ep_2)$ intersect at most in boundary. Indeed, let $\d(\V,\U_1)\leq\ep_2$. Then it follows
\begin{equation}
    \d(\V,\U_2)\geq \d(\U_1,\U_2)-\d(\V,\U_1) \geq \ep_1+\ep_2-\ep_1=\ep_2 \,.
\end{equation}
Therefore $\vol\left(B(\U_1,\ep_1)\cap B(\U_2,\ep)\right)=0$. Furthermore, by definition $B(\U_1,\ep_1)\cup B(\U_2,\ep) \subset B(\U_1,\ep_1+2\ep_2)$. Combining these two facts and  we  get
\begin{equation}
   \vol(B(\U_1,\ep_1))+\vol(B(\U_2,\ep_2)) = \vol\left(B(\U_1,\ep_1)\cup B(\U_2,\ep)\right) \leq \vol\left(B(\U_1,\ep_1+2\ep_2)\right)\,.
\end{equation}
We conclude the proof by using the unitary invariance of the volume (Haar measure) on $\UU(\dim)$.
\end{proof}

\begin{remark}\label{rem:generalizVOLound}
An analogous argument to the one above can be applied to any unitary invariant metric on $\UU(\dim)$ and to manifold of pure states $\S(\dim)$.
\end{remark}

\subsection{Equidistribution on \ensuremath{\UU(\dim)} from approximate designs}

\begin{lemma}
\label{lem:equidistribution-appr-des-preparatory}
Let $\nu=\{\nu_\alpha,\h{U}_\alpha\}$ be a $\delta$-approximate unitary $k$-expander. 
Then for any $\h{U} \in \UU(\dim)$ and for any $\kapls,\kapgr$, and $\ep$ satisfying $0<\kapls<\ep<\kapgr$, we have
\begin{align}
\label{eq:equip-approx-preparatory}
\vol\left (\kapls\right) - \gdirac\left(\ep-\kapls\right) -h_k^\sigma - \delta \|\F_k^\sigma\|_2 
\leq 
\nu(B(\h{U},\ep)) \leq  
\vol(\kapgr) + \gdirac(\kapgr-\ep)+h_k^\sigma  +\delta \|\F_k^\sigma \|_2\,.
\end{align}

\end{lemma}
\begin{proof}
From Lemma 3 of Ref.~\cite{OSH2020} we have that for $\kappa\in[0,2]$
\begin{align}
    \left|\vol(\kappa)-\int_{B(\h{I},\kappa)} \dt \mu(\h{U}) \sum_\alpha \nu_\alpha \F^\sigma_{k,\h{U}_\alpha}(\h{U})\right|
    \leq \delta \|\F_k^\sigma\|_2 \sqrt{\vol(\kappa)}\,.
\end{align}
Since $\vol(\kappa)\leq1$ we choose to drop $\sqrt{\vol(\kappa)}$ obtaining
\begin{align}
\label{eq:vol-vs-appr-des}
    \vol(\kappa)-\delta\|\F_k^\sigma\|_2
    \leq
    \int_{B(\h{I},\kappa)} \dt \mu(\h{U}) \sum_\alpha \nu_\alpha  \F^\sigma_{k,\h{U}_\alpha}(\h{U})
    \leq \vol(\kappa)+\delta \|\F_k^\sigma\|_2\,.
\end{align}
We now repeat the proof of \cref{lem:equidistribution-exact-des} replacing the first equality in Eq.~\eqref{eq:vol-vs-des-1} with the inequality
\begin{align}
\vol(\kappa_{<})-\delta\|\F_k^\sigma\|_2   \leq
    \int_{B(\h{I},\kappa_{<})} \dt \mu(\h{U}) \sum_\alpha \nu_\alpha  \F^\sigma_{k,\h{U}_\alpha}(\h{U})    
\end{align}
and replacing the first equality in 
Eq.~\eqref{eq:vol-vs-des-2} with 
\begin{align}
1- (\vol(\kappa_{>})+\delta \|\F_k^\sigma\|_2  )\leq
    \int_{B(\h{I},\kappa_{>})^c} \dt \mu(\h{U}) \sum_\alpha \nu_\alpha  \F^\sigma_{k,\h{U}_\alpha}(\h{U})\,.
\end{align}

\end{proof}

Before formulating the main result of this section, let us present an estimate that one can obtain from Eq.~\eqref{eq:L2norm}.
\begin{lemma}
\label{lem:delta-2norm}
For $0<\sigma\leq 1/d$, $d\geq7$ and $k\geq d/\sigma$, the following implication holds. If 
\begin{align}
\label{eq:delta}
     \delta\leq
    \left(
    \clower \frac{\ep \sigma}{ 2} \right)^{d^2-1}\,, 
\end{align}
then 
\begin{equation} \label{eq:intermediateINEQ}
\delta\|\F^\sigma_k \|_2  
\leq \frac{1}{6} \vol(\ep)\,.
\end{equation}
\end{lemma}
\begin{proof}
Due to bound in Eq.~\eqref{eq:L2norm}, we need to ensure that 
\begin{align} 
\label{eq:ensure}
    \delta \leq 
    \frac{\frac{1}{6}
(\clower \ep)^{d^2-1} }{8 \times 2^{d^2} \sigma^{-d(d-\frac12)}}\,.
\end{align}
One finds, that for $\sigma\leq 1/d$ and $d\geq 7$ we have 
\begin{align}
    6 \times 8 \times 2^{d^2} \sigma^{-d(d-\frac12)}
    \leq \left(\frac{2}{\sigma}\right)^{d^2-1}\,.
\end{align}
This implies that if 
\begin{align}
    \delta\leq
     \left(
    \clower \frac{\ep \sigma}{2} \right)^{d^2-1}\,,
\end{align}
then Eq.~\eqref{eq:ensure} is satisfied, which ends the proof.
\end{proof}

We now formulate the main proposition. 
\begin{proposition}
\label{lem:equidistribution-appr-des}
Let  $0<\ep< 1/4$, $1/2\leq\alpha<1<\beta\leq 3/2$, $\clower$ be constant from Eq.~\eqref{eq:szarek}, and let $\nu=\{\nu_\alpha,\h{U}_\alpha\}$ be a $\delta$-approximate $k$-expander with 
\begin{align}
    \delta\leq \left( \frac{\clower \Gamma^2 \ep^2}{
    \dim \sqrt{\log\frac{6}{\clower \ep\Gamma}}}\right)^{d^2-1}, \qquad \Gamma=\min\{1-\alpha,\beta-1\}\,,
\end{align}
and 
\begin{align} 
    k\geq  30 \frac{\dim^{5/2}}{\ep\Gamma }
   \eta(\ep\Gamma,\dim)\,,
\end{align}
with 
\begin{align}
\eta(\ep,\dim)=
\sqrt{\log\frac{6}{\clower \ep}}
 \sqrt{\frac{1}{2}\log\frac{6}{\clower \ep} + \log\biggl(\frac{6\dim}{\ep}\log\frac{6}{\clower \ep}\biggr)}\,.
\end{align}
It follows that
\begin{equation}
    \vol(\alpha\ep)\leq \nu(B(\h{U},\ep))\leq \vol(\beta\ep)\,.
\end{equation}
\end{proposition}
\begin{proof}
{\bf Lower bound.}
From left hand side of Eq.~\eqref{eq:equip-approx-preparatory} of \cref{lem:equidistribution-appr-des-preparatory}
we have
\begin{align}
    \vol\left (\kapls\right) - \gdirac\left(\ep -\kapls\right) - h_k^\sigma - \delta \|\F_k^\sigma\|_2
    \leq \nu(B(\h{U},\ep))\,.
\end{align}
We proceed as in the proof of \cref{prop:equi-exact}. 
The only difference is that for $k\geq d/\sigma$, we bound $\delta \|\F_k^\sigma\|_2$ by $1/6\, \vol(\ep-\kapls)$, 
which is possible if $\delta$ satisfies Eq.~\eqref{eq:delta} with $\ep-\kapls$ in place of $\ep$ and with $\sigma=\sigma_\ast(\ep-\kapls)$. Here, as in the proof of \cref{prop:equi-exact}, $\sigma^*$ is given by Eq.~\eqref{eq:sigma-star} and 
$\ep-\kapls=(1-\alpha)\ep/3$). 
Note that the required lower bound for $k$ given by $d/\sigma$ is smaller than one needed to bound $\gdirac(\ep-\kapls)$ + $h_k^\sigma$, we can thus keep the bound for $k$ the same as in \cref{prop:equi-exact} given by $k^*(\ep-\kapls)$  with $k^*$ of Eq.~\eqref{eq:ktilde}. The expression $\ep-\kapls$ we replace by $\ep\Gamma/3$.
For such conditions for $k$, $\sigma$, and $\delta$ we find
\begin{align}
    \nu(B(\h{U},\ep)) \geq
    \vol(\kapls)-\frac56\vol(\ep-\kapls)-\frac16\vol(\ep-\kapls) \geq \vol(\kapls)-\vol(\ep-\kapls)\,.
\end{align}
We have thus obtained the same estimate as Eq.~\eqref{eq:vol-kappa-vol-alpha}
in the proof of \cref{prop:equi-exact}. Therefore,
by the subsequent reasoning from that proposition leads to the required lower bound. 

{\bf Upper bound.}  Similarly as in the proof for lower bound, we have
\begin{align}
\nu(B(\h{U},\ep)) \leq  
\vol(\kapgr) + \gdirac(\kapgr-\ep)+h_k^\sigma  +\delta \|\F_k^\sigma \|_2,
\end{align}
so that 
\begin{align}
    \nu(B(\h{U},\ep)) \leq  
    \vol(\kapgr)+ \frac56 \vol(\kapgr-\ep) + \frac{1}{6} \vol(\kapgr-\ep) \leq
    \vol(\kapgr)+  \vol(\kapgr-\ep).
\end{align}
We have thus arrived at the estimate in Eq.~\eqref{eq:nu_kappa_kappa_ep}, and we can continue exactly as in \cref{prop:equi-exact}, which concludes the proof.
\end{proof}

\subsection{Equidistribution on \ensuremath{\S(\dim)} induced by approximate unitary designs}
We consider the set of states $\S(\dim)$ equipped with measure $\mu_\S$ (induced by the Haar measure on unitaries). The trace distance can be rewritten in terms of the overlap as 
\begin{align}
    \d(\psi,\phi)=\sqrt{1-\tr(\psi \phi)}\,.
\end{align}
The ball $B(\phi,\ep)=\{\phi:\d(\phi,\psi)\leq \ep\}$ can be equivalently described as follows
\begin{equation}
\label{eq:ball-overlap}
    B(\phi,\ep)=\{\psi: \tr(\psi \phi)\geq \sqrt{1-\ep^2}\}\,.
\end{equation}
We introduce the following polynomial approximation of Dirac delta centered around $\phi$
\begin{align}
    \F_{k,\phi}(\psi)=\frac{1}{I_k} \tr(\psi \phi)^k\,,
\end{align}
where 
\begin{align}
\label{eq:Ik}
    I_k = \int \dt \nu_\S(\psi) \tr(\psi \phi)^k = \binom{\dim+k-1}{k}^{-1}\,.
\end{align}
If the index $\phi$ is omitted, this means that  we consider $\phi=\psi_0$, where $\psi_0$ is 
arbitrarily but fixed chosen reference state. Let us further denote
\begin{equation}
    \Jdirac(\ep)=\int_{B^c(\psi_0,\ep)} \dt \nu_\S(\psi)\,  \F_k(\psi)\,. 
\end{equation}
We have the following bound on $\Jdirac(\ep)$:
\begin{lemma}
\label{lem:bound-J}
For $0<\ep\leq1$ and $k\geq \frac{4 d }{\ep^2}$
we have $\Jdirac(\ep)\leq \exp\left(-k \frac{\ep^4}{8}\right) $.
\end{lemma}

\begin{proof}
Directly from the definition of $\Jdirac(\ep)$ and Eq.~\eqref{eq:ball-overlap} we obtain
\begin{align}
    \Jdirac(\ep) = \int_{B^c(\psi_0,\ep)}  \dt \mu_\S(\psi) \F_k(\psi)
    =
    \binom{\dim +k - 1}{k}\int_{\psi:\tr(\psi \psi_0)\leq \sqrt{1-\ep^2}} \dt \mu_\S(\psi) \tr(\psi \psi_0)^k \,.
\end{align}
Passing to the variable $x=\tr(\psi \psi_0)$,
whose distribution is given by  Eq.~\eqref{eq:overlapdistribution} we get
\begin{align}
    \Jdirac(\ep) = \binom{\dim +k - 1}{k} \int_0^{\sqrt{1-\ep^2}} \dt x\, x^k (\dim-1) (1-x)^{d-1}\,.
\end{align}
The distribution appearing in above equation
is known as Beta distribution (supported on interval $[0,1]$) with two parameters
\begin{align}
    p_{\rm Beta}(x)(a,b) = {\mathcal N}(a,b)\, x^{b-1}(1-x)^{a-1}\,,
\end{align}
where $\mathcal N$ is normalization constant.
In our case $a=k+1$, $b = \dim-1$, and so $\mathcal{N}(a,b)$ is equal to $I^{-1}_k$ in Eq.~\eqref{eq:Ik}. 
The average of this distribution is given by $\BE X = \frac{a}{a+b}=\frac{k+1}{\dim+k}$. In Ref.~\cite{beta-tail} ({see also}~\cite{zhang2020nonasymptotic}) the following tail inequality is provided for the Beta distributed random variable $X$
\begin{align}
    \max\big\{
    \Pr (X\geq \BE X+\Delta),~
    \Pr (X\leq \BE X-\Delta)\big\}
    \leq \exp\big(-2(a+b+1) \Delta^2
    \big)\,,
\end{align}
which gives in our case as $\Pr( X\leq \frac{k+1}{\dim+k} -\Delta)\leq \exp(-2(\dim+k+1) \Delta^2\bigr)$. Since $\Jdirac(\ep)=\Pr(X\leq \sqrt{1-\ep^2})$, we 
choose 
\begin{align}
    \Delta= \left(1 - \frac{k+1}{\dim-1} \right)- \sqrt{1-\ep^2}\geq 
     \frac{k+1}{ \dim+k} - \ep^2\,,
\end{align}
where we applied $\sqrt{1-\ep^2}\leq 1-\ep^2/2$, so that we obtain
\begin{align}
    \Jdirac (\ep)\leq \exp\left
    \{-2(\dim+k+1)\left(\frac{\ep^2}{2}
    - \frac{d-1}{d+k}\right)^2\right\}.
\end{align}
We now restrict to $k$ satisfying $k\geq 4 \dim/\ep^2$.  This implies that $\frac{d-1}{d+k}\leq \frac{\ep^2}{4}$ and hence
\begin{align}
   \Jdirac(\ep)\leq \exp\left
    \{-2(\dim+k+1)\left(\frac{\ep^2}{2}
    - \frac{d-1}{d+k}\right)^2\right\} 
    \leq 
        \exp\left
    \{-2(\dim+k+1)\frac{\ep^4}{16}\right\} \leq  e^{-k \frac{\ep^4}{8}}\,.
\end{align}
\end{proof}

Moreover, we have the following estimate for second norm of $\F_k$.

\begin{lemma}
\label{lem:F-bound}
For $k\geq \dim$ we have that
\begin{equation}
    \|\F_k\|_2 \leq \left( \frac{6k}{\dim}\right)^\dim\,.
\end{equation}
\end{lemma}

\begin{proof}
We start by noting that the 2-norm of $\F_k$ can be written as
\begin{align}
\|\F_k\|_2 = \frac{\sqrt{I_{2k}}}{I_k} \leq \frac{1}{I_k} = \binom{d+k-1}{k}\,.
\end{align}
Proceeding, we write
\begin{equation}
    \binom{d+k-1}{k} = \binom{d+k-1}{d-1} \leq \binom{d+k}{d} = \frac{(d+k)(d+k-1)\ldots (k+1)}{d!}\,.
\end{equation}
In the regime where $k$ is larger than $d$, we have
\begin{equation}
    \binom{d+k}{d} = \frac{(k+d)(k+d-1)\ldots(k+1)}{d!} \leq \frac{(2 k)^d}{d!} \leq \left( \frac{2ek}{d}\right)^d\,,
\end{equation}
which gives the desired estimate for $1/I_k$ and hence also for $\|\F_k\|_2$.
\end{proof}

To prove equidistribution we need a relation between $\|\F_k\|_2$ and $\Jdirac(\ep)$ and the volumes of balls. This is contained in the following lemma.
\begin{lemma}
\label{lem:FJ-vol}
For $0<\ep< \frac14 $ and $k$ satisfying 
\begin{align}
    k\geq k_{**}(\ep)=
    16\,\frac{\dim}{\ep^4}\log( 1/\ep)\,,
\end{align}
we have 
\begin{align}
    \Jdirac(\ep)\leq \frac12 \vol_\S(\ep)\,.
\end{align}
Moreover, for $k$ as above and $\delta$ satisfying
\begin{align}
    \delta\leq \left(\frac{\ep}{3 \log( 1/\ep)}\right)^{6d}\,,
\end{align}
we have
\begin{align}
    \delta \|\F_k\|_2\leq \frac12 \vol_\S(\ep)\,.
\end{align}
\end{lemma}

\begin{proof}
{\bf Bound for $\Jdirac$.}
Given the expression for the volume $\vol_\S(\ep)=\ep^{2\dim-2}$ 
and the bound on $\Jdirac$ in \cref{lem:bound-J}, we need to choose $k$ such that 
\begin{align}
\label{eq:dociaganie-J}
    e^{-k \frac{\ep^4}{8}}
    \leq \frac12 
    \left(\frac{\ep}{2}\right)^{2\dim-2}\,.
\end{align}
Taking logarithm of both sides, solving for $k$ and noting that $\ep^2\leq 1/2$ if $\ep\leq 1/4$, we get that
for $k\geq k_{\ast \ast}(\ep)$ the inequality Eq.~\eqref{eq:dociaganie-J} is satisfied. Also note, that $k\geq k_{\ast \ast}$ implies that $k\geq 4\dim/\ep^2 $ as needed in order to
use \cref{lem:bound-J}.

{\bf Bound for $\|\F_k\|_2$}.
We  use the  bound in \cref{lem:F-bound}, with $k$ as in assumptions, and want $\delta$ to satisfy
\begin{align}
    \delta  \left(\frac{d}{6k}\right)^{d}
    \leq\frac12 \ep^{2d-2}\,,
\end{align}
or after inserting $k_{\ast\ast}$ 
\begin{align}
    \delta\leq\frac12\ep^{6d-2} \left(\frac{1}{96\log(1/\ep)}\right)^d\,.
\end{align}
Since we can bound the right hand side as follows
\begin{align}
\frac12\ep^{6d-2} \left(\frac{1}{96\log(1/\ep)}\right)^d\geq     
\frac12\ep^{6d} \frac{1}{2\ep^2} \left(\frac{1}{3\log(1/\ep)}\right)^{6d} \geq \left(\frac{\ep}{3\log(1/\ep)}\right)^{6 d}\,,
\end{align}
where the last estimate holds for $\ep\leq 1/4$, we obtain the required result.
\end{proof}

Finally, we need the following lemma (proven in Ref.~\cite{Varju2013} and also in Ref.~\cite{OSH2020}):
\begin{lemma}
Let $\nu=\{\nu_\alpha,\h{U}_\alpha\}$
be $\delta$-approximate unitary $k$-expander. 
Let $\psi_\alpha=\h{U}_\alpha (\psi_0)$. Then  
\begin{align}
\left|\vol_\S  (\ep)  - \int_{B(\psi_0,\ep)} \sum_\alpha \nu_\alpha \F_{k,\psi_\alpha}(\psi) \dt\mu_\S \right|\leq \delta \| \F_k\|_2\,.
\end{align}
\begin{proof}
For an arbitrary measure $\xi$ and a function $f$ on states, let us denote:
\begin{align}\label{eq:averaging-for-states}
    (\avgstate{\xi} f)(\psi)=\int \dt \xi(\h{U})
    f(\h{U}^{-1} (\psi))\,.
\end{align}
We then notice that 
\begin{align}
    (\avgstate{\mu_\S} \F_k)(\psi) = 1,
    \quad 
    (\avgstate{\mu_\S} \F_k)(\psi) = 
    \sum_\alpha \nu_\alpha \F_{k,\psi_\alpha}(\psi)\,.
\end{align}
Thus we have 
\begin{align}
\label{eq:Ts}
\left|\vol_\S(\ep)  - \int_{B(\psi_0,\ep)}\sum_\alpha \nu_\alpha \F_{k,\psi_\alpha}(\psi) \dt\mu_\S \right|
=\left|\int_{B(\psi_0,\ep)} \big(\avgstate{\mu} \F_k - \avgstate{\nu} \F_k\big)\dt \mu_\S\right| 
=\left|\langle (\avgstate{\mu_\S}-\avgstate{\nu}) \F_k, I_{B(\psi_0,\ep)}\rangle\right|\,.
\end{align}
Here $I_{B(\psi_0,\ep)}$ is indicator function of the ball $B(\psi_0,\ep)$, and scalar product is with respect to square integrable functions on $\S(\dim)$ with measure $\mu_\S$.
We next apply the Cauchy-Schwartz inequality
\begin{align}
| \< (\avgstate{\mu_\S}-\avgstate{\nu} \F_k, I_{B(\psi_0,\ep)}\> | \leq 
\|(\avgstate{\mu_\S}-\avgstate{\nu})\F_k\|_2 \, \|I_{B(\psi_0,\ep)}\|_2 \leq 
\|(\avgstate{\mu_\S}|_k-\avgstate{\nu}|_k)\|\, \|F_k\|_2 \sqrt{\vol_\S(\ep)}\,,
\end{align}
where by $|_k$ we denote the restriction to the 
set of the polynomials of degree $k$. 
Next we notice that our polynomials on $\S(\dim)$ embed isometrically into polynomials on $\UU(\dim)$ so that 
\begin{equation}\label{eq:T-states}
    \|\avgstate{\mu_\S}|_k-\avgstate{\nu}|_k\|
    \leq  \|M_{\mu}|_k-M_\nu|_k\|\,,
\end{equation}
where  $M_\xi$ is defined on unitary channels as in \eqref{eq:averaging-operator}, and with Haar measure $\mu$  in place of $\mu_\S$.
In terms of notation from Ref.~\cite{OSH2020},
$\|M_{\mu}|_k-M_\nu|_k\|=
\|M_{\mu,k}-M_{\nu,k}\|$
which in turn is equal to $\delta$
since $\nu$ is a $\delta$-approximate $k$-design.
Returning to Eq.~\eqref{eq:Ts}, we see that this ends the proof.
\end{proof}
\end{lemma}

We can now turn to the topic of equidistribution for states. In analogy to the unitary channels case, we obtain the following lemma.

\begin{lemma}
\label{lem:equidistribution-appr-des-preparatory-states}
Let $\nu=\{\nu_\alpha,\U_\alpha\}$ be a unitary $\delta$-approximate $k$-expander. Consider the measure   $\nu_\S=\{\nu_\alpha,\U_\alpha(\psi_0)\}$ on the set of states $\S(\dim)$.
Then for any $\h{\psi} \in \S(\dim)$, and for any  $\kapls,\kapgr$ and $\ep$ satisfying $0<\kapls<\ep<\kapgr$, we have
\begin{align}
\vol_\S\left (\kapls\right) - \Jdirac\left(\ep-\kapls\right)  - \delta \|\F_k\|_2   \leq 
\nu_\S(B(\h{\psi},\ep)) \leq  
\vol(\kapgr) + \Jdirac(\kapgr-\ep) +\delta \|\F_k\|_2
\end{align}
\end{lemma}

As we see the only difference  (in comparison with unitary designs) is that there is no $h$ function.
This is because our function is now positive, and the $h$ function vanishes. Thus the proof of this lemma is the same as the proof of  \cref{lem:equidistribution-appr-des-preparatory} with the function $h$ set to zero.  

\begin{proposition}
\label{lem:equidistribution-appr-des-states}
Let $0<\ep\leq 1$, and
let $\nu=\{\nu_\alpha,\U_\alpha\}$ be a unitary $\delta$-approximate $k$-expander
with  
\begin{align} 
\label{eq:delta-states}
    \delta\leq  \left(\frac{ \ep\Gamma}{9 \log\bigl(3 /(\ep\Gamma)\bigr)}\right)^{6d}
\end{align}
and 
\begin{align}
\label{eq:k-states}
    k\geq   1296\,\frac{d}{(\ep\Gamma)^4}\log\bigl(3/(\ep\Gamma)\bigr)
\end{align}
with $\Gamma=\min\{\beta-1,1-\alpha\}$.
Consider the measure $\nu_\S=\{\nu_i,\U_i(\psi_0)\}$ on the set of states $\S(\dim)$. We then have for any $\psi$
\begin{align}
      \vol_\S(\alpha \ep)\leq \nu_\S(B(\psi,\ep))\leq \vol_\S(\beta\ep)\,.
\end{align}
\end{proposition}

\begin{proof}
We know from \cref{lem:FJ-vol} 
that for $k$ satisfying Eq.~\eqref{eq:k-states} and $\delta$
satisfying Eq.~\eqref{eq:delta-states}
\begin{equation}
\begin{split}
    \Jdirac(\ep-\kapls)\leq \frac12
    \vol_\S\left(\ep-\kapls\right), &\quad 
    \Jdirac\left(\kapgr-\ep\right)\leq \frac12 \vol_\S\left(\kapgr-\ep\right),\\
    \delta \|\F_k\|_2\leq \frac12 \vol_\S\left(\kapgr-\ep\right),
    &\quad 
    \delta \|\F_k\|_2\leq \frac12 \vol_\S\left(\ep-\kapls\right).
\end{split}
\end{equation}
We apply these inequalities to \cref{lem:equidistribution-appr-des-preparatory-states} and get 
\begin{align}
\vol_\S\left (\kapls\right) -
\vol_\S\left (\ep-\kapls\right)
\leq 
\nu_\S(B(\h{\psi},\ep)) \leq  
\vol_\S(\kapgr) + 
\vol_\S\left (\kapgr-\ep\right).
\end{align}
which are analogues of estimates Eqs.~\eqref{eq:nu_kappa_kappa_ep} and \eqref{eq:vol-kappa-vol-alpha} from the proof of \cref{prop:equi-exact} concerning unitaries.  Then using \cref{fact:vol-2} (or actually its immediate modification for states), and knowing that diameter of $\S(\dim)$ is equal to 1 we obtain the desired estimate in the same way as in the proof of \cref{prop:equi-exact}.
\end{proof}

\section{Technical results concerning packing and covering numbers, distances and volumes in the set of unitary channels}\label{app:technicalapp}

\subsubsection*{Packing and covering numbers}

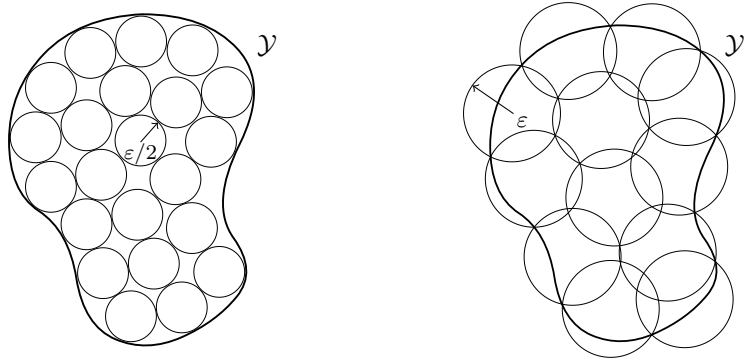
\begin{figure}[h]
    \centering
    \begin{tikzpicture}[scale=0.6,baseline=-1mm]
    \node at (0,0) {\includegraphics[width=0.2\linewidth]{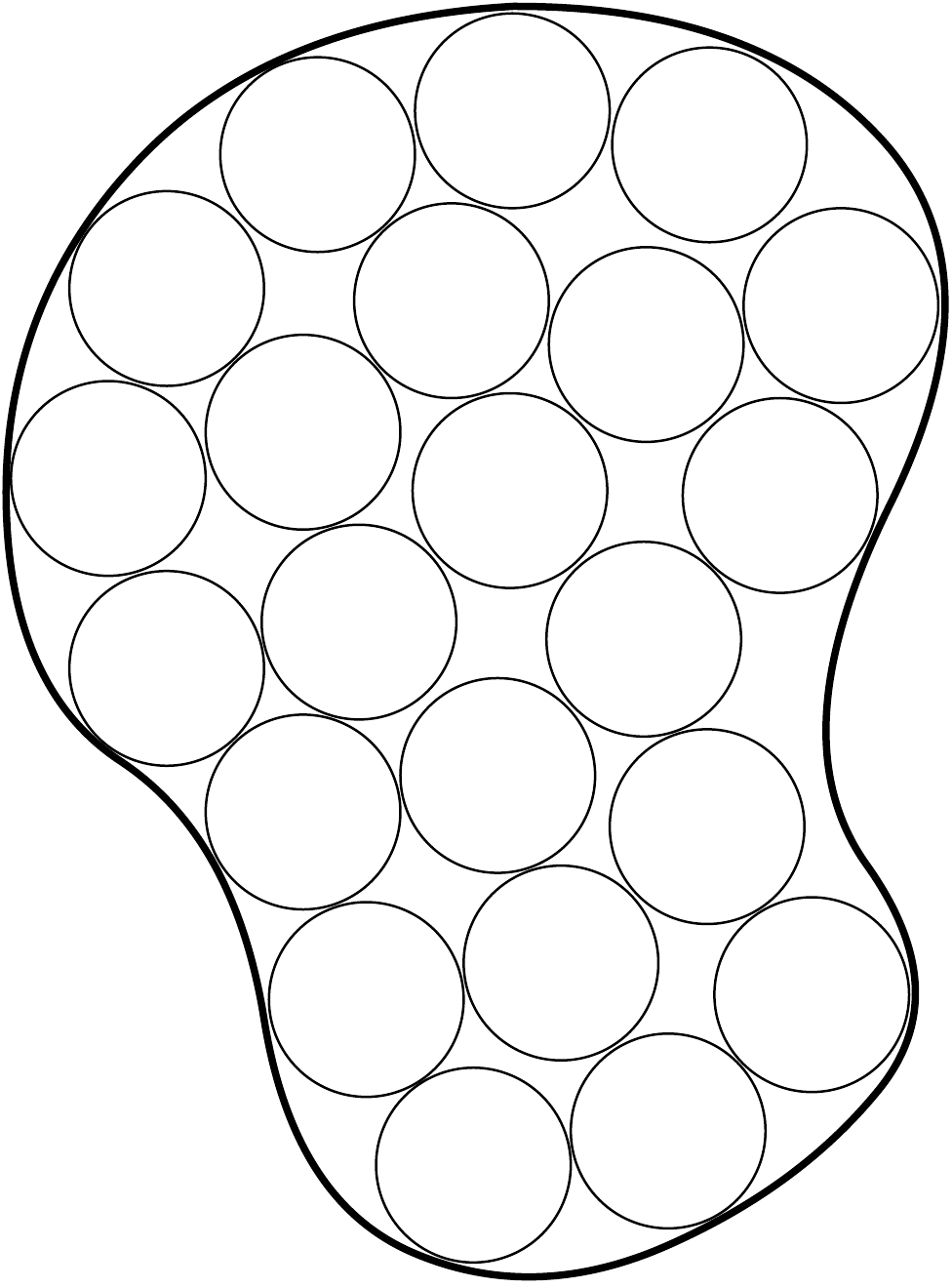}};
    \draw[->] (0.2,0.85) -- (0.58,1.28);
    \node at (0.2,0.6) {{\footnotesize $\ep/2$}};
    \node at (3,3) {$\Y$};
    \end{tikzpicture}
    \hspace*{2cm}
    \begin{tikzpicture}[scale=0.6,baseline=-1mm]
    \node at (0,0) {\includegraphics[width=0.22\linewidth]{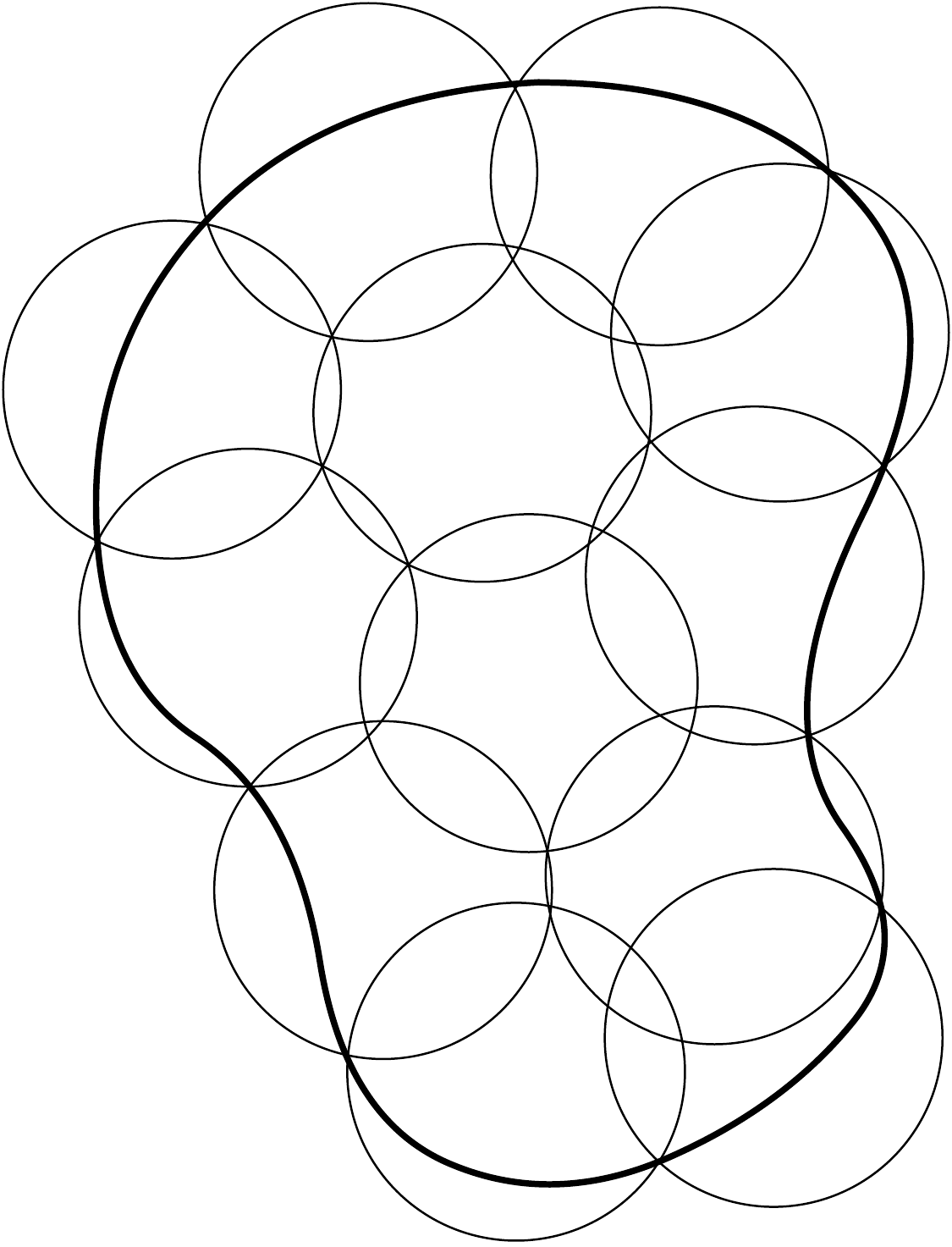}};
    \draw[->] (-1.9,1.48) -- (-2.8,2.1);
    \node at (-1.7,1.3) {\small $\ep$};
    \node at (3,3) {$\Y$};
    \end{tikzpicture}
    \caption{Illustrations of the packing and covering numbers. On the left: a diagram of a maximal packing of balls of radius $\ep/2$ in a subset $\Y$ of a metric space $\X$. On the right: a minimal covering of the same subset, built from balls of radius $\ep$. 
    }
   \label{fig:pack-cov}
\end{figure}

 Throughout this work we will make use of the \emph{packing} and \emph{covering} numbers; see, for instance, Ref.~\cite{SzarekBook}. They are defined as follows: Let $\Y$ be a subset $\Y\subset \X$ of a metric space $\X$, equipped with distance measure $\d$. We will consider $\X=\UU(\dim)$ or $\X=\S(\dim)$ with distances defined in the previous subsection. The packing number $\npack(\Y,\ep)$ quantifies the maximal number of points in $\Y$ that are pairwise distant. The covering number counts the minimal number of balls of size $\ep$ centered at points in $\Y$ needed to cover the whole set. For an illustration of these concepts see \cref{fig:pack-cov}. Formally, we have
\begin{align}
    \npack(\Y,\ep) &:= \max\big\{|S|: \forall_{x,y\in S}\ \d(x,y)\geq \ep, ~S\subset \Y\big\}\\
    \ncov(\Y,\ep) &:= \min\left\{|S|:Y\subset\bigcup_{x\in S} B(x,\ep), ~S\subset \Y\right\}.
\end{align}
Note that dependence of above numbers on the metric $\d$ is suppressed. 

Covering and packing numbers are related as follows:
\begin{align}
\label{eq:pack-cov}
    \ncov(\Y,2\ep)\leq \npack(\Y,\ep) \leq \ncov(\Y,\ep)\,.
\end{align}
Moreover, we have
\begin{equation}
\label{eq:npack-sum-rule}
\npack\left(\bigcup_i \Y_i,\ep\right)\leq \sum_i\npack(\Y_i,\ep)\,.
\end{equation}
In the cases where $\X=\Y=\UU(\dim)$ and $\X=\Y=\S(\dim)$, both the metric and the measure are \emph{unitary invariant} and therefore
\begin{equation}
    \npack(\UU(\dim),\ep)\leq \vol(\ep)^{-1}\ ,\ \npack(\S(d),\ep)\leq \vol_\S(\ep)^{-1} \ .
\end{equation}
Similarly, we have 
\begin{equation}
    \ncov(\UU(\dim),\ep)\geq \vol(\ep)^{-1}\ ,\ \ncov(\S(d),\ep)\geq \vol_\S(\ep)^{-1} \ .
\end{equation}
Analogous bounds can be derived if we want to control the covering and packing numbers of a measure $\nu$ that is not invariant, but where we have control over the measure it assigns to balls of radius of size of order $\ep$ (like in the definition of approximate equidistribution). In the following lemma we collect bounds for the covering and packing numbers for unitary channels and states that result from Eqs.~\eqref{eq:szarek} and \eqref{eq:pureSTATESvol} and bounds given in Eq.~\eqref{eq:szarek}.
\begin{lemma}[Bounds on covering and packing numbers for unitary channels and states]
We have the following bounds on packing and covering numbers of $\UU(\dim)$ and $\S(\dim)$ equipped with distances defined in Eq.~\eqref{eq:distancesDEF}:
\begin{equation}
\label{eq:szarek-cov}
\left(\frac{1}{\ep\cupper}\right)^{\dim^2-1}  \leq   \ncov(\UU(\dim),\ep) \leq \left(\frac{2}{\ep \clower}\right)^{\dim^2-1}\,,
\end{equation}
\begin{equation}
\label{eq:szarek-pack}
\left(\frac{1}{2\ep\cupper}\right)^{\dim^2-1}  \leq   \npack(\UU(\dim),\ep) \leq \left(\frac{1}{\ep \clower}\right)^{\dim^2-1}\,,
\end{equation}
\begin{equation}
\label{eq:cover-pure}
\left(\frac{1}{\ep}\right)^{2\dim-2}  \leq   \ncov(\S(\dim),\ep) \leq \left(\frac{2}{\ep}\right)^{2\dim-2} \,,
\end{equation}
\begin{equation}
\label{eq:pack-pure}
\left(\frac{1}{2\ep}\right)^{2\dim-2}  \leq   \npack(\S(\dim),\ep) \leq \left(\frac{1}{\ep}\right)^{2\dim-2} \,.
\end{equation}
\end{lemma}

\subsubsection*{Distances and volumes}

\begin{lemma}\label{lem:innerProddist}
If two unitary channels $\U$ and $\V$ have $\d(\U,\V)\leq \ep$, then  Hilbert-Schmidt inner product of the corresponding unitary matrices satisfies $|\tr(UV^\dagger)|^2\geq d^2(1-\ep^2)$.
\end{lemma}
\begin{proof} The proof of this claim follows from a similar statement in Ref.~\cite[Lemma 3]{complexitygrowth2019}. We proceed by lower bounding the distance in terms of the diamond distance as
\begin{equation}
    \d(\U,\V) \geq \frac{1}{2}\|\U-\V\|_\diamond \geq \frac{1}{2} \big\| (\U\otimes\I - \V\otimes\I)(\ketbra{\Omega}) \big\|_1 = \sqrt{1- \frac{1}{d^2} |\tr(UV^\dagger)|^2}\,,
\end{equation}
where $\ket\Omega$ is again the maximally entangled state on two copies of the system. It thus follows that $|\tr(UV^\dagger)|^2\geq d^2(1-\ep^2)$ is a necessary condition for $\d(\U,\V)\leq \ep$.
\end{proof}

\begin{lemma}[Volumes of balls in $\UU(\dim)$ according to approximate unitary expanders] \label{lem:volUnitaryDesignsMoments}
Let $\nu$ be $\delta$-approximate unitary $k$-expander. For any unitary channel $\V\in\UU(d)$  and any $\ep>0$, we then have
\begin{equation}\label{eq:unitaryBOUND}
    \nu\big(B(\V,\ep)\big)
    \leq\frac{k! + \dim^{2k}\delta}{\dim^{2k}(1-\ep^2)^{k}}\ .
\end{equation}
\end{lemma}

\begin{proof}
We first note that if $\d(\U,\V)\leq \ep$, then it follows from \cref{lem:innerProddist} that 
\begin{align}
    \left|\tr(UV^\dagger)\right|^2\geq 
    \dim^2(1-\ep^2)\,.
\end{align}
Therefore 
\begin{align}
    \nu \left(B(\V,\ep) \right)
    \leq
    \nu\left(\{\U:
    \left|\tr(UV^\dagger)\right|^2\geq 
    \dim^2(1-\ep^2)\}\right) 
    =
    \nu\left(\{\U:
    \left|\tr(UV^\dagger)\right|^{2k}\geq 
    \dim^{2k}(1-\ep^2)^{k}\}\right) 
\end{align}
Applying Markov's inequality, we find that
\begin{align}
\label{eq:ball-bound}
    \nu \left(B(\V,\ep) \right)
    \leq\frac{\BE_{\U\sim \nu}\left[\big|\tr(UV^\dagger)\big|^{2k}\right]}{\dim^{2k}(1-\ep^2)^{k}}\,.
\end{align}
We now want to bound this quantity for approximate unitary designs. First we note that we may rewrite the quantity in the expectation as \
\begin{align}
    \big|\tr(UV^\dagger)\big|^2 &= \dim^2 \,\tr \big( \ketbra{\Omega} \U\V^{-1} \otimes \h{I}(\ketbra{\Omega})  \big)\\
    &= \dim^2\bra{\Omega}(\U \otimes \I)\big((\V^{-1}\otimes \h{I})(\ketbra{\Omega})\big) \ket\Omega\,,
    \label{eq:trUV}
\end{align}
where $\ket\Omega = \frac{1}{\sqrt{\dim}} \sum_j \ket{j}\otimes \ket{j}$ is the maximally entangled state on two copies of the system and $\{\ket j\}_{j=1}^\dim$ is the computational basis of $\C^\dim$.

Let $\nu$ be an $\delta$-approximate unitary $k$-expander, it follows from Eq.~\eqref{eq:trUV} that
\begin{align}
    \BE_{U\sim \nu}\big[|\tr(UV^\dagger)|^{2k}\big] 
    &= \BE_{U\sim \nu}\big[|\tr(UV^\dagger)|^{2k}\big] - \BE_{U\sim \mu}\big[|\tr(UV^\dagger)|^{2k}\big] + \BE_{U\sim \mu}\big[|\tr(UV^\dagger)|^{2k}\big]\\
    &= \dim^{2k} \tr\Big(\ketbra{\Omega}^{\otimes 2k} \big(\Phi^{(k)}_\nu - \Phi^{(k)}_\mu\big) (\V^{-1}(\ketbra{\Omega}))^{\otimes k} \Big) + \BE_{U\sim \mu}\big[|\tr(U)|^{2k}\big]\\
    &\leq \dim^{2k} \Big\|\Big(\Phi^{(k)}_\nu - \Phi^{(k)}_\mu\Big) (V^\dagger\ketbra{\Omega}V)^{\otimes k} \Big\|_2 
    + k!\\
    &\leq \dim^{2k} \big\|\Phi^{(k)}_\nu - \Phi^{(k)}_\mu \big\|_{2\ra 2} + k!\\
    &\leq \dim^{2k} \delta + k!\,,
\end{align}
where we use the fact that for any $k$ the Haar expectation of moments of traces is $\BE_{U\sim \mu}\big[|\tr(U)|^{2k}\big]\leq k!$, with equality if $k\leq \dim$. Furthermore, in the third line we use H\"older's inequality, in the fourth line we employ the definition of the superoperator 2-norm, and in the last line we use the equality between $k$-fold channels and moment operators $\|\Phi^{(k)}_\nu - \Phi^{(k)}_\mu \|_{2\ra 2} = \|M^{(k)}_\nu - M^{(k)}_\mu \|_\infty$ and then \cref{def:unitarydesign}. Subsequently, it follows that
\begin{equation}
    \nu\big(B(\V,\ep)\big)
    \leq\frac{k! + \dim^{2k}\delta}{\dim^{2k}(1-\ep^2)^{k}}\,.
\end{equation}
\end{proof}

\begin{lemma}[Volumes of balls in $\S(d)$ according to measures induced by approximate designs]\label{lem:volStatesDesignsMoments}
Let $\nu$ be $\delta$-approximate unitary $k$-expander in $\UU(\dim)$ and let $\nu_\S$ be the induced measure on the set of pure quantum states $\S(d)$ (see \cref{sec:technical}). For $\ep>0$ and for a pure state $\phi\in\S(\dim)$ we have
\begin{equation} \label{eq:stateBOUND}
    \nu_\S \big(B(\phi,\ep) \big) \leq  \frac{k!+d^k\delta}{d^k(1-\ep^2)^k}\,.
\end{equation}
\end{lemma}
\begin{proof}
First, we note that for two pure states $\psi$ and $\phi$, we have $\d(\psi,\phi) = \sqrt{1-\tr(\psi \phi)}$, and thus for a distribution $\nu_\S$ on the set of pure state
\begin{equation}\label{eq:chainbounds}
    \nu_\S \left(B(\phi,\ep) \right) \leq
    \nu_\S\left(\{\psi : \tr(\psi\phi) \geq 1-\ep^2\}\right) 
    =\nu_\S\left(\{\psi :
    \left|\tr(\psi\phi )\right|^k\geq (1-\ep^2)^k\}\right) \leq\frac{\BE_{\psi\sim \nu_\S}\left[|\tr(\psi\phi)|^k\right]}{(1-\ep^2)^k}\,,
\end{equation}
where in the last step we used Markov's inequality. If the distribution on states $\nu_\S$ is a $\delta$-approximate state design, then it follows that
\begin{align}
    \BE_{\psi\sim \nu_\S}\left[|\tr(\psi\phi)|^k\right] &= \BE_{\psi\sim \nu_\S}\left[|\tr(\psi\phi)|^k\right] - \BE_{\psi\sim \mu_\S}\left[|\tr(\psi\phi)|^k\right] + \BE_{\psi\sim \mu_\S}\left[|\tr(\psi\phi)|^k\right]\\
    &= \tr \left[\phi^{\otimes k} (\Phi_\nu^{(k)} - \Phi_\mu^{(k)})( \phi_0^{ \otimes k})\right]  + \binom{d+k-1}{k}^{-1} \tr\left( \phi^{\otimes k} \Pi_\vee\right)\,, \\
    &\leq \big\|\Phi^{(k)}_\nu - \Phi^{(k)}_\mu \big\|_{2\ra 2} 
    + \binom{d+k-1}{k}^{-1}\ ,
\end{align}
where we used the fact that the $k$-fold average of Haar random states is proportional to the projector onto the symmetric subspace $\Pi_\vee$,
and in the last line used the same reasoning as in the proof of \cref{lem:volUnitaryDesignsMoments}. We then use $\|\Phi^{(k)}_\nu - \Phi^{(k)}_\mu \|_{2\ra 2} = \|M^{(k)}_\nu - M^{(k)}_\mu \|_\infty \leq \delta$ to upper bound the expression in Eq.~\eqref{eq:chainbounds}. To conclude, we use the fact that $\binom{d+k-1}{k}\geq \frac{d^k}{k!}$.
\end{proof}

It is interesting to further comment on the forms of \cref{lem:volUnitaryDesignsMoments} and \cref{lem:volStatesDesignsMoments}. Let us consider the respective Haar volumes, for which the lemmas apply with $\delta=0$ and for all $k$. For large enough $k$, \cref{lem:volUnitaryDesignsMoments} fails to improve as the numerator grows. However for $k=O(d^2)$, we recover the correct $d$ dependence of $\vol(\ep)$, albeit with a much weaker $\ep$ dependence which does not diminish as we take $\ep\ra 0$. On the other hand, \cref{lem:volStatesDesignsMoments} is quite different; before taking the final bound in the proof we had for states $\nu_\S(B(\phi,\ep))\leq \binom{d+k-1}{k}^{-1}(1-\ep^2)^{-k}$, where the inverse binomial coefficient continues to decrease for larger $k$. From the expression we recover the correct $d$ dependence of $\vol_\S(\ep)=\ep^{2(d-1)}$, but the bound for states appears to know much more about the dependence on the radius of balls in the space of states.

We conclude this section by proving \cref{prop:ringVOL}, an important technical proposition that allows one to pinpoint the timescales at which a complexity recurrence takes place for the case of quantum channels. We start by stating a useful relation between the distance  $\d(\U,\h{I})$ and the length of the shortest arc containing the eigenvalues of a unitary matrix $U\in\mathbb{U}(\dim)$ that gives rise to a channel $\U\in\UU(\dim)$.
\begin{lemma}\label{lem:distARCtranslate}
Let $\mathrm{Eig}(U)=\lbrace{\exp(\ii\alpha_1),\exp(\ii\alpha_2),\ldots,\exp(\ii\alpha_d)\rbrace}$ be eigenvalues of $U\in\mathbb{U}(\dim)$ corresponding to unitary channel $\U\in\UU(\dim)$.  Furthermore, let $\arc(X)$ denote the length of the shortest arc containing a subset $X\in S^1$. Assuming that $\d(\U,\h{I})\leq \sqrt{2}$, we then have
\begin{equation}
    \d(\U,\h{I})=\kappa \quad\Longleftrightarrow\quad \arc(\mathrm{Eig}(U))= 4\arcsin\left(\frac{\kappa}{2}\right)\,.
\end{equation}
\end{lemma}
The proof of the above follows from elementary geometric considerations given in Proposition 6 and illustrated in Figure 3 of \cite{OSH2020}.

\begin{proposition}[Upper bound for the volume of an annulus in $\UU(\dim)$]\label{prop:ringVOL}
Let $\lambda\geq1$ and $\kappa>0$, such that $\lambda \kappa\leq \sqrt{2}$, and let $A$ be an annulus in $\UU(d)$
\begin{equation}
    A(\kappa,\lambda \kappa)=\lbrace{\U\in\UU(\dim)\ |\ \kappa\leq \d(\U,\h{I})\leq \lambda \kappa   \rbrace}\,.
\end{equation}
Then we have 
\begin{equation}\label{eq:unitRINGbound}
    \vol\left(A(\kappa,\lambda \kappa) \right) \leq (2 \lambda \kappa)^{\dim^2-1}(\lambda -1)\,. 
\end{equation}
\end{proposition}
\begin{proof}
We first recall that for subset $\mathcal{A}\subset \UU(\dim)$ we have $\vol(\mathcal{A})=\mu_{\mathbb{U}(\dim)}(\mathcal{A}')$, where $\mu_{\mathbb{U}(\dim)}$ is the Haar measure on $\mathbb{U}(\dim)$ and $\mathcal{A}'=\lbrace{V \in \mathbb{U}(\dim)| \V\in \mathcal{A} \rbrace}$. Consequently by setting $f(x)\coloneqq 4\arcsin\left(\frac{x}{2}\right)$ and using \cref{lem:distARCtranslate} we get
\begin{equation}
     \vol\left(A(\kappa,\lambda \kappa)\right)= \mu_{\mathbb{U}(\dim)} \left( \lbrace{V\in\mathbb{U}(\dim)\ |\ f(\kappa)\leq \arc(\Eig(V)) \leq f(\lambda \kappa)   \rbrace} \right) \,.
\end{equation}
Since $\arc(\Eig(V))$ is a class function on $\mathbb{U}(\dim)$, \ie $\arc(\Eig(WVW^\dagger))=\arc(\Eig(V))$ for all $V,W\in\mathbb{U}(\dim)$, we can use Weyl integration formula \cite{WallachBook} to rewrite the above expression in terms of integral over phases
\begin{equation}
     \vol\left(A(\kappa,\lambda \kappa)\right) = N_d \int_{ \bm{\alpha}\in[0,2\pi]^d :  f(\kappa) \leq\arc(\bm{\alpha}) \leq f(\lambda \kappa)  } \prod_{1\leq i<j\leq d}|\exp(\ii\alpha_i)- \exp(\ii\alpha_j)|^2   \dt \bm{\alpha}\,,
\end{equation}
where $\bm{\alpha}=(\alpha_1,\ldots,\alpha_d)$, $\dt \bm{\alpha} = \dt\alpha_1 \ldots \dt \alpha_d $, and $N_d=\frac{1}{d! (2\pi)^d}$. Using $|\exp(\ii\alpha_i)- \exp(\ii\alpha_j)|\leq 2\lambda\kappa$ (see Figure 3 of \cite{OSH2020}) we obtain
\begin{equation}\label{eq:ARCprobBOUND}
    \vol\left(A(\kappa,\lambda \kappa)\right) \leq \frac{(2\lambda \kappa)^{d(d-1)}}{d!} \Pr_{\bm{\alpha}}\left[ f(\kappa)\leq \arc(\bm{\alpha})\leq f(\lambda\kappa)\right]\,,
\end{equation}
where $\Pr_{\bm{\alpha}}$ denotes probability with respect to phases $\bm{\alpha}=(\alpha_1,\ldots,\alpha_d)$ uniformly and independently distributed on $S^1$. The distribution of the random variable $\arc(\alpha)$, the length of the shortest arc containing randomly and independently distributed points on a unit circle, has been studied previously in \cite[Section 3.6]{rao1969some}, where it was shown that for $r\leq \pi$
\begin{equation}\label{eq:arcDIST}
    \Pr_{\bm{\alpha}}\left[\arc(\bm{\alpha})\leq r\right] = d\left(\frac{r}{2\pi} \right)^{d-1}\,.
\end{equation}
Since $f(x)\leq x$ for $x\in[0,\sqrt{2}]$ we can apply the above formula to Eq.~\eqref{eq:ARCprobBOUND} and obtain 
\begin{equation}\label{eq:ineqSEMIfin}
     \vol\left(A(\kappa,\lambda \kappa)\right) \leq \frac{(2\lambda\kappa)^{d(d-1)}}{(2\pi)^{d-1}(d-1)!}\left(f(\lambda\kappa)^{d-1}-f(\kappa)^{d-1}\right)\,.
\end{equation}
To simplify the above formula we use that for $x>y\geq0$ we have the following inequalities
\begin{equation}\label{eq:simpleINEQ}
    x^{d-1} -y^{d-1} \leq  (d-1) x^{d-2} (x-y)\ ,\  \ f(x)-f(y)\leq 2 \sqrt{2} (x-y)\,,
\end{equation}
where the second bound follows from integral representation of $\arcsin$: $\arcsin(x)=\int_0^x \dt t \frac{1}{\sqrt{1-t^2}}$. Application of Eq.~\eqref{eq:simpleINEQ} in Eq.~\eqref{eq:ineqSEMIfin} finally gives
\begin{equation}
    \vol\left(A(\kappa,\lambda \kappa)\right) \leq \frac{(2\lambda\kappa)^{d(d-1)} f(\lambda \kappa)^{d-2}}{(2\pi)^{d-1}(d-2)!}\left(f(\lambda\kappa)-f(\kappa)\right) \leq \frac{(2\lambda\kappa)^{d^2-1} (\sqrt{2})^{d-1}}{(2\pi)^{d-1}(d-2)!\lambda}(\lambda-1) \leq (2\lambda \kappa)^{d^2-1}(\lambda-1)\,,
\end{equation}
where in the last inequality we used $\lambda\geq 1$.
\end{proof}

\section{Proofs for random quantum circuits and unitary designs}
\label{app:rqcproofs}

In this Section we prove the first half of \cref{prop:RQChighdegree} (equation \eqref{eq:time-for-gap-rqc}), which is the following.

\begin{proposition}[Spectral gap of $G$-local quantum circuits]
\label{th:gapGlocalCIRC}
Let $\delta>0$. Then $\Grqc$-local random quantum circuits on $n$ qudits with local dimension $q$ and depth $t$ satisfy $\|M_{\nu_\Grqc} -M_\mu \|_\infty \leq \delta  $ for:
\begin{equation}\label{eq:glocal-gap}
    t\geq  c_2 n^6 d^2 \log\left(\frac{1}{\delta}\right)
\end{equation}
where $c_2 = 10^5$  as in \cref{prop:RQChighdegree}.
\end{proposition}

We also describe a generalization of the random quantum circuit model that allows the use of a finite universal gateset instead of Haar random gates. This is formally captured by the following definition, which generalizes Definition \ref{def:glocRQC}. 

\begin{definition}[$(\Grqc, \G)$-local random quantum circuits]\label{def:glocRQC-gates}
Consider $n$ qudits defined on a graph $\Grqc$ which contains a Hamiltonian path.
Let $\nu_{\Grqc, \G}$ be the distribution for a single step of the random circuit, choosing a pair $(i,j)$ uniformly at random from the set of edges of $\Grqc$, and applying a random 2-site gate $g_{i,j}$, drawn uniformly at random from $\G$, to the pair of qudits $(i,j)$. Here $\G$ is a set of universal 2-qudit gates, which does \emph{not} necessarily contain inverses. Depth $t$ $(\Grqc, \G)$-local random quantum circuits are defined by the probability distribution  $\nu_{t}^{(\Grqc,\G)}  := (\nu_{\Grqc, \G})^{*t}$. 
\end{definition}

In this setup, we prove a statement analogous to \cref{th:gapGlocalCIRC}, but operating at the level of $k$-designs rather than the spectral gap.

\begin{proposition}[$(\Grqc, \G)$-local RQCs form approximate unitary designs]
\label{prop:GRQCdesigns}
Let $\delta>0$ and $4k\leq d^{2/5}$ (where $d=q^n$). Then $(\Grqc, \G)$-local random quantum circuits on $n$ qudits with local dimension $q$ form $\delta$-approximate unitary $k$-designs when the circuit depth is
\begin{equation}\label{eq:propGRQCdesigns-eq1}
    t\geq c(\G) n^3 \log^4(k) k^{9.5} \log(1/\delta)\,,
\end{equation}
where $c(\G)$ is a constant that depends on the gate set $\G$. 
For arbitrarily high moments, i.e.\ with no restriction on $k$, $(\Grqc, \G))$-local random quantum circuits on $n$ qudits form $\delta$-approximate unitary $k$-designs when the circuit depth is
\begin{equation}
\label{eq:time-crqc}
    t\geq c(\G) n^7 \log^2(k) \dim^2 \log(1/\delta)\,.
\end{equation}
\end{proposition}

Both propositions involve bounding the spectral gap of either the averaging operator or the $k$-moment operator for the model defined on a general graph $G$ that contains a Hamiltonian path. We first show that we can reduce the problem to studying the gap of the model on a 1D chain.

The spectral gap of the moment operators has a convenient reinterpretation as the spectral gap of a local Hamiltonian \cite{Zni08,BrownViola2010tdesign,BH13}. Moreover, this Hamiltonian is frustration-free, powerful techniques from many-body physics may be employed to lower bound the spectral gap, as in \cite{BH13,BHH2016,HHJ20}. We start by reviewing the Hamiltonian and its relation to the gap of the random circuit moment operators. 

For $n$ qudits on a graph $\Grqc$, we can define the corresponding local Hamiltonian acting on $n$ subsystems each of dimension $q^{2k}$
\begin{equation}
    H_{n,k}^{(\Grqc)} = \sum_{(i,j)\in \Grqc} h_{i,j} \with h_{i,j} := (\iden-P_H)_{i,j}\,,
\end{equation}
where the local term $h_{i,j}$ acts nontrivially only on sites $i$ and $j$, and $P_H$ is the $k$-fold Haar moment operator the 2-sites: $P_H = \int_{\UU(q^2)} \dt\mu(\U)\, U^{\otimes k}\otimes \bar{U}^{\otimes k}$. In the Hamiltonian $H_{n,k}^{(\Grqc)}$ we sum over all edges of the graph $\Grqc$. This Hamiltonian is translation-invariant, has ground state energy zero, and is frustration-free. Moreover, the dimension of the ground space is ${\rm dim}\ker (H^{(\Grqc)}_{n,k}) = k!$. Most importantly, the spectral gap of the Hamiltonian is directly related to the spectral gap of a single time step of the random quantum circuit with Haar-random local gates. In one dimension we have $g(\nu_n,k) = 1 - \Delta(H^{({\rm 1D})}_{n,k})/n$ as in Ref.~\cite{BHH2016}, which for a general interaction graph $\Grqc$ becomes $g(\nu_{\UU(q^2),\Grqc},k) = 1 - \Delta(H^{(\Grqc)}_{n,k})/|E|$, where $|E|$ is the cardinality of the edge set of the graph $\Grqc$. 

Note that the Hamiltonians $H^{\rm (1D)}_{n,k}$ and $H^{(\Grqc)}_{n,k}$ have the same ground space, spanned by permutations. As the graph $\Grqc$ contains a Hamiltonian path, up to relabeling of the vertices, $\Grqc$ contains the 1D chain. The $\Grqc$-local Hamiltonian can then be obtained by adding projectors to the 1D Hamiltonian. The operator inequality $H^{(\Grqc)}_{n,k}\geq H^{\rm (1D)}_{n,k}$ then implies that $\Delta(H^{(\Grqc)}_{n,k}) \geq \Delta(H^{\rm (1D)}_{n,k})$. Since $|E| \leq n^2$, we thus have:
\begin{equation}\label{eq:replace-G-with-1D}
    g(\nu_{\UU(q^2),\Grqc},k) = 1 - \frac{\Delta(H^{(\Grqc)}_{n,k})}{|E|} \leq 1 - \frac{\Delta(H^{\rm (1D)}_{n,k})}{|E|} \leq 1 - \frac{1}{n} \cdot \frac{\Delta(H^{\rm (1D)}_{n,k})}{n}
\end{equation}
which implies that the spectral gap can shrink by a factor of at most $n$ when replacing a general graph $G$ with a 1D chain.


Using the amplification property $g(\nu^{*t},k) = g(\nu,k)^t$, it follows that 1D random quantum circuits $\nu_{\UU(q^2),{\rm 1D}}$ with Haar random local gates form $\delta$-approximate designs when
\begin{equation}
        t \geq \frac{n}{\Delta(H^{\rm (1D)}_{n,k})} \log (1/\delta)\,.
        \label{eq:trqcgap}
\end{equation}

We now continue to proving the two propositions.

\begin{proof}[Proof of \cref{th:gapGlocalCIRC}]
First, we note that by \eqref{eq:spectral-gap} bounding the spectral gap for $M_{\nu}$ is equivalent to providing a bound on the gaps of $k$-moment operators $M_{\nu,k}$ which is independent of $k$. This follows from an exponentially small spectral gap, which in turn follows from a number of Lemmas in Ref.~\cite{BHH2016} and an improved bound in Ref.~\cite{Haferkamp22}. Using a method for bounding the mixing time of Markov chains, a technique called path-coupling and specifically a version for random walks on the unitary group, Lemmas 19 and 20 in Ref.~\cite{BHH2016} prove a $k$-independent lower bound on the spectral gap of $H^{\rm (1D)}_{n,k}$ of
\begin{equation}
\label{eq:Jonas}
    \Delta(H^{\rm (1D)}_{n,k}) \geq \frac{1}{n}\frac{1}{(e(q^2+1))^n}\,.
\end{equation}
The spectral gap is exponentially small, but holds for all moments $k$. For local random quantum circuits on $n$ qudits, the design depth is related to the spectral gap as in Eq.~\eqref{eq:trqcgap}. Noting that $q\geq 2$, this gap gives that RQCs generate expanders for all $k$ when the circuit depth is $t\geq n^2 d^{3.78}\log(1/\delta)$. More recently, Ref.~\cite{Haferkamp22} gave an improved lower bound for an exponentially-small but $k$-independent gap, proving that the spectral gap of $H^{\rm (1D)}_{n,k}$
\begin{equation}
    \Delta(H^{\rm (1D)}_{n,k}) \geq \frac{1}{\crqc n^4}\frac{1}{d^2}\,,
    \label{eq:expgapbound}
\end{equation}
where $\crqc = 10^5$. By applying this to \eqref{eq:trqcgap} and using the inequality \eqref{eq:replace-G-with-1D} we obtain \eqref{eq:glocal-gap}, which finishes the proof.
\end{proof}

For the proof of \cref{prop:GRQCdesigns} we will need to make use of the following theorem.

\begin{theorem}[RQCs form unitary designs \cite{BHH2016}]
\label{thm:RQCdesigns}
For $\delta>0$ and $4k\leq d^{2/5}$, local random quantum circuits on $n$ qudits form $\delta$-approximate unitary $k$-designs when the circuit depth is
\begin{equation}
    t\geq c_1 n \lceil \log (4k)\rceil^2 k^{9.5} \log(1/\delta)\,.
\end{equation}
where the constant is taken to be $c_1=4\times 10^7$. \end{theorem}
We note that there is evidence that the polynomial dependence on $k$ can likely be substantially improved \cite{NHJ19,HHJ20}. Specifically, it is known that if one takes the local dimension to be large (e.g.\ $q\geq 6k^2$) the spectral gap is constant and the design depth scales linearly in $k$. Quite recently, the exponent in \cref{thm:RQCdesigns} was improved from $k^{9.5}$ to $k^{4+o(1)}$ via an improved RQC spectral gap \cite{Haferkamp22}. 



\begin{proof}[Proof of \cref{prop:GRQCdesigns}]

For universal gate sets containing inverses and comprised of algebraic entries, the spectral gap of a single step of the walk is bounded independently of $k$ by results of \cite{Bourgain2011}, so in fact we obtain the same result as in \cref{th:gapGlocalCIRC} (using the same proof) at the expense of a constant $c(\G)$ which depends on the gate set. Following the results of Refs.~\cite{Varju2013,OSH2020}, we can drop the restrictions on the universal gate set at the additional\label{eq:propGRQCdesings-eq1} expense of polynomial factors (Result 5 in \cite{OSH2020}). Specifically, we have for the spectral gaps of the distributions for a single step of the walks:

\begin{equation}
    \big(1-g\big(\nu_{(\Grqc, \G)},k\big)\big) \geq \frac{1}{c(\G)n\log^2(k)}\big(1-g\big(\nu_{(\UU(q^2),\Grqc)},k\big)\big)\,.
\end{equation}
Using \eqref{eq:replace-G-with-1D}, we conclude is that $(G,\G)$-local random quantum circuits on $n$ qudits form $\delta$-approximate unitary $k$-designs, in operator norm, when the circuit depth is
\begin{equation}
    t\geq c(\G) \log^2(k) \frac{n^3}{\Delta(H^{\rm (1D)}_{n,k})}\log(1/\delta)\,,
\end{equation}
Using the bounds on the 1D spectral gap from Ref.~\cite{BHH2016} (see \cref{thm:RQCdesigns}) for \eqref{eq:propGRQCdesigns-eq1} and the exponentially small gap in Eq.~\eqref{eq:expgapbound} for \eqref{eq:time-crqc}, the claims in the proposition then follow.
\end{proof}

\section{Technical properties of the Stochastic Local Hamiltonian model}\label{app:slh}

This Appendix contains a number of technical properties of the SLH model needed to establish results analogous to those for random quantum circuits. First, we prove that the SLH model enjoys virtually the same spectral gap as the random quantum circuit model based on a specific gateset. This is proved in \cref{th:gap-SLH}, basically following the steps laid out in \cite{Onorati17}. Next, in \cref{prop:rec-SLH} we prove a recurrence bound necessary for the results on saturation and recurrence to hold for the SLH model. The proof relies on a certain stability result (appearing in main text as \cref{thm:continuity}), which is proved in the last subsection of this Appendix using matrix martingale methods.

\subsubsection*{Spectral gap for Stochastic Local Hamiltonians}

This subsection contains the proof of the second half of \cref{prop:RQChighdegree}, namely the following statement.

\begin{proposition}[Spectral gap of the Stochastic Local Hamiltonian Model]
\label{th:gap-SLH}
Let $\delta>0$. Then the SLH model on $n$ qudits with local dimension $q$ after time $t$ satisfies $\|M_{\nuslh} -M_\mu \|_\infty \leq \delta  $ for:
\begin{equation}\label{eq:slh-gap}
    t\geq  4c_2 n^4 d^2 \log\left(\frac{1}{\delta}\right)
\end{equation}
where $c_2 = 10^5$  as in \cref{prop:RQChighdegree}.
\end{proposition}

\begin{proof}
We will first prove the statement for the case when the interaction graph $G$ is the 1D chain. Obviously it suffices to prove the following inequality:
\begin{equation}
    \|M_{\nuslh}-M_\mu\|_\infty
    \leq \exp\biggl(-   \frac{t}{4c_2 n^4d^2}\biggr)
\end{equation} 
Throughout the proof, the operators : $M_{\nuslh,k}$ and $M_{\nuslh}$ will be denoted by $M_{SLH(t),k}$ and $M_{SLH(t)}$, respectively.

By \eqref{eq:spectral-gap}, it is enough to provide an upper bound for $\|M_{SLH(t),k}-M_{\mu,k}\|_\infty$ independent of $k$. We closely follow the proofs from \cite{Onorati17} (Section 5), which we sketch for reader's convenience. First, we represent the moment operator at time $t$ in terms of a composition of moment operators for times $\Delta t$:
\begin{align}
    M_{SLH(t),k}=M_{SLH(\Delta t),k}^{t/{\Delta t} } 
\end{align}
which holds because SLH is a Markov process. We thus have: 
\begin{align}
\label{eq:Mslh}
    \|M_{SLH(t),k}-M_{\mu,k}\|_\infty=
    \lim_{\Delta t\to 0}\|M_{SLH(\Delta t),k}^{t/\Delta t}-M_{\mu,k}\|_\infty=
    \lim_{\Delta t\to \infty}\|M_{SLH(\Delta t),k}-M_{\mu,k}\|^{t/\Delta t}_\infty
    \end{align}
    where the last equality comes from the fact that since $\mu$ is the Haar measure, $M_{\mu,k}=M^n_{\mu,k}$ for any power $n$ and $M_{\mu,k}$ is a projector onto the eigenspace of $k!-1$ largest eigenvalues of $M_{SLH(t),k}$ (which are equal to $1$). 
    
    Now, the moment operator with time $\Delta t$ will be approximated by the moment operator for random quantum circuits, with gates $G(\theta ^{(e)}_{\Delta t})$ defined in \cite{Onorati17}  before Eq. (72). In the original paper $e$ denotes the edge of the graph. Here we consider the 1D chain, so $e$ denotes just a pair of neighboring qudits. We denote the corresponding moment operator by $\Mgates$. Thus $\Mgates$ is the moment operator for a random circuit where one picks at random a neighboring pair of qudits and applies to the pair a gate taken from the ensemble $G(\theta ^{(e)}_{\Delta t})$. Now,  Eq. (71) in \cite{Onorati17} says that: 
    \begin{align}
        M_{SLH(\Delta t),k}=\Mgates+O(\Delta t^2)
    \end{align}
    We then have 
    \begin{align}
    \label{eq:Mgates}
    &\|M_{SLH(\Delta t),k}-M_{\mu,k}\|_\infty  \leq 
    \|M_{SLH(\Delta t),k}-\Mgates\|_\infty +\|\Mgates-M_{\mu,k}\|_\infty \\ \nonumber  
    &= \|\Mgates-M_{\mu,k}\|_\infty+O(\Delta t^2).
    \end{align}
    Let us denote:
    \begin{align}
    \label{eq:gap-notation}
    \gap(\nu,k)=1- \|M_{\nu,k}-M_{\mu,k}\|_\infty    
    \end{align}
    Let us denote by $m_{\Delta t,k}$ the moment operator for just two qudits, with measure given by the ensemble $G(\theta ^{(e)}_{\Delta t})$. 
    We can then apply the
    estimate from  Lemma 16 of \cite{Onorati17} (appearing implicitly in \cite{BHH2016}, Corollary 7)
    that relates the gap for random circuits composed from an ensemble of gates with the gap 
    of random quantum circuits composed from Haar random gates and the gap of the ensemble of gates itself:
    \begin{align}
    1- \|\Mgates-M_{\mu,k}\|_\infty\equiv 
    \gap(\gates,k) \geq \gap(G(\theta^{(e)}_{\Delta t}),k)\,\gap({\rm RQC}, \mu,k)
    \end{align}
    where according to the notation in \eqref{eq:gap-notation}:
    \begin{align}
        \gap(G(\theta^{(e)}_{\Delta t}),k)=1-\|m_{\Delta t,k}-m_{\mu,k}\|_\infty
    \end{align}
 
Regarding the first gap, it was shown in \cite{Onorati17} that:
\begin{align}\label{eq:slh-gap-E}
\gap_1\equiv\gap(G(\theta^{(e)}_{\Delta t}),k)= \frac{1}{2} |E| \Delta t+O(\Delta t^2)
\end{align} 
(in 1D case $|E|=n-1$, where $n$ is number of qudits). The other gap is estimated in \cite{BHH2016}.  Here we use more optimal estimate from \cite{Haferkamp22}. mentioned in Section \ref{app:rqcproofs}, which gives: 
\begin{align}
\gap_2\equiv    \gap(RQC,\mu,k)\geq \frac{1}{c_2 n^5} \frac{1}{d^2}
\end{align}
with $c_2=10^5$.

We thus arrive at the formula:
\begin{align}
    \|\Mgates-M_{\mu,k}\|_\infty\leq 
    1- \gap_1\gap_2
\end{align}
which by \eqref{eq:Mslh} and \eqref{eq:Mgates} implies:
\begin{align}
\label{eq:Mslh2}
    &\|M_{SLH(t),k}-M_{\mu,k}\|_\infty\leq
    \lim_{\Delta t\to \infty}(1-\gap_1\gap_2
  )^{t/\Delta t}=\\ \nonumber 
    &=\lim_{\Delta t\to \infty}\biggl(1- \Delta t \,(n-1)\frac{1}{2}\cdot \frac{1}{c_2 n^5 d^2}
  \biggr)^{t/\Delta t} \leq 
  \exp\biggl(-  \frac{t}{4c_2 n^4 d^2}\biggr)
    \end{align}
This finishes the proof for the case of the 1D chain. To prove the proposition for general graphs $G$, we appeal to the same reasoning as in the proof of \cref{prop:GRQCdesigns} (equation \eqref{eq:replace-G-with-1D}), which implies that the spectral gap for a general graph with $|E|$ edges is at most $|E|/n$ times , so the two occurences of $|E|$ cancel out and we end up with the same result as for the 1D case.
\end{proof}

\subsubsection*{Bounds for recurrence for the SLH model}

We first need a technical lemma:
\begin{lemma}[Prob. of recurrence in continuous walk versus one in discrete walk]
\label{lem:reccur-AD}
Let $\h{U}_t$ be the SLH random walk. 
Consider an interval $[t_1,t_2]$ and let $\Deltat=(t_2-t_1)/K$, where $K$ is a natural number.
We then have: 
\begin{align}
\label{eq:large-compl}
   \Prob(\exists_{t\in[t_1,t_2]} \complhalf(\U_t) \leq r) \leq \Prob(\exists_{t\in\{t_1+ \Deltat,\ldots,t_1+ K\Deltat\} }\,\compl(\U_t)\leq r) + K \cdot \Prob(\max_{t\in[0,\Deltat]} \d(\U_t,I)> \ep/2)
\end{align}    
\end{lemma}
\begin{proof}
Let us define the following events:
\begin{align}
    &A_r^\ep=\{\forall_{t\in\{t_1+ \Deltat,\ldots,t_1+ K\Deltat\} }\,\compl(\U_t) \geq r\} \\
    &D^{\ep/2}=\{\max_{t\in[0,\Deltat]} \d(\U_t,I)\leq \ep/2\} \\
    &D_l^{\ep/2} =\{\forall_{t_\in[t_1+(l-1)\Deltat,t_1+l \Deltat]} 
\d(\U_t,\U_{t_1+l\Deltat})\leq \ep/2\} \\
&B_l^{\ep/2}=\{\exists_{t\in[t_1+l\Deltat,t_1+(l+1)\Deltat]} \complhalf(\U_t) \leq r \} \\ 
   & B^{\ep/2}= \bigcup_{l}B_l^{\ep/2}= \{\exists_{t\in[t_1,t_2]} \complhalf(\U_t) \leq r \}
\}
\end{align}
Let us first note that:
\begin{align}\label{eq:bda}
    B_l^{\ep/2}\cap D_l^{\ep/2} \cap A_{r+1}^\ep=\emptyset
\end{align}
Indeed, by the definition of complexity, if $\compl(U) > r$ and $\d(V,U)\leq \ep/2$, then $\complhalf(V) > r$. Now, $A_{r+1}^\ep$ implies that the unitary $\U_{t_1+l\Deltat}$ 
has $\ep$-complexity at least $r+1$. The event $D_l^{\ep/2}$ implies that any unitary from the interval 
 $[t_1 + (l-1) \Deltat, t_1+l \Deltat ]$ 
 is $\ep/2$ close to the above unitary, hence
 by the definition of complexity it must be more than $\ep/2$ far from any
 word of length $r$. 
 Therefore it
must have $\ep/2$-complexity larger than $r$, hence it  cannot be in the set $B_l^{\ep/2}$.

We now proceed with bounding $Pr(B^{\ep/2})$:
\begin{align}
    &\Prob(B^{\ep/2})\leq Pr(\overline A_{r+1}^{\ep}) + \Prob(B^{\ep/2}\cap A_{r+1}^{\ep})
    \leq 
    \Prob(\overline A_{r+1}^{\ep}) + \sum_{l=1}^K \Prob(B_l^{\ep/2}\cap A_{r+1}^{\ep}) 
    \leq \\
    &\Prob(\overline A_{r+1}^{\ep})+\sum_{l=1}^K\bigl( \Prob(\overline D_l^{\ep/2}) + \Prob(B_l^{\ep/2}\cap D_l^{\ep/2} \cap A_{r+1}^{\ep}) \bigr)=
    \Prob(\overline A_{r+1}^{\ep})+ 
    \sum_{l=1}^K \Prob(\overline D_l^{\ep/2})
\end{align}
where the last equality follows by \eqref{eq:bda}.
Finally, due to Markovianity we have: 
\begin{align}
    \Prob(D_l^{\ep/2}) = \Prob(D^{\ep/2})
\end{align}
which gives: 
\begin{align}
    \Prob(B^{\ep/2})\leq \Prob(\overline A_{r+1}^{\ep})+ K \cdot \Prob(\overline D^{\ep/2}).
\end{align}
as desired.
\end{proof}

We are in position to prove bound on probability of recurrence of SLH random walk.
\begin{proposition}[Probability of recurrence for SLH model]
\label{prop:rec-SLH}
     Let $t_2\geq t_1\geq \tau(SLH,\ep)$ where $\tau(SLH,\ep)$ is the equidistribution time for the  SLH model (\cref{def:equidistibTIME}). For the SLH model 
    we then have: 
    \begin{align}
    \label{eq:prop-rec-SLH}
    \Pr\big(\exists_{t\in[t_1,t_2]} \compl(\U_t) \leq r \big)\leq    
    64(t_2-t_1) m \cdot d^2 \left(|\gset|^{r+2} + 1\right) (2\cupper \beta \ep)^{d^2-2}\,.
    \end{align}
\end{proposition}
\begin{proof}
Let $s \geq 1$ be a number to be chosen later. We divide the interval $[t_1,t_2]$ into $K$ intervals of length $\Deltat = \frac{\ep}{2 m s}$ each.
We then use
\cref{thm:continuity}
with $x=\frac{m \Deltat s}{2}$, which gives:
\begin{align}
    \Prob\left(\max_{t\in[0,\Deltat]}\d(\U_t,I)>m \Deltat s\right)\leq 2 d \exp\left(-\frac{m \Deltat s^2}{8}\right)
\end{align}
By plugging $\Deltat$ we then get
\begin{align}    \Prob\left(\max_{t\in[0,\Deltat]}\d(\U_t,I)>\frac{\ep}{2}\right)\leq 2 d \exp\left(-\frac{\ep s}{16}\right)
\end{align}
By choosing $\Deltat$ as above 
we have $K=(t_2-t_1)2 m s/\ep$.
Next, from \cref{lem:bouns-rec-events}
we get: 
\begin{align}   \Prob(\exists_{k\in\{1,\ldots,K\} }\compl(\U_{\tau+k\Deltat})\leq r+1)\leq
    K |\gset|^{r+2} \vol(\beta \ep)\,,
\end{align}
Using \cref{lem:reccur-AD} we thus get:
\begin{align}
    Pr\left(\exists_{t\in[t_1,t_2]} \complhalf(\U_t \leq r) \right) \leq 
    K\left(|\gset|^{r+2} \vol(\beta \ep)+
    2d \exp\left(-\frac{\ep s}{16}\right)\right)
\end{align}
Now, we set:
\begin{align}
s=\frac{16}{\ep}\left(\max\left\{(d^2-1) \log\frac{1}{\cupper \beta\ep}, 0\right\} + \log(2d)\right)
\end{align}
where clearly $s \geq 1$. If $\cupper \beta\ep < 1$, then since $K=(t_2-t_1)2 m s /\ep$, by using \eqref{eq:szarek} we arrive at:
\begin{align}
    Pr\left(\exists_{t\in[t_1,t_2]} \complhalf(\U_t \leq r) \right) \leq 
    32(t_2-t_1)  m \left((d^2-1) \log\frac{1}{\cupper \beta\ep} + \log(2d)\right) \left(|\gset|^{r+2} + 1\right) (\cupper \beta \ep)^{d^2-1}
\end{align}
By using $\log(x) \leq x$ combined with $\cupper \beta\ep < 1$ and $\log(2d) < (d^2-1) \frac{1}{\cupper \beta\ep}$, we end up with:
\begin{align}
    Pr\left(\exists_{t\in[t_1,t_2]} \complhalf(\U_t \leq r) \right) \leq 
    64(t_2-t_1) m \cdot d^2 \left(|\gset|^{r+2} + 1\right) (\cupper \beta \ep)^{d^2-2}
\end{align}
which gives the required estimate Eq.~\eqref{eq:prop-rec-SLH} after substituting $\ep/2 \to \ep$. The case $\cupper \beta\ep \geq 1$ is handled similarly by estimating $\vol(\beta\ep) \leq 1 \leq (\cupper\beta\ep)^{d^2-2}$ and $\log(2d) < d^2$.
\end{proof}

\subsubsection*{Stability properties for Stochastic Local Hamiltonians}

We present the necessary technical results to prove \cref{thm:continuity}. We assume basic familiarity with stochastic processes, i.e. the notion of an SDE in Ito's form, martingales and quadratic variation of a process.

It will be convenient to use an alternative description of the SLH process $U_t$ from \eqref{eq:slh-definition} using a stochastic differential equation (SDE).
\begin{proposition}\label{prop:sde}
    The SLH process $U_t$ satisfies the following SDE with $U_0 = Id$:
    \begin{equation}\label{eq:sde}
            dU_t =  U_t d\Theta_t - \frac{m}{2} U_t dt
    \end{equation}
where $\Theta_t$ is the Brownian motion in the Lie algebra arising as the $\Delta t \to 0$ limit of increments in \eqref{eq:theta} and $m = \abs{E}$ is the number of edges in the interaction graph.
\end{proposition}

\begin{proof}
By basic properties of the construction of $U_t$ (follows e.g. from \cite{McKean}, Chapter 4) the process will satisfy an SDE:
\begin{equation}
    dU_t = U_t d\Theta_t - \frac{1}{2}U_t Q dt
\end{equation}
where $\Theta_t$ is the underlying Brownian motion in the Lie algebra and $Q$ is the quadratic variation of $\Theta_t$, which for a complex valued matrix process is equal to $d\Theta_t \cdot d\Theta^{\dagger}_t$. By combining equations \eqref{eq:theta}, \eqref{eq:small-theta} and using the covariance structure described by \eqref{eq:covariance}, we easily see that:
\begin{equation}\label{eq:slh-Q}
    Q = \sum_{e,e',\mu, \mu'}A^{(e)}_{\mu}A^{(e')}_{\mu'}  \delta_{e,e'} (\kappa^{-1})_{\mu,\mu'}
\end{equation}
Since $\kappa$ is the Killing metric, the sum over $(e,\mu), (e',\mu')$ is then simply equal to the sum over $e$ of the local Casimir elements for the Lie algebra $i\cdot\mathfrak{u}(q^2)$, which are equal to $I$ each, which finishes the proof.
\end{proof}

From standard properties of Ito processes (e.g. Dynkin's martingale formula) it follows that the process:
\begin{equation}\label{eq:martingale}
    Z_t := U_t - I + \frac{m}{2} \int_{0}^{t}U_s ds
\end{equation}
is a (matrix-valued) martingale. We will now use matrix martingale concentration tools from \cite{bacry} and \cite{tropp} to prove concentration for this martingale, which will result in the proof of \cref{thm:continuity}.

Following \cite{bacry}, for an $n \times m$ matrix $X$ let $\dilation(X)$ be the symmetric extension of $X$ given by the $n+m \times n+m$ matrix:
\begin{equation}
\dilation(X) :=
\begin{pmatrix}
0 & X \\
X^{\dagger} & 0
\end{pmatrix}
\end{equation}

\begin{lemma}\label{lm:qv}
For the martingale $Z_t$ defined in \eqref{eq:martingale}, the quadratic variation of the sum of columns of the symmetric extension of $Z_t$ is equal to:
\begin{equation}
\sum_{j=1}^{2d}\qv{\dilation (Z)_{\blackdot,j}}_t = m t \cdot I
\end{equation}
where $I$ is the $2d \times 2d$ identity matrix.
\end{lemma}

\begin{proof}
We mimic the computations on page 21 of \cite{bacry}. We will need to compute a subset of quadratic covariations between entries of $Z_t$, namely:
\begin{equation}\label{eq:sumj}
    \sum_{j=1}^{2d}d [Z_{kj}, \overline{Z_{lj}}]
\end{equation}
The complex conjugation appears since, unlike in \cite{bacry}, the entries are complex valued. Since the drift part of \eqref{eq:martingale} has quadratic variation zero (since it is continuous), the q.v. of $Z_t$ is the same as the q.v. of $U_t$, which we now compute.  Note that the expression in equation \eqref{eq:sumj} (with $Z_t$ replaced by $U_t$) is exactly equal to $(dU_t \cdot dU_t^{\dagger})_{kl}$. This matrix is easily computed using the SDE \eqref{eq:sde} and proceeding as in the proof of Proposition \ref{prop:sde}:
\begin{equation}
dU \cdot dU^{\dagger} = Q dt
\end{equation}
where $Q$ is given as in Eq.~\eqref{eq:slh-Q}. By putting these results together we obtain:
\begin{equation}
dU \cdot dU^{\dagger} =  m I dt
\end{equation}
so by equations from page 21 of \cite{bacry}:
\begin{equation}
    \sum_{j=1}^{2d}\qv{\dilation(Z)_{\blackdot,j}}_t = m t \cdot I
\end{equation}
\end{proof}

\begin{corollary}\label{cor:martingale}
For any $t \geq 0$, we have
\begin{equation}\label{eq:st}
\BE [S_t(\xi)] \leq 2d \qquad{\rm for}\qquad 
S_t(\xi) := \tr \exp\left( \xi \dilation (Z_t) - \frac{\xi^2}{2}m t I \right) 
\end{equation}

\end{corollary}

\begin{proof}
The proof of this claim follows from the last statement on page 21 in \cite{bacry}, since in our case by Lemma \ref{lm:qv} we have $V_t = mt I$.
\end{proof}

\begin{proof}[Proof of \cref{thm:continuity}]
First, note that directly from the definition \eqref{eq:distancesDEF} if $U, V$ are any unitaries corresponding to channels $\h{U}, \h{V}$, then $\d(\h{U}, \h{V}) \leq \norm{U - V}_{\infty}$. Thus, it suffices to prove the theorem for the distance induced by the operator norm $\norm{\cdot}_{\infty}$ instead of the distance $\d$.

We mimic the proof of Theorem 2.3. from \cite{tropp}. Let $Z_t$ be the martingale defined in \eqref{eq:martingale} and let $\dilation(Z_t)$ be its symmetric extension. Since $\norm{U_t}_{\infty}=1$, we have:
\begin{equation}
\norm{Z_t}_{\infty} = \norm{U_t - I + \frac{m}{2}\int_{0}^{t}U_s ds}_{\infty} \geq \norm{U_t - I}_{\infty} - \frac{mt}{2} 
\end{equation}
so if $\norm{U_t - I}_{\infty} > \frac{m t}{2} + x$, then $\norm{Z_t}_{\infty} > x$, so it suffices to bound the latter probability. Since the spectral norm of $Z_t$ and its symmetric extension $\dilation (Z_t)$ are the same, we aim to bound the latter.

Since $U_t$ is a continuous process, it suffices to prove the inequality \eqref{eq:maximal-inequality} for any dense countable subset of $[0,s]$, which we denote by $\{t_k\}_{k=0}^{\infty}$. Let $\kappa$ be the stopping time defined as
\begin{equation}
\kappa := \inf\{t_k \geq 0 \ \vert \ \lambda_{max}(\dilation(Z_{t_k})) \geq x \}
\end{equation}
let $E_k$ be the event that $\lambda_{max}(\dilation(Z_{t_k})) \geq x$ and let $E = \cup_{k}E_k$. On the event $E$ the stopping time $\kappa$ is finite and by Lemma 2.2. of \cite{tropp} applied to $S_{\kappa}(\xi)$ (defined in Eq.~\eqref{eq:st}) we have:
\begin{equation}
S_{\kappa}(\xi) \geq \exp\left( \xi x - \frac{\xi^2}{2}mt \right)
\end{equation}
This follows by putting $w=m s$ in the Lemma and noting that in our case the process $W_t$ is simply $mt\cdot I$. We then apply the chain of inequalities as in the proof of Theorem 2.3 of \cite{tropp} and use Corollary \ref{cor:martingale}, to arrive at:
\begin{equation}
\Pp(E) \leq 2d \cdot \exp\left(-\xi x + \frac{\xi^2}{2}m t\right)
\end{equation}
Minimizing the right hand side over $\xi$ we arrive at:
\begin{equation}
\Pp\left( \exists_k \norm{\dilation(Z_{t_k})}_{\infty} > x \right) \leq 2d \cdot \exp\left( -\frac{x^2}{2ms}\right)
\end{equation}
which finishes the proof.
\end{proof}

\bibliographystyle{utphys}
\bibliography{comprefs}

\end{document}